\documentclass[11pt, letterpaper]{article}
\RequirePackage{bm}
\usepackage{graphicx}

\usepackage[authoryear]{natbib}
\usepackage{delarray}
\usepackage{hyperref}
\usepackage{amsthm}
\usepackage{rotating}
\usepackage{url}
\usepackage{colortbl,color}
\usepackage{anysize}
\usepackage{multirow,amssymb,amsmath, amsfonts}
\usepackage{comment}
\usepackage{caption}
\usepackage{subcaption}
\usepackage{enumitem}
\usepackage{float}

\usepackage{ dsfont }
\usepackage{varwidth}

\usepackage{algpseudocode}
\usepackage{tikz}

\newtheorem{thm}{Theorem}[section]

\newtheorem{coro}{Corollary}[section]
\newtheorem{lemma}{Lemma}[section]

\newtheorem{definition}{Definition}[section]

\newtheorem{prop}{Proposition}[section]

\def\boxit#1{\vbox{\hrule\hbox{\vrule\kern6pt\vbox{\kern6pt#1\kern6pt}\kern6pt\vrule}\hrule}}

\def\argmin{\mathop{\rm argmin}}
\def\tr{\mathop{\rm tr}}

\def\sgn{\mathop{\rm sign}}

\allowdisplaybreaks

\usepackage{tikz}
\newcommand*\circled[1]{\tikz[baseline=(char.base)]{
            \node[shape=circle,draw,inner sep=.6pt] (char) {#1};}}

\def\expv{\mathbb{E}}
\def\cov{\mathop{\rm cov}}
\def\var{\mathop{\rm var}}

\newcommand{\Xb}{\bm{X}}
\newcommand{\Zb}{\bm{Z}}
\newcommand{\Ub}{\bm{U}}
\newcommand{\Sb}{\bm{S}}
\newcommand{\Yb}{\bm{Y}}
\newcommand{\Wb}{\bm{W}}

\newcommand{\fb}{\bm{f}}

\newcommand{\bB}{{\bf B}}
\newcommand{\bx}{{\bf x}}

\newcommand{\bv}{{\bf v}}
\newcommand{\ba}{{\bf a}}

\newcommand{\bY}{{\bf Y}}

\newcommand{\bI}{{\bf I}}

\newcommand{\bM}{{\bf M}}

\newcommand{\bW}{{\bf W}}
\newcommand{\bH}{{\bf H}}

\newcommand{\bA}{{\bf A}}
\newcommand{\bu}{{\bf u}}

\newcommand{\bzero}{{\bf 0}}

\newcommand{\bSigma}{\boldsymbol{\Sigma}}
\newcommand{\hbSigma}{\widehat{\bSigma}}

\newcommand{\bmu}{\boldsymbol{\mu}}
\newcommand{\hbmu}{\widehat{\bmu}}
\newcommand{\bOmega}{\boldsymbol{\Omega}}
\newcommand{\hbOmega}{\widehat{\bOmega}}

\newcommand{\bLambda}{\boldsymbol{\Lambda}}

\newcommand{\T}{{\rm T}}

\newcommand{\bGamma}{\boldsymbol{\Gamma}}

\newcommand{\real}{I\kern-0.37emR}
\definecolor{annotxt}{rgb}{0,0,1}
\def\argmin{\mathop{\rm argmin}}

\def\bbr{ \mathbb{R}}
\newcommand{\beq}{\begin{equation}}
\newcommand{\eeq}{\end{equation}}
\newcommand{\beqr}{\begin{eqnarray}}
\newcommand{\eeqr}{\end{eqnarray}}
\newcommand{\beqrn}{\begin{eqnarray*}}
\newcommand{\eeqrn}{\end{eqnarray*}}

\newcommand{\sn}{\sum_{i=1}^n}
\makeatletter

\newcommand{\Rmnum}[1]{\expandafter\@slowromancap\romannumeral #1@}
\makeatother
\newcommand{\e}{\mathbb{E}}

\newcommand{\bbS}{\mathbb{S}}
\newcommand{\rI}{\mathrm{I}}
\newcommand{\rM}{\mathrm{M}}
\newcommand{\rB}{\mathrm{B}}

\newcommand{\E}{\mathbb{E}}

\begin{document}
\title{Higher Moment Estimation for Elliptically-distributed Data: Is it Necessary to Use a Sledgehammer to Crack an Egg?
}

\author{Zheng Tracy Ke\footnote{Department of Statistics, Harvard University, Cambridge, MA 02138.}, ~Koushiki Bose\footnote{Department of Operations Research and Financial Engineering, Princeton University, Princeton, NJ 08544.}, ~Jianqing Fan\footnote{Department of Operations Research and Financial Engineering, Princeton University, Princeton, NJ 08544.  Fan's research is supported by NSF grants DMS-1712591 and DMS-1662139 and NIH
grant R01-GM072611.}}


\date{}

\maketitle

\begin{abstract}
Multivariate elliptically-contoured distributions are widely used for modeling economic and financial data.
We study the problem of estimating moment parameters of a semi-parametric elliptical model in a high-dimensional setting. Such estimators are useful for financial data analysis and quadratic discriminant analysis.


For low-dimensional elliptical models, efficient moment estimators can be obtained by plugging in an estimate of the precision matrix. Natural generalizations of the plug-in estimator to high-dimensional settings perform unsatisfactorily, due to estimating a large precision matrix.
Do we really need a sledgehammer to crack an egg? Fortunately, we discover that moment parameters can be efficiently estimated without estimating the precision matrix in high-dimension.

We propose a marginal aggregation estimator (MAE) for moment parameters. The MAE only requires estimating the diagonal of covariance matrix and is convenient to implement. With mild sparsity on the covariance structure, we prove that the asymptotic variance of MAE is the same as the ideal plug-in estimator which knows the true precision matrix, so MAE is asymptotically efficient. We also extend MAE to a block-wise aggregation estimator (BAE) when estimates of diagonal blocks of covariance matrix are available. The performance of our methods is validated by extensive simulations and an application to financial returns.
\end{abstract}


\section{Introduction}
The classical multivariate statistics is largely motivated by relaxing the Gaussian assumption, which is not satisfied in many applications.
There is an extensive literature in finance on the tail-index estimates of stock returns;
while being unimodal and symmetric, the empirical returns exhibit leptokurtosis, which means that they have heavier tails and flatter peaks than those of normal data \citep{fama, Bollerslev, keller, frahm03, cizekbook}.
Empirical evidence of the violation of Gaussian assumption has also been observed in genomics \citep{liu03, posekany,hardin} and in bioimaging \citep{ruttimann}. The family of multivariate elliptically contoured distributions \citep{kelker}, which we shall call elliptical distributions in short, provides a natural generalization of multivariate Gaussian distributions. Recently, many statistical methods for elliptically distributed data have been proposed, including works on covariance matrix estimation \citep{fan2018}, graphical modeling \citep{liu2012b}, classification \citep{fan2013quadro}, etc.

The elliptical distributions are typically used as a semi-parametric model.
Given a mean vector $\bmu = (\mu_1,\ldots ,\mu_p)^\T \in \bbr^p$, a covariance matrix $\bSigma = (\sigma_{jk} )_{1\leq j, k\leq p } \in \bbr^{p\times p}$ and a probability characteristic function $\phi: [0,\infty)\to \mathbb{R}$, we say a random vector $\Yb=(Y_1, \ldots, Y_p)^\T$ has an elliptical distribution $\mathcal{E}(\bmu,\bSigma,\phi)$ if
\beq \label{Ydecomposition}
    \Yb =  \bmu+\xi \, \bSigma^{1/2} \Ub,
\eeq
where $\Ub$ is a random vector that is uniformly distributed on the unit sphere $\bbS^{p-1}$, and independent of $\Ub$, $\xi$ is a nonnegative random variable  whose characteristic function is $\phi$. For model identifiability, we normalize $\xi$ such that
\beq \label{normalization}
\expv( \xi^2 ) = p.
\eeq
Under \eqref{Ydecomposition}-\eqref{normalization},  $\bmu$ and $\bSigma$ are the mean vector and covariance matrix of $\Yb$, respectively. The variable $\xi$ determines which sub-family the distribution belongs to. When $\xi^2$ is a chi-square random variable, it belongs to the multivariate Gaussian sub-family, and when $\xi^2$ follows an $F$-distribution, it belongs to the multivariate $t$ sub-family or multivariate Cauchy sub-family. For most applications, the sub-family of the elliptical distribution is unknown, leaving the distribution of $\xi$  unspecified.

Although full knowledge of the distribution of $\xi$ is often not required, an estimate of its moment parameters is useful to statistical analysis and for understanding the tail of the distributions. One application is in quadratic classification. When data from two classes both follow elliptical distributions but have unequal covariance matrices, \cite{fan2013quadro} showed that an estimate of $\mathbb{E}(\xi^4)$ is desired for building a quadratic classifier. Another application is to capture the tail behavior of financial returns by estimating the leptokurtosis. Modeling the returns of a set of financial assets by an elliptical distribution, the leptokurtosis equals to $\{ p (p+2) \}^{-1} \expv( \xi^{4} )-1$, so the problem reduces to estimating $\mathbb{E}(\xi^4)$.


For any $m\geq 1$, define the  $m$-th scaled even moment of $\xi$ by
\beq \label{moment}
\theta_m \equiv p^{-m} \expv ( \xi^{2m} ).
\eeq
The first scaled even moment $\theta_1$ is $1$. In this paper, we are interested in estimating $\theta_m$ for any fixed $m\geq 2$, given independent and identically distributed (i.i.d.) samples $\Yb_1,\cdots, \Yb_n$ from \eqref{Ydecomposition}.

\subsection{The plug-in estimators}
We consider an ideal case where $(\bmu,\bSigma)$ are known.
Given $iid$ samples $\Yb_1,\cdots, \Yb_n$ from an unknown elliptical distribution, each $\Yb_i$ has a decomposition  $\Yb_i = \bmu + \xi_i\bSigma^{1/2}\Ub_i$, and $\xi_1,\ldots,\xi_n$ are $iid$ copies of $\xi$. Using the fact that $\Ub_i$ takes values on the unit sphere, we observe $\xi_i^2 = (\Yb_i-\bmu)^\T \bOmega(\Yb_i-\bmu)$ for $i =1, \ldots, n$, where $\bOmega\equiv \bSigma^{-1}$. Hence, in the ideal case, $\xi_1, \ldots, \xi_n$ are directly observed. It motivates the following estimator of $\theta_m$:
\beq\label{theta-oracle}
\widehat{\theta}_m^{\, \rI}(\bmu,\bOmega) =\frac{1}{n p^m}\sum_{i=1}^n (\xi_i^2)^{m}= \frac{1}{np^{m}}\sum_{i=1}^n \{ (\Yb_i-\bmu)^\T \bOmega(\Yb_i-\bmu) \}^m.
\eeq
We call $\widehat{\theta}_m^{\, \rI}(\bmu,\bOmega)$ the {\it Ideal Estimator}. The ideal estimator is not feasible in practice, and a natural modification is to plug in estimates of $(\bmu,\bOmega)$. This gives rise to the plug-in estimator:
\beq  \label{theta-literature}
  \widehat{\theta}^{\, \rI}_m(\hbmu,\hbOmega)=\frac{1}{np^m} \sum_{i=1}^n \{ (\Yb_{i}-\hbmu)^\T \widehat{\bOmega}(\Yb_{i}-\hbmu) \}^{m},
\eeq
This estimator was proposed by  \cite{maruyama} in the setting of a fixed dimension, where they used the sample mean to estimate $\bmu$ and the inverse of the sample covariance matrix to estimate $\bOmega$. In the modern high-dimensional settings where $p$ grows with $n$, one can on longer use the inverse of sample covariance matrix to estimate $\bOmega$; \cite{fan2013quadro} proposed plugging in an estimator of $\bOmega$ from high-dimensional sparse precision matrix estimation methods, with  stringent structural assumptions on $\bOmega$.

\begin{figure}[t]        \centering
                \includegraphics[width=.46\textwidth,height=.42\textwidth]{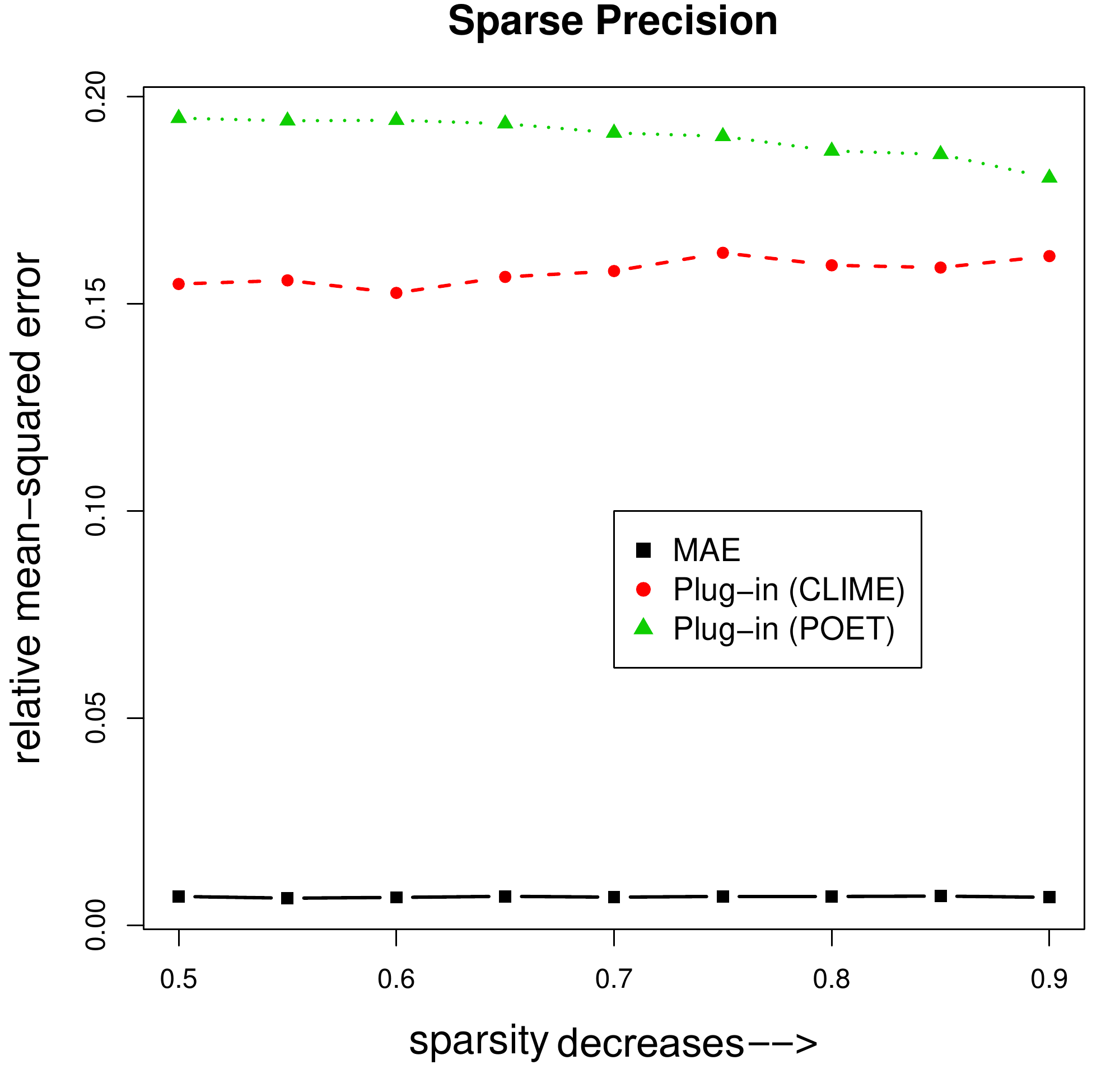}
                \includegraphics[width=.46\textwidth,height=.42\textwidth]{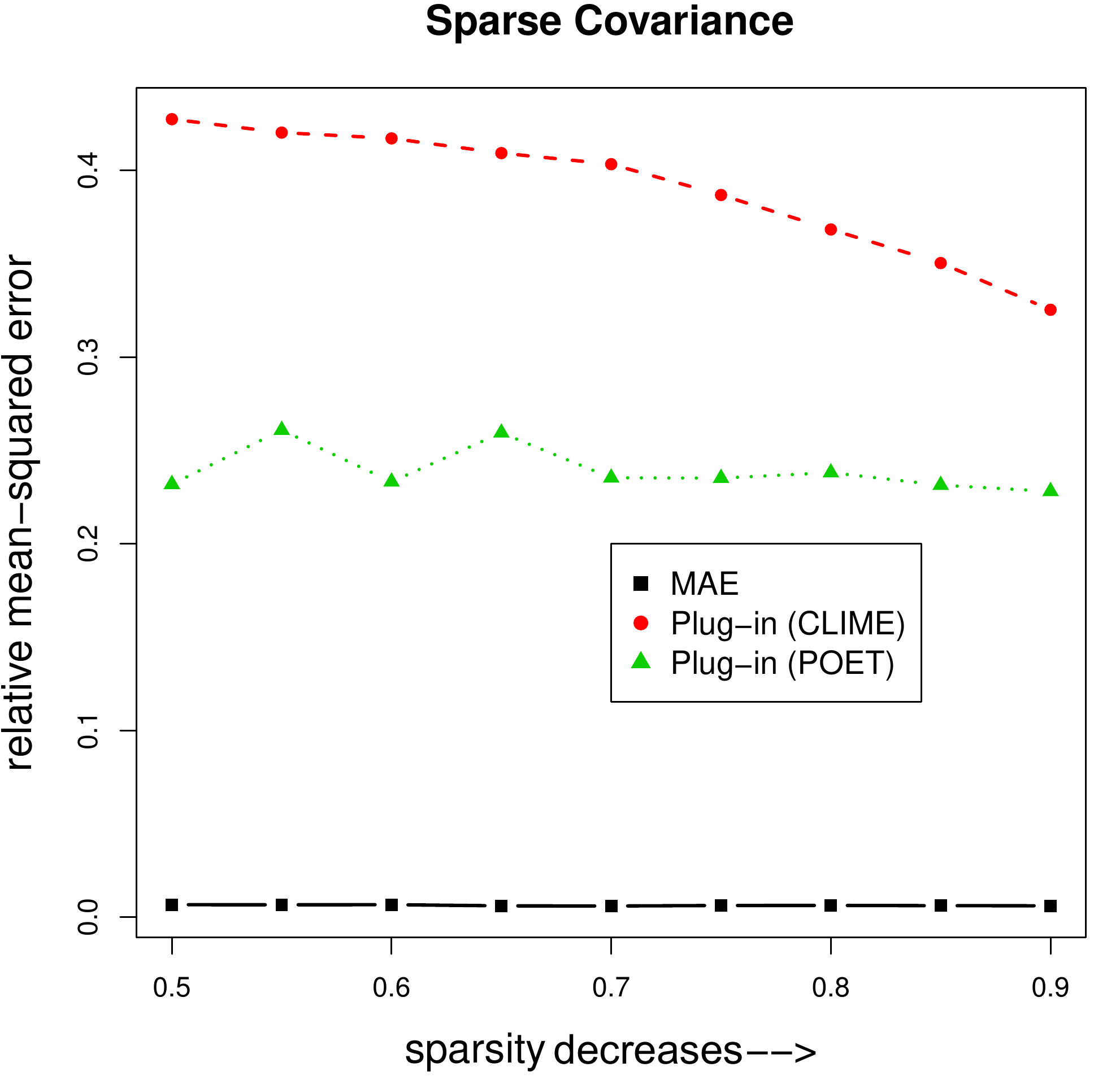}
                \caption {Plug-in estimators and MAE ($p=100$, $n=20$, true distribution is multivariate Gaussian). For plug-in estimators, we use two estimators of $\bOmega$, CLIME \citep{CLIME} and POET \citep{fanmincheva}. CLIME requires $\bOmega$ to be sparse and POET assumes $\bSigma$ has a low-rank plus sparse structure. When these structural assumptions are violated, the plug-in estimator of $\theta_2$ has a poor performance ($y$-axis is log of squared errors). In contrast, MAE always outperforms the plug-in estimators.
}\label{fig:intro_plot}
\end{figure}

However, the plug-in estimators perform unsatisfactorily for high-dimensional settings due to the difficulty of estimating $\bOmega$. Existing methods of estimating $\bOmega$ only perform well under stringent conditions, such as the sub-Gaussian assumption on the distribution and/or structural assumptions on $\bOmega$ (e.g., sparsity). Especially, the structural assumption on $\bOmega$ is critical for the success of these methods. Figure~\ref{fig:intro_plot} shows the performance of the plug-in estimator when the structural assumption required by $\hbOmega$ is violated. We consider two estimators of $\bOmega$, the CLIME estimator \citep{CLIME} which requires sparsity of $\bOmega$, and the POET estimator \citep{fanmincheva} which assumes a factor structure with sparse covariance of the idiosyncratic component. On the left panel of Figure~\ref{fig:intro_plot}, we generate elliptical data with a sparse covariance matrix, $\bSigma_{i,j}=a^{|i-j|}$, $1\leq i,j\leq p$, where $a$ controls the sparsity level and varies in $\{0.5,0.55,\ldots,0.85,0.9\}$. Here, the structural assumption of POET is not satisfied, and the associated plug-in estimator of $\theta_2$ performs unsatisfactorily. On the right panel, we generate data with a sparse precision matrix $\bOmega$, where each entry of the upper triangle of $\bOmega$ has a probability of $a$ to be nonzero,\footnote{We generate $\bOmega$ using {\it fastclime.generator}($\cdot$) in the R package {\it clime}, where the {\it graph} argument is set  ``random".} with $a$ chosen from $\{0.5,0.55,\ldots,0.85,0.9\}$. The assumption of CLIME is violated, so the associated plug-in estimator of $\theta_2$ has a unsatisfactory performance.



In fact, the philosophy of plug-in estimators is problematic. Estimating large precision matrices is a well-known difficult problem (even for Gaussian data), as one needs to estimate a large number of parameters. On the other hand, our problem only involves estimating one single parameter $\theta_m$. Intuitively, the latter should be much easier than the former. The plug-in estimators are realy using ``a sledgehammer to crack an egg."

\subsection{The marginal aggregation estimator (MAE)} \label{subsec:MAE}
Is it possible to avoid using the ``sledgehammer" of precision matrix estimation? We show that this is possible by a new marginal aggregation estimator.
In model \eqref{Ydecomposition}, letting $\widetilde{U}_j$ be the $j$-th coordinate of $\widetilde{\Ub}  \equiv  \bSigma^{1/2}\Ub $, we have
\beq \label{Ycoordinate}
Y_j  = \mu_j  + \xi \, \widetilde{U}_j , \qquad  j =1 , \ldots, p.
\eeq
Our key observation is that each individual coordinate of $\Yb$ contains information of $\xi$. It motivates us to construct an estimator of $\theta_m$ using only one coordinate of samples. Let  $\sigma_{jj}$ be the $j$-th diagonal of $\bSigma$.  We notice that \eqref{Ycoordinate} implies $\xi^{2m}=( Y_j -\mu_j )^{2m}/ \widetilde{U}_j^{2m}$.
The random variable $\widetilde{U}_j$ is unobserved, but its distribution is known once $\sigma_{jj}$ is given. It can be shown that (see Proposition~\ref{lem:beta-dist})
\beq \label{cmp}
\expv ( \widetilde{U}_j^{2m} ) = p^{-m}  c_{m}\, \sigma_{jj}^m, \qquad \mbox{where}\quad c_{m} = (2m-1)!! \, (p/2)^m \frac{ \Gamma(p/2)}{  \Gamma(p/2+m)}.
\eeq
Inspired by \eqref{Ycoordinate}-\eqref{cmp},
we introduce an estimator of $\theta_m$ using the marginal data $Y_{1j}, \ldots, Y_{nj}$:
\beq \label{theta-marginal}
 \widehat{\theta}^{\, \rM}_{ m, j}(\mu_j,\sigma_{jj}) = \frac{1}{n p^m }\sum_{i=1}^n \frac{  ( Y_{ij} - \mu_j )^{2m}}{ \expv(\widetilde{U}_j^{2m}) } = \frac{1}{ c_{m} \, \sigma_{jj}^m  } \frac{1}{n}\sum_{i=1}^n \  ( Y_{ij}- \mu_j )^{2m} .
\eeq
We call $ \widehat{\theta}^{\, \rM}_{ m, j}(\mu_j,\sigma_{jj})$ the {\it Marginal Estimator}. It  only requires knowledge of $(\mu_j, \sigma_{jj})$ and successfully avoids precision matrix estimation. For each $1\leq j\leq p$, we can define a marginal estimator and we will show that all marginal estimator contains the same amount of information about $\theta_m$ (see Theorem~\ref{thm:normality}). All these marginal estimators are unbiased, so taking their average gives rise to a new unbiased estimator:
 \beq \label{theta-1}
 \widehat{\theta}^{\, \rM}_m(\bmu,\mathrm{diag}(\bSigma))  = \frac{1}{p}\sum_{j=1}^p \widehat{\theta}^{\, \rM}_{m, j} (\mu_j,\sigma_{jj})  = \frac{1}{c_{m} \, n p}\sum_{j=1}^p \left\{ \frac{1}{\sigma_{jj}^m  }\sum_{i=1}^n   ( Y_{ij}- \mu_j )^{2m} \right\} .
  \eeq
We call $\widehat{\theta}^{\, \rM}_m(\bmu,\mathrm{diag}(\bSigma))$ the {\it Marginal Aggregation Estimator (MAE)}. The ``aggregation" of marginal estimators helps reduce the asymptotic variance.
Our proposed estimator is a natural plug-in version of \eqref{theta-1} given~by
\beq  \label{theta-2}
 \widehat{\theta}^{\, \rM}_m(\hbmu,\mathrm{diag}(\hbSigma)) = \frac{1}{c_{m} \, n p} \sum_{j=1}^p\left\{  \frac{1}{\widehat{\sigma}_{jj}^m}\sum_{i=1}^n  ( Y_{ij}-\widehat{\mu}_j )^{2m}   \right\} ,
\eeq
where $c_{m}$ is as in \eqref{cmp}, $\hbmu$ is an estimator of $\bmu$, and $\{\widehat{\sigma}_{jj} \}_{j=1}^p$ are the estimators of $\{ \sigma_{jj} \}_{j=1}^p$.

Compared with the plug-in estimator \eqref{theta-literature}, MAE is numerically more appealing, as it only needs to estimate the diagonal entries of $\bSigma$. Back to the example in Figure~\ref{fig:intro_plot}, we implement MAE using sample mean as $\hbmu$ and sample covariance matrix as  $\hbSigma$. MAE significantly outperforms the plug-in estimators, even when the structural assumptions of the plug-in estimators are satisfied.

\subsection{Organization of the paper}
In Section~\ref{sec:theory}, we study the theoretical properties of MAE. Under mild regularity conditions, we show that MAE is unbiased and root-$n$ consistent, regardless of the structure of $\bSigma$. We also show that MAE is asymptotically efficient, with an asymptotic variance matching that of the ideal estimator when $(\bmu,\bSigma)$ are given. We also discuss how to construct a confidence interval of $\theta_m$.

In Section~\ref{sec:BAE}, we generalize the idea of MAE to develop estimators of $\theta_m$ that use a small subset of the coordinates. We introduce the block-wise estimator and the blockwise aggregation estimator (BAE), analogous to the marginal estimator and MAE. These ideas help further reduce the estimation errors in the second order.


Section~\ref{sec:simulations} validates the theoretical insight by extensive simulations.
Section~\ref{sec:realized_xi} gives an application of MAE to time series data. We consider an extension of model \eqref{Ydecomposition} to multivariate time series:
\[
\Yb_t = \bmu_t + \bB  \fb_t + \xi_t \bSigma_t^{1/2}\Ub_t, \qquad t = 1,\cdots, T,
\]
where $\bmu_t$ is the time-varying mean, $\fb_t \in \mathbb{R}^K$ is a vector of $K$ observed factors, and $\bB$ is a $p \times K$ matrix of factor loadings. We extend MAE to a method for estimating the realized $\xi_t$.
Its application to stock returns provides a new index that captures information of {\it whole market}.
Section~\ref{sec:Discu} contains conclusions and discussions. All the proofs are relegated to the appendix.


\medskip
\noindent
{\sc Notation}: Throughout this paper, for any vector $\bv$ and matrix $\bM$, we let $\|\bv\|$ denote the Euclidean norm of $\bv$ and let  $\|\bM\|$, $\|\bM\|_F$ and $\|\bM\|_{\max}$ denote its spectral norm, Frobenius norm and entry-wise maximum norm, respectively.
We use $\widehat{\theta}^{\, \rM}_{ m, j}(\mu_j,\sigma_{jj})$, $\widehat{\theta}^{\, \rM}_{m}(\bmu,\mathrm{diag}(\bSigma))$, $\widehat{\theta}^{\, \rI}_{m}(\bmu,\bOmega)$ and $\widehat{\theta}^{\, \rB}_{m}(\bmu,\mathrm{diag}_{\cal A}(\bSigma))$ to denote the Marginal Estimator, MAE, Ideal Estimator, and BAE (to be introduced), respectively, with given $(\bmu,\bSigma)$;  when $(\bmu,\bSigma)$ are replaced by $(\hbmu,\hbSigma)$, it means we plug in  estimators of the mean vector and covariance matrix. We frequently use notations $(\theta_m,c_m,\eta_m,r_m)$, where $\theta_m$ is defined in \eqref{moment}, $c_m$ is defined in \eqref{cmp}, $\eta_m$ and $r_m$ are defined in Definition~\ref{def:r+eta}. For all settings in this paper, $\eta_m$ is a constant, $(\theta_m,c_m,r_m)$ depend on $p$ but are at the constant scale.

\section{Theoretical properties of MAE} \label{sec:theory}
We study the asymptotic properties of MAE defined in \eqref{theta-2}, assuming both $(n,p)$ tend to infinity.
First, we study the consistency of MAE. The following theorem shows that, when the distribution is marginally sub-Gaussian, if we plug in the sample mean and sample covariance matrix as $(\hbmu,\hbSigma)$, then MAE is always root-$n$ consistent.
\begin{thm}[Root-$n$ consistency] \label{thm:root-n}
Under model \eqref{Ydecomposition}, suppose $\log^2(p)=o(n)$ and $\max_{1\leq j\leq p}\|Y_j-\mu_j\|_{\psi_2}\leq C$, where $\|\cdot\|_{\psi_2}$ denotes the sub-Gaussian norm.\footnote{For a random variable $X$, its sub-Gaussian norm is defined as $\|X\|_{\psi_2}=\sup_{k\geq 1}k^{-1}(\mathbb{E}|X|^{k})^{1/k}$.} Given iid samples $\{\Yb_i\}_{i=1}^n$, consider the MAE in \eqref{theta-2}, where $(\hbmu,\hbSigma)$ are the sample mean vector and sample covariance matrix. Then,
\[
|\widehat{\theta}^{\, \rM}_m(\hbmu,\mathrm{diag}(\hbSigma))-\theta_m|= O_{\mathbb{P}}(n^{-1/2}).
\]
\end{thm}

The root-$n$ consistency of MAE requires {\it no} conditions on either $\bSigma$ or $\bOmega$. It confirms our previous insight that estimating moment parameters is an ``easier" statistical problem than estimating large matrices. On the other hand, the plug-in estimators only perform well when the assumed structural assumptions (e.g., sparsity) on $\bSigma$ or $\bOmega$ are satisfied.

Many distributions in the elliptical family are heavy-tailed and don't satisfy the marginal sub-Gaussianity assumption. In these cases, we prefer to use robust estimators of $\bmu$ and $\bSigma$ \citep{FLW2017, sun2016adaptive}. They are M-estimators with robust loss functions or rank-based estimators. Compared to the sample mean and sample covariance estimators, these robust estimators lead to sharper bounds of $\|\hbmu-\bmu\|_\infty$ and $\|\hbSigma-\bSigma\|_{\max}$ in the case of heavy-tailed data.
The next theorem studies MAE with general mean/covariance estimators.

\begin{thm}[Consistency, with general mean/covariance estimators] \label{thm:consistency}
Under model \eqref{Ydecomposition}, suppose $\log^2(p)=o(n)$ and $\theta_{2m}\leq C$. Given iid samples $\{\Yb_i\}_{i=1}^n$, consider the MAE in \eqref{theta-2}. We assume the estimators $(\hbmu,\hbSigma)$ satisfy $\max_{1\leq j\leq p}|\widehat{\mu}_j-\mu_j|\leq \alpha_n$ and $\max_{1\leq j\leq p}|\widehat{\sigma}_{jj}-\sigma_{jj}|\leq \beta_n$ with probability $1-o(1)$, where $\alpha_n\to 0$ and $\beta_n\to 0$ as $n,p\to\infty$. Then, for any $\epsilon>0$, with probability $1-\epsilon$, there is a constant $C_\epsilon>0$ such that
\[
\bigl|\widehat{\theta}^{\, \rM}_m(\hbmu,\mathrm{diag}(\hbSigma))-\theta_m\bigr|\leq C_\epsilon \bigl( n^{-1/2} + \max\{\alpha_n, \beta_n\}\bigr).
\]
\end{thm}

The typical error rate of robust estimators is $\alpha_n\asymp \sqrt{\log(p)/n}$ and $\beta_n\asymp\sqrt{\log(p)/n}$ \citep{FLW2017, sun2016adaptive}, so the associated MAE satisfies  $|\widehat{\theta}^{\, \rM}_m-\theta_m|=O_{\mathbb{P}}(\sqrt{\log(p)/n})$. Compared with the rate in Theorem~\ref{thm:root-n}, the extra $\sqrt{\log(p)}$ factor here is a price paid for heavy tails.

Next, we study the asymptotic variance of MAE. By Theorem~\ref{thm:root-n}, MAE is already rate-optimal. We would like to see whether it also achieves the optimal  ``constant". We shall compare its asymptotic variance with that of the Ideal Estimator  \eqref{theta-oracle}.
Since the Ideal Estimator knows the true $(\bmu,\bSigma)$, for a fair comparison, we consider MAE with true $(\bmu,\bSigma)$.
\begin{definition} \label{def:r+eta}
For any $k\geq 1$, let $\eta_k=\mathbb{E}[N(0,1)^{2k}]$ and $r_k=(\mathbb{E}\xi^{2k})/(\mathbb{E}\chi_p^{2k})$, where $\chi^2_p$ denotes the chi-square distribution with $p$ degrees of freedom.
\end{definition}
The quantities $r_k$ capture the difference between moments of an elliptical distribution and moments of a multivariate Gaussian distribution with matching mean and covariance matrix. It depends on $p$ but is at the constant scale under our settings.
\begin{thm}[Variance] \label{thm:var}
Under model \eqref{Ydecomposition}, suppose $\log^2(p)=o(n)$ and $\theta_{2m}\leq C$.
Given iid samples $\{\Yb_i\}_{i=1}^n$, consider the MAE in \eqref{theta-1} where $(\bmu,\bSigma)$ are given. Let $\bLambda=[\mathrm{diag}(\bSigma)]^{-1/2}\bSigma[\mathrm{diag}
(\bSigma)]^{-1/2}$ be the correlation matrix. There is a constant $C_m>0$, independent of the distribution of $\xi$, such that
\[
\frac{\var\bigl( \widehat{\theta}^{\, \rM}_m(\bmu,\mathrm{diag}(\bSigma)) \bigr)}{\theta_m^2} \leq \frac{1}{n}\frac{r_{2m}-r_m^2}{r_m^2} +\frac{1}{np}\frac{r_{2m}}{r^2_m} \frac{\eta_{2m}-\eta_m^2}{\eta_m^2}+ \frac{C_m}{n}\frac{r_{2m}}{r^2_m\eta_m^2}
\frac{\|\bLambda-\bI\|_F^2}{p^2}.
\]
When $m=2$, the equality holds with $C_m=72$.
\end{thm}

The upper bound for the variance has three terms: The first term is $O(n^{-1})$; as we shall see, this term matches with the variance of the benchmark estimator. The second term is $O(n^{-1}p^{-1})$ and is negligible for diverging $p$. The third term is caused by correlations among different marginal estimators $\widehat{\theta}^{\,\rM}_{m,j}$. This term is negligible as long as $\|\bLambda-\bI\|_F^2=o(p^2)$; consider a special case where $\|\bSigma\|$ is bounded, then $\|\bLambda-\bI\|_F^2=O(p)$; so the requirement of $\|\bLambda-\bI\|_F^2=o(p^2)$ is mild. Indeed, if requires that the sparsity of correlation coefficients: $\sum_{i \not= j} \lambda_{ij} = o(p^2)$, where $\bLambda = (\lambda_{ij})$.
The next proposition confirms that the asymptotic variance of MAE is the same as the asymptotic variance of the Ideal Estimator:
\begin{prop}[Comparison with benchmark] \label{prop:ideal}
Let $\{\Yb_i\}_{i=1}^n$ be iid samples of model \eqref{Ydecomposition}. Suppose $\theta_{2m}\leq C$. For the Ideal Estimator in \eqref{theta-oracle},
\[
\frac{\var\bigl( \widehat{\theta}^{\, \rI}_m(\bmu,\bOmega) \bigr)}{\theta_m^2} = \frac{1}{n}\frac{r_{2m}-r_m^2}{r_m^2} +\frac{1}{np}\frac{r_{2m}}{r_m^2}2m^2\bigl[1+O(p^{-1})\bigr].
\]
As a result, if $\|\bLambda-\bI\|_F^2=o(p^2)$, where $\bLambda=[\mathrm{diag}(\bSigma)]^{-1/2}
\bSigma[\mathrm{diag}(\bSigma)]^{-1/2}$ is the correlation matrix, then
\[
\frac{ \var\bigl( \widehat{\theta}^{\, \rM}_m(\bmu,\mathrm{diag}(\bSigma)) \bigr)  }{ \var\bigl( \widehat{\theta}^{\, \rI}_m(\bmu,\bOmega) \bigr) }\to 1.
\]
\end{prop}

Last, we construct confidence intervals of $\theta_m$. Since MAE is the average of $p$ strongly dependent marginal estimators, its asymptotic normality is hard to approach. We instead use the marginal estimator in \eqref{theta-marginal} to construct confidence intervals.

\begin{thm}[Asymptotic normality] \label{thm:normality}
Under model \eqref{Ydecomposition}, suppose $\log^2(p)=o(n)$ and $\max_{1\leq j\leq p}\|Y_j-\mu_j\|_{\psi_2}\leq C$, where $\|\cdot\|_{\psi_2}$ denotes the sub-Gaussian norm. Given iid samples $\{\Yb_i\}_{i=1}^n$, consider the Marginal Estimator in \eqref{theta-marginal} for an arbitrary $1\leq j\leq p$, where $(\widehat{\mu}_j,\widehat{\sigma}_{jj})$ are the sample mean and sample variance of $\{Y_{ij}\}_{i=1}^n$. Then,
\[
\frac{\sqrt{n}\Bigl(\widehat{\theta}^{\, \rM}_{ m, j}(\widehat{\mu}_j,\widehat{\sigma}_{jj})-\theta_m\Bigr)}{\sqrt{ \frac{c_{2m}}{c_m^2}\widehat{\theta}_{2m} -\widehat{\theta}_m^2 }} \to_d N(0,1),
\]
where $c_k = (2k-1)!! \, (p/2)^k \frac{ \Gamma(p/2)}{  \Gamma(p/2+k)}$ for $k\geq 1$, and $(\widehat{\theta}_{2m},\widehat{\theta}_m)$ are consistent estimators of $(\theta_{2m},\theta_m)$.
\end{thm}

This theorem shows somewhat surprisingly that all marginal estimator contains the same amount of information about $\theta_m$. Given consistent estimators $(\widehat{\theta}_{2m}, \widehat{\theta}_{m})$, the asymptotic level-$\alpha$ confidence interval of $\theta_m$ is
\beq\label{conf.interval}
\widehat{\theta}^{\, \rM}_{ m, j}\pm  \frac{q_{1-\alpha/2}}{\sqrt{n}}\sqrt{ \frac{c_{2m}}{c_m^2}\widehat{\theta}_{2m} -\widehat{\theta}_m^2 },
\eeq
where $q_{1-\alpha/2}$ is the $(1-\alpha/2)$-quantile of a standard normal. It doesn't matter which of $1\leq j\leq p$ we use, as these marginal estimators have the same asymptotic variance. For the estimators $(\widehat{\theta}_{2m}, \widehat{\theta}_{m})$, we suggest using MAE. 

If we only need a point estimator but not a confidence interval, we prefer MAE to the Marginal Estimator, as MAE has a smaller variance in many scenarios. For example, when $\|\bLambda-\bI\|_F^2=o(p^2)$, by plugging in the true $(\bmu,\bSigma)$,
\[
\frac{\var( \widehat{\theta}^{\, \rM}_m)}{\theta_m^2}\sim \frac{\var\bigl( \widehat{\theta}^{\, \rI}_m)}{\theta_m^2}\sim \frac{1}{n}\frac{r_{2m}-r_m^2}{r_m^2}, \qquad
\frac{\var(\widehat{\theta}^{\, \rM}_{ m, j})}{\theta_m^2} \sim \frac{1}{n}\frac{(\eta_{2m}/\eta_m^2)r_{2m}-r_m^2}{r_m^2}.\footnote{This expression combines the asymptotic variance in Theorem~\ref{thm:normality} and the fact that $c_m\theta_m=r_m\eta_m$}
\]
Since $\eta_{2m}> \eta_m^2$, the latter variance is strictly larger.  In contrast, MAE is first-order efficient.

\section{Extension to blockwise aggregation} \label{sec:BAE}
In the construction of MAE, each marginal estimator only uses one coordinate of the samples. It is convenient to implement and gives rise to an estimator that is first-order efficient, provided that the third term in Theorem~\ref{thm:var} is negligible. It turns out that, the second order term in the variance can be improved upon by using blockwise aggregation, and so is the third term, which is related to the correlation structure.  Our simulation studies below show that the improvement is real.
This motivates us to extend the marginal estimator to a blockwise estimator that uses a small number of coordinates of the samples and takes into account their correlation structures. We then generalize MAE to BAE --- an aggregation of many blockwise estimators.

\begin{figure}[t]
        \centering
                \includegraphics[width=.55\textwidth]{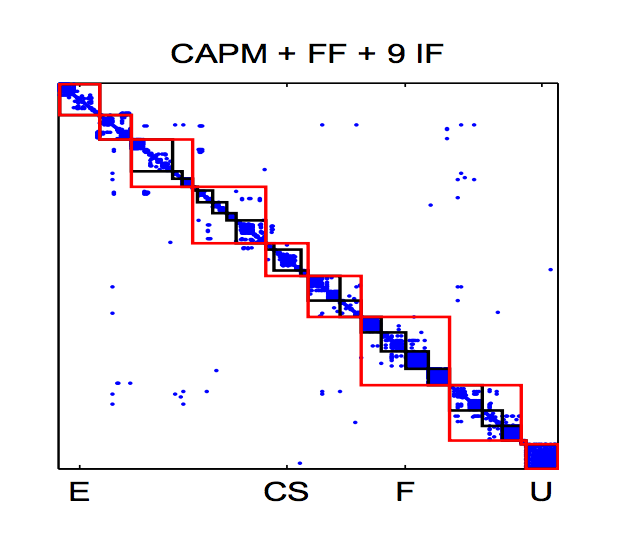}
              \caption{  Estimated $\bSigma$ after factor-removal from S\&P500 returns in 2010---2012. Red squares: Sector blocks. Black squares: Industry groups. (From \cite{furger})  }
              \label{fig:cov-stocks}
\end{figure}

BAE can be applied to settings where the covariance matrix is approximately blockwise diagonal after row/column permutation. Figure~\ref{fig:cov-stocks} gives such an example, where the S\&P 500 stocks divide into many small-size blocks according to sectors or industries of stocks and the stock returns within each block are correlated but admits block structure after taking out the market factor. BAE can take advantage of the within-block correlations and further improve MAE in the second order term.

\subsection{A block-wise aggregation estimator (BAE)} \label{subsec:Block-Estimator}
We fix a block $J\subset\{1,2,\ldots,p\}$ and let $K=|J|$. For any vector $\bv\in\mathbb{R}^p$ and matrix $\bM\in\mathbb{R}^{p \times p}$, let $\bv_J$ be the subvector of $\bv$ containing the coordinates indexed by $J$ and let $\bM_{J J}$ be the submatrix of $\bM$ containing the entries indexed by $ J \times J$. By \cite{fang}, when $\Yb$ follows an elliptical distribution \eqref{Ydecomposition}, the subvector $\Yb_J$ satisfies that
\beq\label{FangZhang1}
 \Yb_J \overset{(d)}{=} \bmu_J +  B^{1/2} \xi   \cdot \bSigma_{J J}^{1/2} {\Ub}_K,
 \eeq
where $B$ is a random variable that follows a beta distribution Beta$\,( \frac{K}{2} , \frac{p-K}{2} )$, the random vector $ {\Ub}_K $ follows a uniform distribution on the unit sphere $\bbS^{K-1}$, and $(\xi, B , {\Ub}_K)$ are mutually independent. Since $\| {\Ub}_K \| = 1$,
\[
\xi^{2m} =\frac{ \{ (\Yb_J - \bmu_J )^\T  \bSigma_{J J}^{-1} (\Yb_J - \bmu_J ) \}^m}{B^m}.
\]
The random variable $B$ is not directly observable, but its expectation is known:
\begin{prop} \label{lem:beta-dist}
For each $m\geq 1$ and $1\leq K \leq p$, define $c^*_{m,K} = p^m \expv ( B^m )$ with $B\sim {\rm Beta}\,( \frac{K}{2}, \frac{p-K}{2})$. Then,
\[ 
c^*_{1,K} = K, \qquad  c_{m,K}^* = p\times \frac{ K+2m-2}{p+2m-2}  \times  c^*_{m-1,K} \qquad\mbox{for }m\geq 2.
\]
\end{prop}
\noindent
Replacing $B^m$ by its expectation, we immediately have an estimator of $\theta_m$ based on $\{\Yb_{i,J}\}_{i=1}^n$:
\begin{align} \label{theta-blockwise}
\widehat{\theta}^{\, \rB }_{m , J}(\bmu_J,\bSigma_{JJ}) &=\frac{1}{np^m}\sum_{i=1}^n \frac{ \{ (\Yb_{i,J} - \bmu_J )^\T  \bSigma_{J J}^{-1} (\Yb_{i,J} - \bmu_J ) \}^m}{\E B^m}\cr
& = \frac{1}{n c_{m,K}^*} \sum_{i=1}^n \{  (\Yb_{i, J}-\bmu_J)^\T \bSigma_{J J}^{-1}(\Yb_{i, J}-\bmu_J) \}^{m}.
\end{align}
We call $\widehat{\theta}^{\, \rB }_{m , J}(\bmu_J,\bSigma_{JJ})$ the {\it Blockwise Estimator}. Now, given a collection of blocks ${\cal A}=\{J_1,J_2,\ldots,J_N\}$, we can define a blockwise estimator for each $J\in{\cal A}$ and then take their average:
\beq \label{theta-BAE}
\widehat{\theta}^{\, \rB}_m\bigl(\bmu,\;\mathrm{diag}_{\cal A}(\bSigma)\bigr) =\frac{1}{ |{\cal A}|} \sum_{J\in {\cal A}}\widehat{\theta}^{\, \rB }_{m , J}(\bmu_J,\bSigma_{JJ}).
\eeq
We call $\widehat{\theta}^{\, \rB}_m\bigl(\bmu,\;\mathrm{diag}_{\cal A}(\bSigma)\bigr)$ the {\it Blockwise Aggregation Estimator (BAE)}. Here $\mathrm{diag}_{\cal A}(\bSigma)$ denotes the collection of diagonal blocks $\bSigma_{JJ}$ with $J\in{\cal A}$. Our final estimator is a plug-in version of BAE by plugging in an estimator $\hbmu$ and estimators of those diagonal blocks of $\bSigma$.

Since BAE only estimates the small-size diagonal blocks of $\bSigma$ and does not need to estimate $\bOmega$, it inherits a nice property of MAE: root-$n$ consistency is guaranteed with no conditions on $\bSigma$ or $\bOmega$.

\begin{thm}[Root-$n$ consistency] \label{thm:root-n(BAE)}
Fix $m\geq 2$ and $K\geq 1$. Under model \eqref{Ydecomposition}, suppose $\log^{2m}(p)=o(n)$ and $\max_{1\leq j\leq p}\|Y_j-\mu_j\|_{\psi_2}\leq C$. We assume the minimum eigenvalue of any $K\times K$ diagonal block of $\bSigma$ is lower bounded by $C$. Let ${\cal A}$ be a collection of nonrandom, non-overlapping blocks such that the size of each block is bounded by $K$.
Given iid samples $\{\Yb_i\}_{i=1}^n$, consider the BAE in \eqref{theta-BAE}, where $(\bmu,\bSigma)$ are estimated by the sample mean vector and sample covariance matrix. Then,
\[
|\widehat{\theta}^{\, \rB}_m(\hbmu,\mathrm{diag}_{\cal A}(\hbSigma))-\theta_m|=O_{\mathbb{P}}(n^{-1/2}).
\]
\end{thm}

\begin{thm}[Consistency, with general mean/covariance estimators] \label{thm:consistency(BAE)}Fix $m\geq 2$ and $K\geq 1$. Under model \eqref{Ydecomposition}, we assume $\log^{2m}(p)=o(n)$, $\theta_{2m}\leq C$, and the minimum eigenvalue of any $K\times K$ diagonal block of $\bSigma$ is lower bounded by $C$. Let ${\cal A}$ be a collection of nonrandom, non-overlapping blocks where the size of blocks is bounded by $K$. Given iid samples $\{\Yb_i\}_{i=1}^n$, consider the BAE in \eqref{theta-BAE}, where $(\hbmu,\hbSigma)$ satisfy $\|\hbmu-\bmu\|_\infty\leq \alpha_n$ and $\max_{J\in{\cal A}}\|\widehat{\bSigma}_{JJ}-\bSigma_{JJ}\|\leq \beta_n$ with probability $1-o(1)$, with $\alpha_n\to 0$ and $\beta_n\to 0$ as $n,p\to\infty$. Then, for any $\epsilon>0$, with probability $1-\epsilon$, there is a constant $C_\epsilon>0$ such that
\[
\bigl|\widehat{\theta}^{\, \rB}_m(\hbmu,\mathrm{diag}_{\cal A}(\hbSigma))-\theta_m\bigr|\leq C_\epsilon \bigl( n^{-1/2}+\max\{\alpha_n,\beta_n\}\bigr).
\]
\end{thm}

We note that MAE is a special case of BAE, with all block size equal to $1$. The motivation of generalizing MAE to BAE is to better take advantage of correlation structures, and this is revealed by comparing the asymptotic variances of two methods; see Section~\ref{subsec:VarCompare} below. To implement BAE, we need to determine the collection of blocks, and in Section~\ref{subsec:implement} we discuss how to select blocks.

\subsection{Variance comparison} \label{subsec:VarCompare}
We compute the asymptotic variance of BAE and compare it with the asymptotic variances of MAE and Ideal Estimator. Same as before, in the variance calculation we assume $(\bmu,\bSigma)$ are given.
\begin{definition} \label{def:h-factor}
For each $k\geq 1$, let $h_{m}(k)= \frac{k\cdot\mathrm{var}(\chi_k^{2m})}{(\mathbb{E}\chi_k^{2m})^2}$, where $\chi^2_k$ denotes the chi-square distribution with $k$ degrees of freedom. Given a collection of blocks ${\cal A}$, let $\bar{h}_m({\cal A})  = \frac{p}{|{\cal A}|^2}\sum_{J\in {\cal A}}\frac{h_m(|J|)}{|J|}$.
\end{definition}
\begin{thm}[Variance of BAE] \label{thm:BAEideal}
Let $\{\Yb_i\}_{i=1}^n$ be $iid$ samples of model \eqref{Ydecomposition}. Fix $m\geq 2$ and suppose $\theta_{2m}\leq C$. There exists a constant $\widetilde{C}_m>0$, independent of the distribution of $\xi$, such that
for any collection ${\cal A}$ of non-overlapping blocks,
\[
\frac{\var\bigl( \widehat{\theta}_m^{\, \rB}\bigl(\bmu,\mathrm{diag}_{\cal A}(\bSigma)\bigr) \bigr)}{\theta_m^2} \leq  \frac{1}{n}\frac{r_{2m} - r_m^2}{r_m^2} + \frac{1}{np} \frac{r_{2m}}{r_m^2} \bar{h}_m({\cal A}) + \frac{C_m}{n}\frac{1}{|{\cal A}|^2} \sum_{\substack{I,J\in {\cal A}\\  I\neq J}} \|\bSigma_{II}^{-1/2}\bSigma_{IJ}\bSigma_{JJ}^{-1/2}\|^2.
\]
\end{thm}

\begin{figure}[tp]        \centering
                \includegraphics[width=.42\textwidth,height=.38\textwidth]{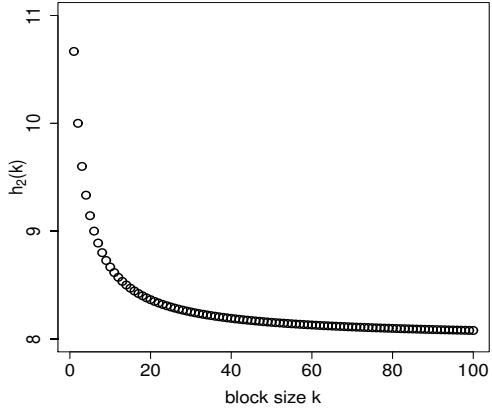}
                  \caption {Plot of $\bar{h}_m({\cal A})$ when all blocks have an equal size $k$ ($m=2$; the x-axis represents $k$). As the block size increases, $\bar{h}_m({\cal A})$ decreases, suggesting a variance reduction.}\label{fig:h_behavior}
\end{figure}

The upper bound of the variance has three terms:
\begin{itemize}
\item The first term is $O(n^{-1})$, which also appears in the variance of MAE and Ideal Estimator. It is the dominating term of the variance.
\item The second term is $O(p^{-1}n^{-1})$, where the constant in front of it is related to a quantity $\bar{h}_m({\cal A})$. We call $\bar{h}_m({\cal A})$ the {\it block-division factor}, as it is only a function of ${\cal A}$. To see how this factor changes with block size, let's consider a special case where all blocks have an equal size $k$ and $p$ is a multiple of $k$. Then,
\[
\bar{h}_m({\cal A}) = h_m(k)= \frac{k\cdot\mathrm{var}(\chi_k^{2m})}{(\mathbb{E}\chi_k^{2m})^2}.
\]
It is a monotone decreasing function of $k$ (see Figure~\ref{fig:h_behavior}). Hence, increasing the block size leads to a reduction of this term, which indicates that the second order efficiency of MAE can be improved with $m > 1$.

\item The last term comes from the correlations among estimators associated with different blocks. It doesn't exist for the Ideal Estimator, but both MAE and BAE have this extra term.
For MAE, all off-diagonal entries of $\bSigma$ contribute to this term. However, for BAE, only off-diagonal blocks contribute. Especially, when $\bSigma$ is blockwise diagonal with respect to ${\cal A}$, this extra term becomes zero. Again, increasing the block size leads to a reduction of this term.
\end{itemize}

From MAE to BAE, we can see that the dominating term in the variance bound remains the same, but the other two terms are reduced and the performance still improves. However, we cannot use too large blocks, because BAE needs to invert an estimate of $\bSigma_{J,J}$ and the error of estimating $\hbSigma_{J,J}$ increases as the block size increases.

We now give a more thorough comparison of four estimators, the Ideal Estimator (IE) $\widehat{\theta}^{\,\rI}_{m}$, the Marginal Estimator (ME) $\widehat{\theta}^{\,\rM}_{m,j}$, the MAE $\widehat{\theta}^{\,\rM}_m$, and the BAE $\widehat{\theta}^{\,\rB}_m$; see Table~\ref{tb:compare4}. We conclude that
\begin{itemize} \itemsep -1pt
\item IE has the optimal variance, but it works unsatisfactorialy in the real case of unknown $(\bmu,\bSigma)$, as it requires estimating $\bOmega$.
\item ME avoids estimating $\bOmega$ and works in the real case, but its asymptotic variance is non-optimal.
\item MAE aggregates a number of ME's and achieves the optimal variance when $\|\bLambda-\bI\|_F^2=o(p^2)$.
\item Compared with MAE, BAE relaxes the condition of $\|\bLambda-\bI\|_F^2=o(p^2)$ and reduces the second-order term of the variance.
\end{itemize}
From ME to BAE, we have used two methodological ideas: to aggregate ``local" estimators and to use a block of coordinates in each ``local" estimator. Both help reduce the variance of the estimator, with the first idea playing a more significant role.

\begin{table}[!hb]
\centering
\caption{Variance comparison of estimators (known $(\bmu,\bSigma)$; $^{**}$ means the constant is optimal).} \label{tb:compare4}
\scalebox{.95}{\begin{tabular}{ c | c| c | c | c}
\hline
& IE & ME & MAE & BAE\\
\hline
dominating term &  $\frac{1}{n}\bigl(\frac{r_{2m}}{r_m^2}-1\bigr)^{**}$ & $\frac{1}{n}\bigl(\frac{r_{2m}\eta_{2m}}{r_m^2\eta_m^2}-1\bigr)$
& $\frac{1}{n}\bigl(\frac{r_{2m}}{r_m^2}-1\bigr)^{**}$
& $\frac{1}{n}\bigl(\frac{r_{2m}}{r_m^2}-1\bigr)^{**}$\\
2nd-order term & $\frac{1}{np}\frac{r_{2m}}{r_m^2}h_m(p)^{**}$
& ---
& $\frac{1}{np}\frac{r_{2m}}{r_m^2}h_m(1)$
& $\frac{1}{np}\frac{r_{2m}}{r_m^2}h_m(k)$\\
correlation term & $0^{**}$ & $0^{**}$ & $\frac{C}{np}\sum_{1\leq i\neq j\leq p}|\Lambda_{jj}|^2$
& $\frac{C}{np}\sum_{I\neq J\in{\cal A}}\|\bLambda_{I,J}\|_F^2$\\
\hline
\end{tabular}}
\end{table}

{\bf Remark 1}. IE and MAE are special cases of BAE with equal-size blocks of $k=1$ and $k=p$, respectively. We note that $h_m(1)=\frac{\eta_{2m}-\eta_m^2}{\eta_m^2}$ and $h_m(p)=2m^2[1+O(p^{-1})]$, so Theorem~\ref{thm:BAEideal} matches with the variance bounds of MAE (Theorem~\ref{thm:var}) and the IE (Proposition~\ref{prop:ideal}).

\begin{figure}[tb]
        \centering
                \includegraphics[width=.48\textwidth]{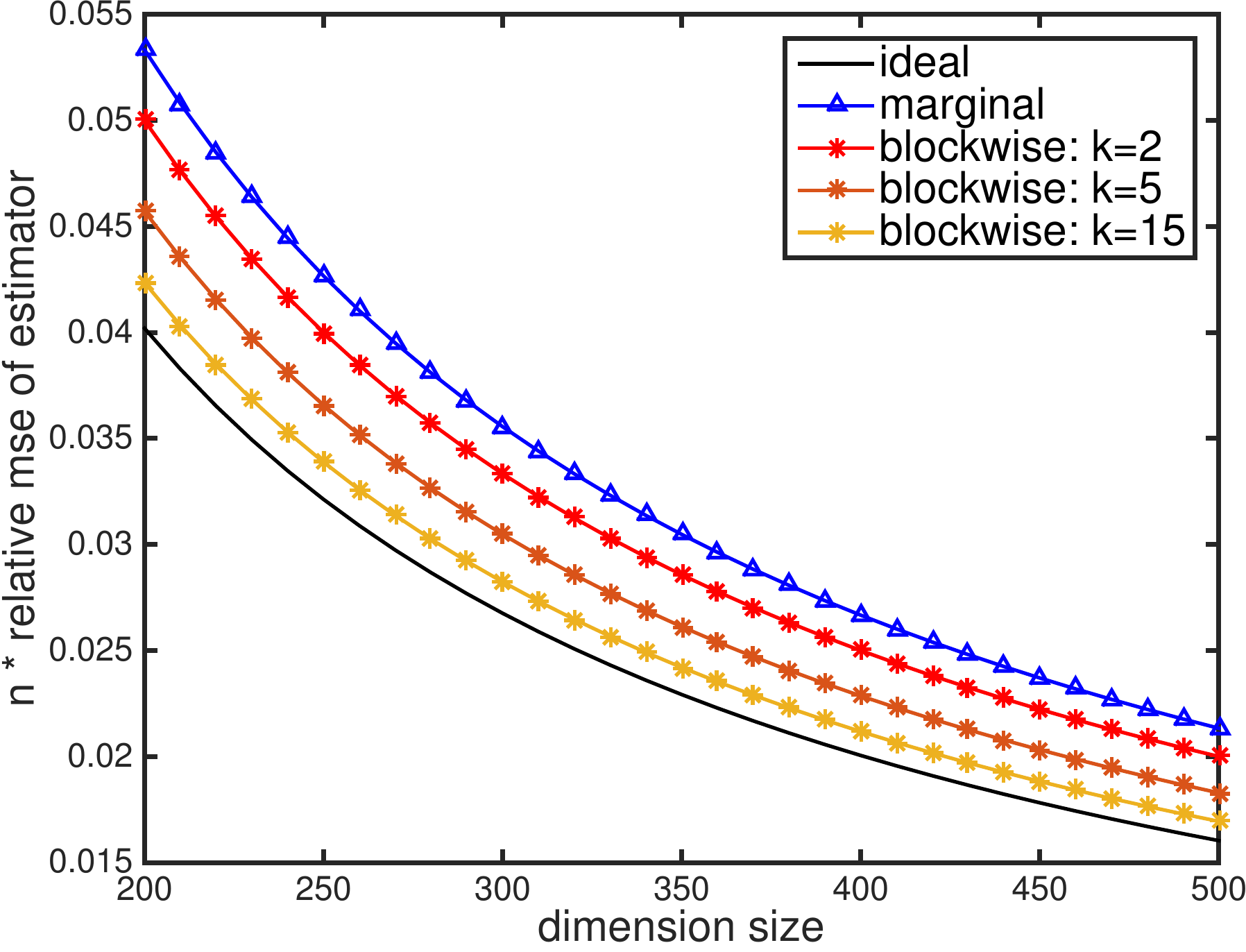}
                \includegraphics[width=.48\textwidth]{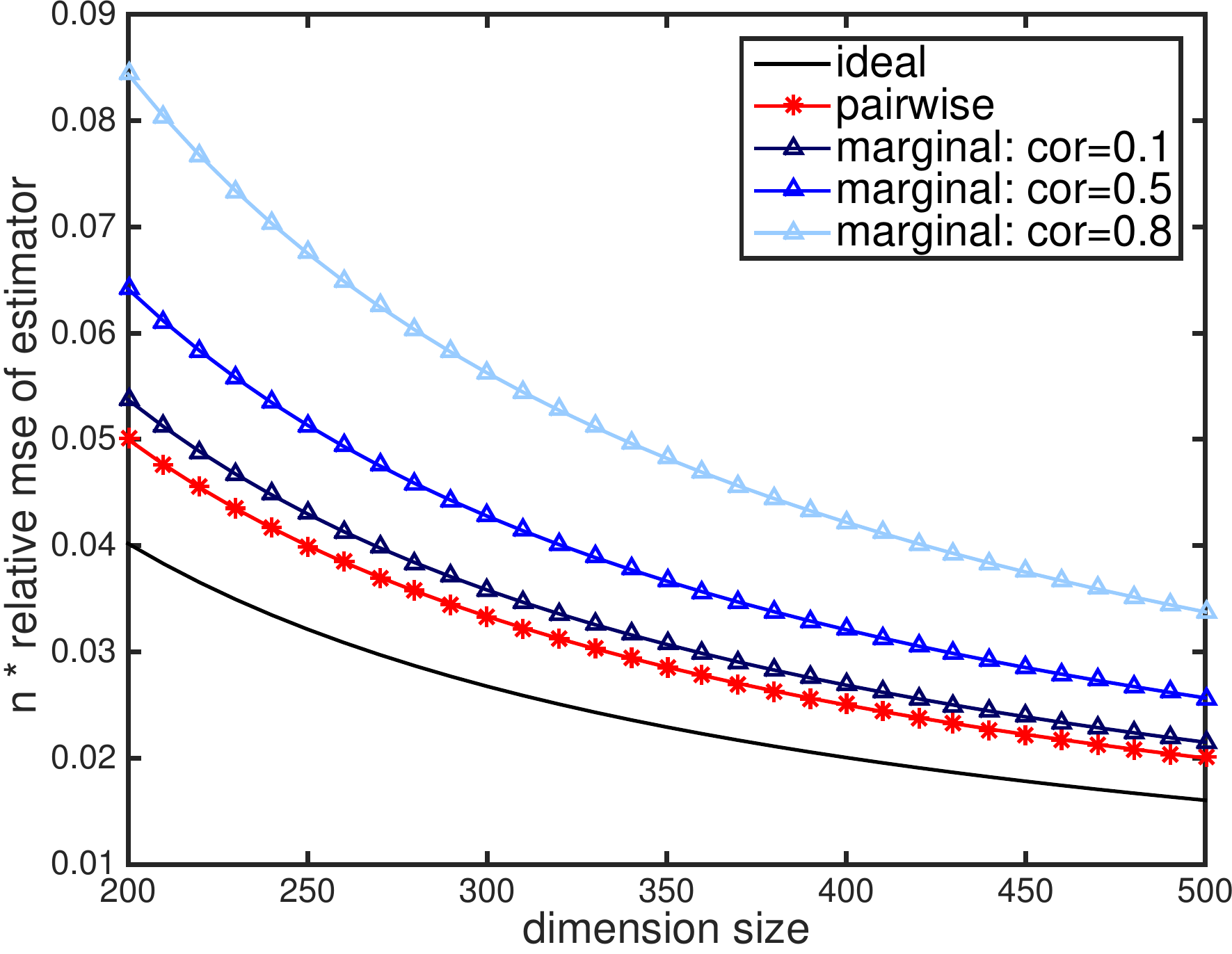}
                \caption{Comparison of IE, MAE and BAE for multivariate Gaussian distributions (y-axis is $\var(\widehat{\theta}^2_2 /\theta_2^2$). Left: $\bSigma=\bI_p$. Right: $\bSigma$ is a blockwise diagonal matrix with $2\times 2$ blocks whose diagonals are $1$ and off-diagonals are $\rho$, where $\rho$ takes values in $\{0.1,0.5,0.8\}$. The pairwise estimator refers to BAE with $k=2$. Curves are from theoretical calculations (see Corollary~\ref{coro:Gaussian} in the appendix). The variance of IE and BAE is independent of $\rho$, so there is only one curve for all values of $\rho$.}
                \label{fig:mse_toy}
\end{figure}

{\bf Remark 2} (multivariate Gaussian). Let's consider a special case where the data are multivariate Gaussian but the user doesn't know and still applies the estimators in this paper. For Gaussian distributions, the first term in the variance bound disappears, so the estimators considered here all have a faster rate of convergence as $O(p^{-1}n^{-1})$. This is the only case where a large $p$ helps, i.e., ``dimensionality is a blessing." Moreover, the difference between MAE and BAE is more prominent, as the second term in the variance bound is now dominating. Figure \ref{fig:mse_toy} displays the error bound according to Theorem~\ref{thm:BAEideal} for the case of $\bSigma=\bI$ and $\bSigma$ being a blockwise diagonal matrix with $2\times 2$ blocks whose off-diagonal element is $\rho$. The results favor BAE, especially for the blockwise $\bSigma$ with large within-block off-diagonals.


\subsection{Construction of blocks} \label{subsec:implement}
We provide two approaches of selecting the blocks.  The first approach works well when the true $\bSigma$ is approximately block-wise diagonal, such as example on the returns of the S\&P 500 components(see Figure~\ref{fig:cov-stocks}). The second approach is a random scheme and works for general settings.

\begin{figure}        \centering
                \includegraphics[width=.45\textwidth, height=.33\textwidth, trim=20 10 0 0, clip=true]{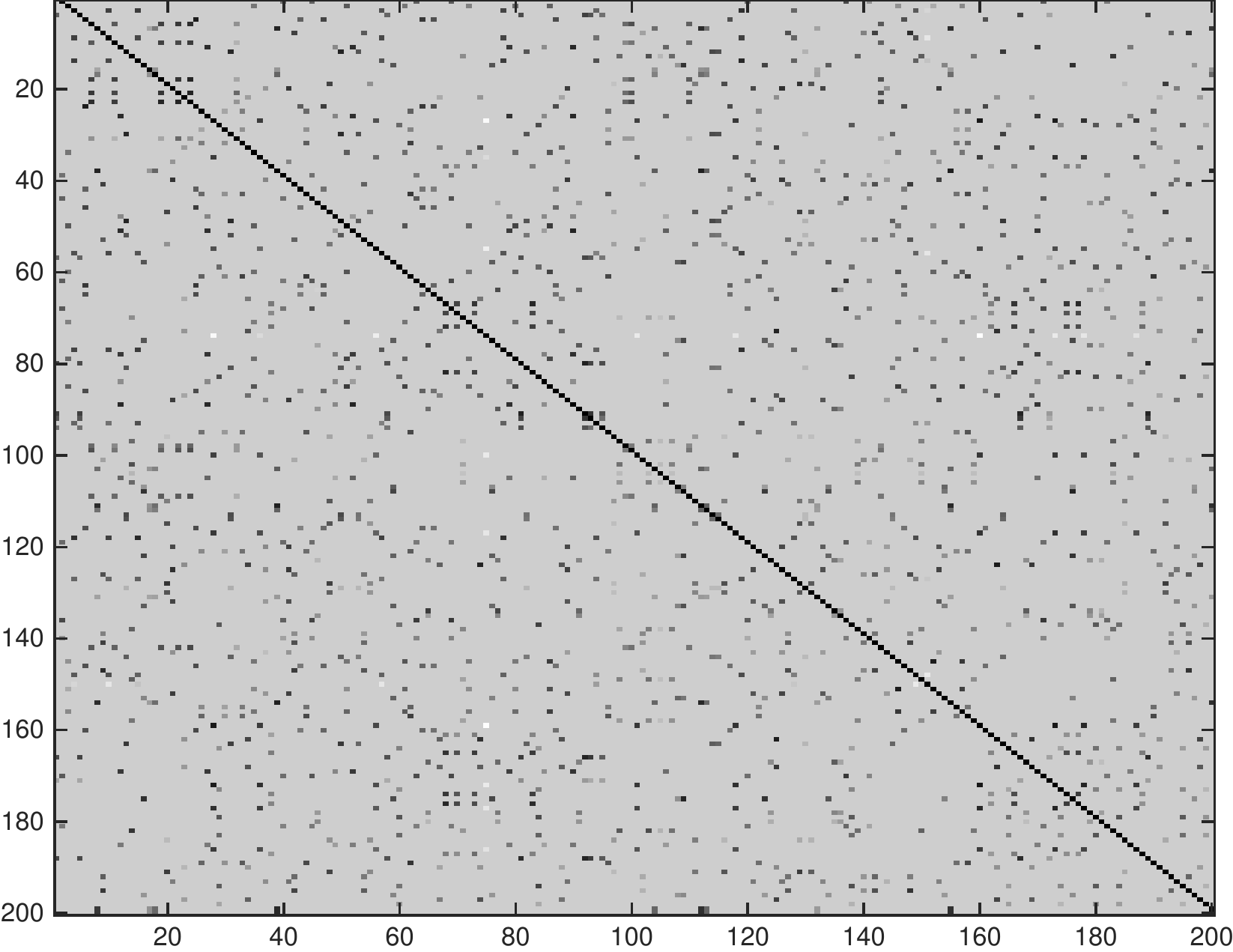} $\;\;$
                \includegraphics[width=.45\textwidth, height=.33\textwidth, trim=20 10 0 0, clip=true]{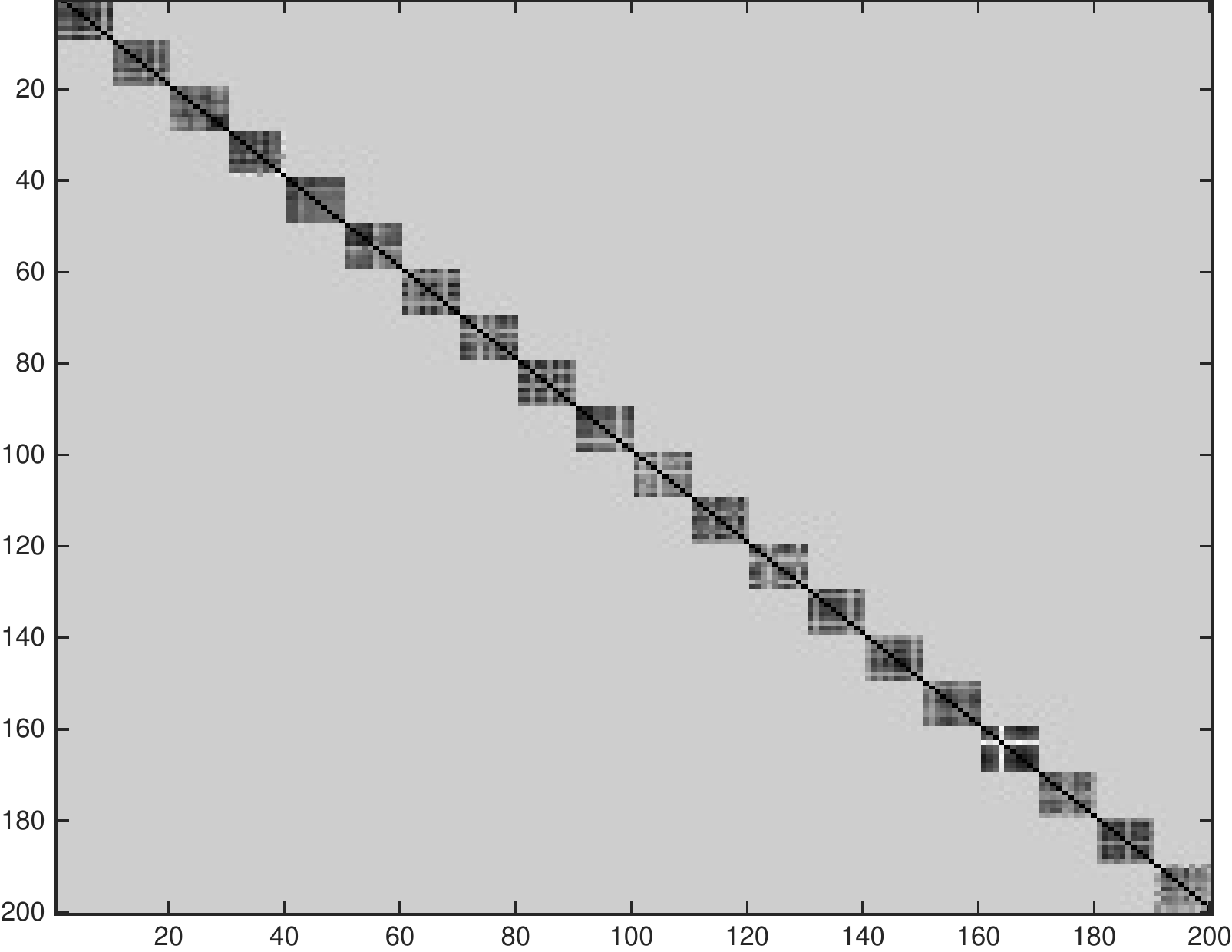}
                \caption{Construction of a blockwise correlation matrix by thresholding. Left panel: Graph of the original correlation matrix. Right panel: Transformation into a block diagonal correlation matrix.}\label{fig:sparse-to-block}
\end{figure}

\medskip
\noindent
{\it BAE1: Constructing blocks from a raw estimate of $\bSigma$.}
Let $\widetilde{\bSigma}$ be a raw estimate of $\bSigma$; for example, it can be the sample covariance matrix or the robust estimator of $\bSigma$ in Section~\ref{sec:simulations}. Fixing a threshold $t\in (0,1)$, we define a graph ${\cal G}_t$ with nodes $\{1,2\cdots, p\}$, where there is an undirected edge between nodes $i$ and $j$ if and only if the estimated absolute correlation exceeds $t$, namely,
$$
    |\widetilde{\Sigma}(i,j)|/\sqrt{\widetilde{\Sigma}(i,i)\widetilde{\Sigma}(j,j)}>t, \qquad
  \mbox{for $1\leq i<j\leq p$}.
$$
The nodes of this graph uniquely partitions into components (a component of a graph is a maximal connected subgraph). We propose using
\[
{\cal A}=\{ \mbox{all components of }{\cal G}_t \}.
\]
See Figure \ref{fig:sparse-to-block} for an illustration of this procedure.

This approach guarantees that all blocks are non-overlapping. Numerical evidence suggests that it performs well with an appropriate choice of $t$, especially when the true $\bSigma$ is blockwise diagonal. However, the threshold $t$ is a tuning parameter, and it can be inconvenient to select $t$ in a data-driven fashion. Below, we introduce a tuning-free approach.

\medskip
\noindent
{\it BAE2: Randomly selecting pairs as blocks.}
In this approach, we let
\[
{\cal A}=\bigl\{ \mbox{$p$ pairs uniformly drawn from $\{(i,j): 1\leq i<j\leq p\}$ without replacement}\bigr\}.
\]
This approach is designed for block size equal to $2$, and the obtained blocks may overlap. Although it sounds ad-hoc, this approach has an appealing numerical performance.  When the number of pairs are sampled sufficiently large, by the law of large numbers, it approaches the all pairwise aggregation estimator and this explains why the approach has an appealing numerical performance.  This approach can easily be extended to blocks of any size that is smaller than $n$ so long as the estimated covariance matrix for each block can be easily inverted and estimated well.

\section{Simulations}\label{sec:simulations}
We investigate the performance of estimators on extensive simulations. To have realistic  simulation settings, we use a $\bSigma$ calibrated from stock returns. The calibration procedure is the same as that in \cite{xiaofeng} and \cite{fanmincheva}. Fix $p$. We take the daily returns of $p$ companies in S\&P 500 index with the largest market capitalization from July 1st, 2013 to June 29th, 2018 (data were downloaded from the COMPUSTAT website). We fit the Fama-French three-factor model to the excess returns $\{\Yb_t\}_{t=1}^T$:
$$
    \bY_t = \ba + \bB \fb_t + \bu_t,
$$
where $\bB\in\mathbb{R}^{p\times 3}$ is the factor loading matrix, $\fb_t\in\mathbb{R}^3$ denotes the Fama-French factors with covariance matrix $ \cov(  \fb_t )\in\mathbb{R}^{3\times 3}$ and $\bu_t$ is the idiosyncratic component.  This factor model induces a covariance structure for $\bY_t$:
\[
\bSigma_Y= \cov(  \Yb) =\bB \cov(  \fb_t )\bB^{\T}+\bSigma_u,
\]
where  $\bSigma_u$ is the covariance matrix of idiosyncratic noise $\bu_t$. We downloaded the factors $\{\fb_t\}_{t=1}^T$ from the Kenneth French data library and used the method in \cite{fanmincheva} with the recommended threshold (for estimating sparse $\bSigma_u$) to get and estimate $\hbSigma_Y$. We then use $\hbSigma_Y$ as the true $\bSigma$ to generate data from model \eqref{Ydecomposition}.

When implementing the estimators, we plug in two different estimators of $(\bmu,\bSigma)$. The first choice is to use sample mean and sample covariance matrix. The second choice is to use robust M-estimators, called adaptive Huber estimator  \citep{FLW2017, sun2016adaptive}, which are designed for heavy-tailed data. These estimators lead to better large-deviation bounds. In detail,
for a tuning parameter $\tau>0$ chosen by cross-validation, we estimate $\bmu$ by $(\widehat{\mu}_1,\ldots,\widehat{\mu}_p)^{\T}$, where
\[
	\widehat{\mu}_j  =  \argmin_{ \beta  \in \bbr} \sn \ell_{\tau}( Y_{ij}  - \beta ),  \qquad\mbox{with}\quad \ell_\tau(u) = \begin{cases}
		\frac{1}{2} u^2,  &~\mbox{ if } |u| \leq \tau, \\
		\tau |u| - \frac{1}{2} \tau^2, &~\mbox{ if } |u| > \tau,
	\end{cases}
\]
the Huber loss. We estimate $\bSigma$ by $(\widehat{\sigma}_{jk})_{1\leq j,k\leq p}$, where
\begin{align*}
\widehat{\sigma}_{jj}   =
	\widehat{\beta}_j -  \widehat{\mu}_j^2\cdot 1\{\widehat{\beta}_j  > \widehat{\mu}_j^2 \},\qquad   &\mbox{ with }\quad \hat{\beta}_j = \argmin_{\beta >0} \sn \ell_{ \tau_{jj} } \bigl(Y_{ij}^2 - \beta \bigr),\cr
	\widehat{\sigma}_{jk} = \widehat{\beta}_{jk} -  \widehat{\mu}_j \widehat{\mu}_k, \qquad &\mbox{with}\quad \widehat{\beta}_{jk} = \argmin_{\beta \in \bbr} \sn \ell_{\tau_{jk} } \bigl(Y_{ij}Y_{ik} - \beta \bigr).
\end{align*}
Here, each tuning parameter $\tau_{jk}$ is selected via cross-validation using the data $\{(Y_{ij},Y_{ik})\}_{i=1}^n$.

\begin{figure}[!t]
        \centering
                \includegraphics[width=.48\textwidth,height=.42\textwidth]{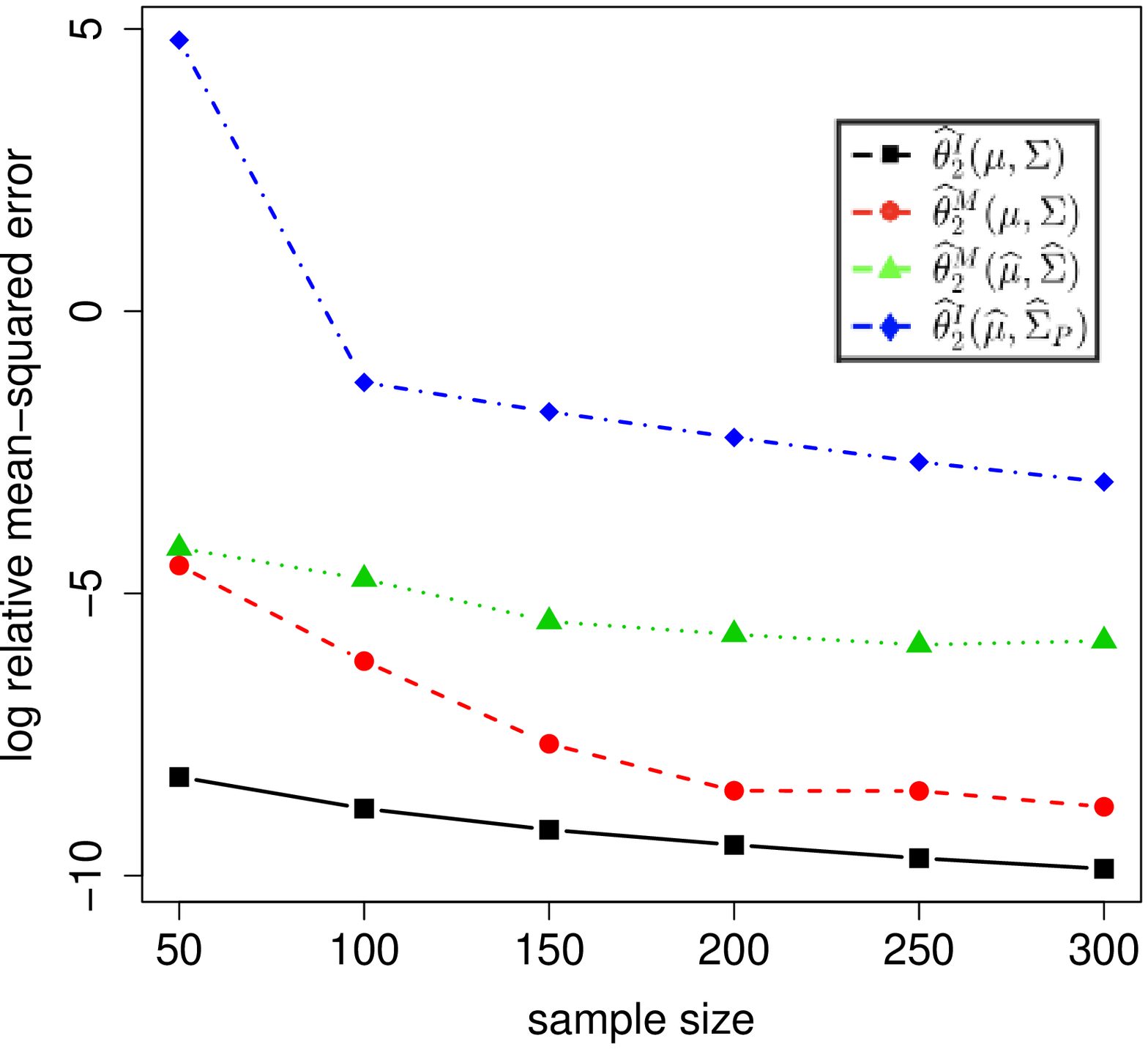}
                \includegraphics[width=.48\textwidth,height=.42\textwidth]{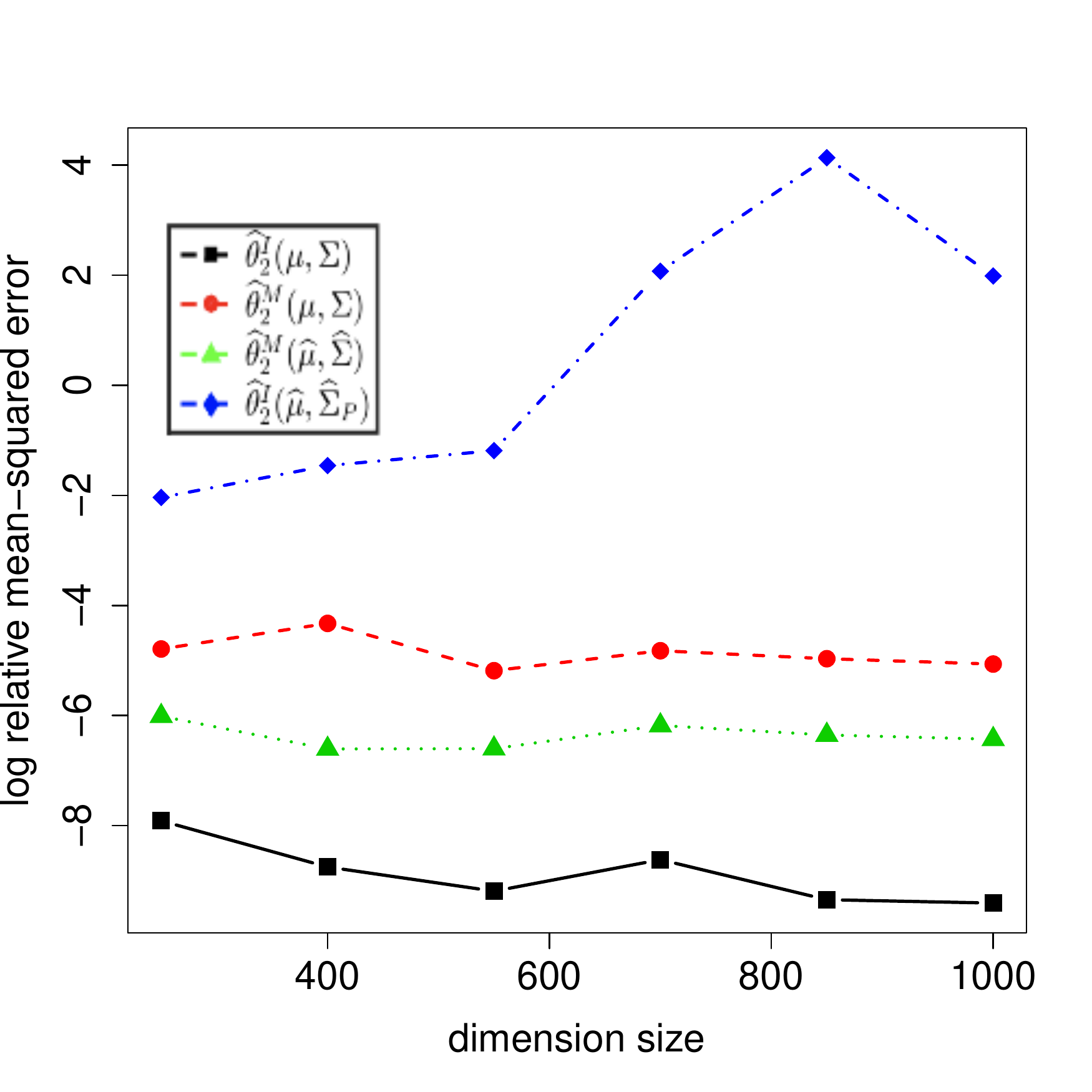}
                 \includegraphics[width=.48\textwidth,height=.42\textwidth]{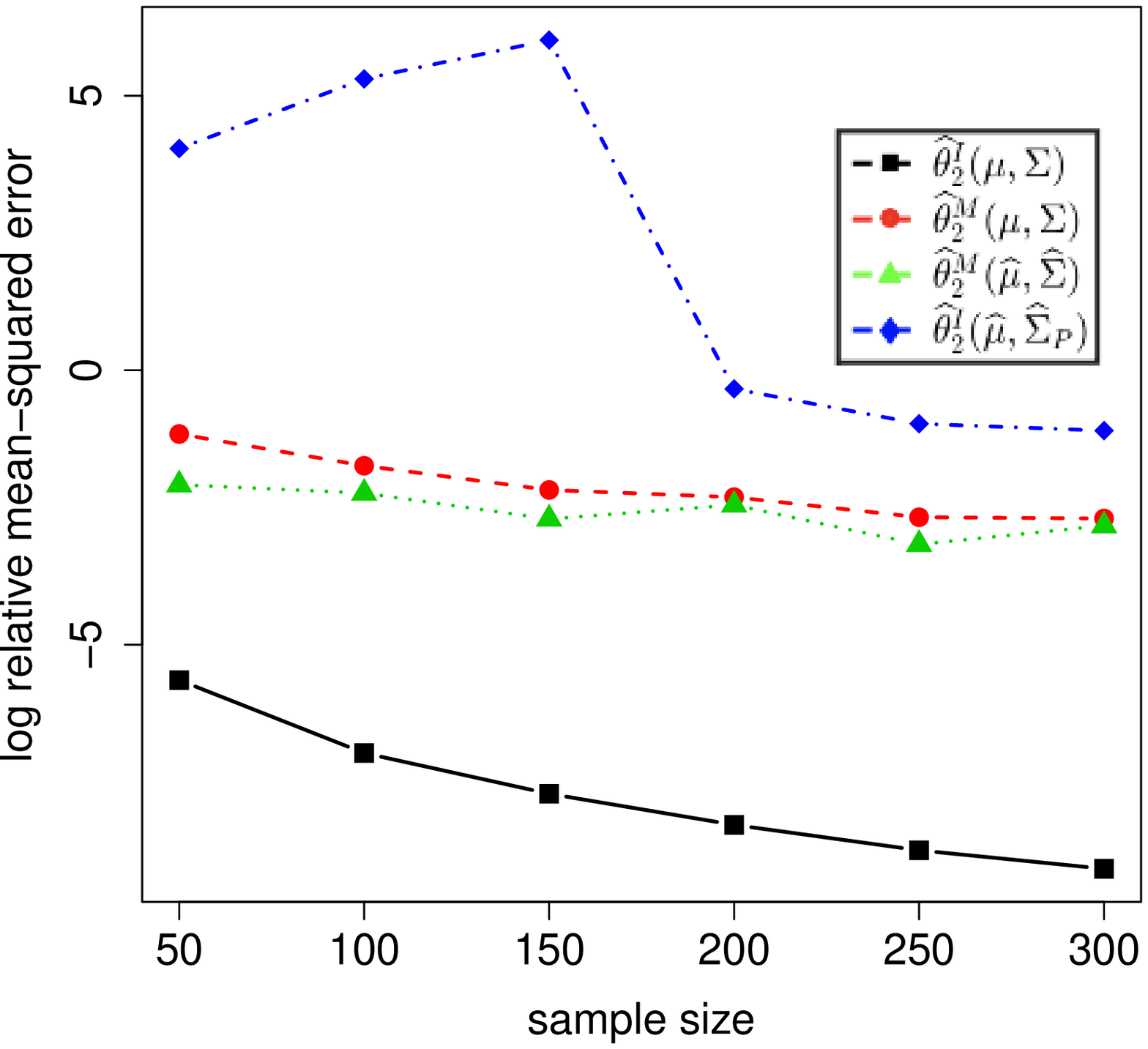}
                \includegraphics[width=.48\textwidth,height=.42\textwidth]{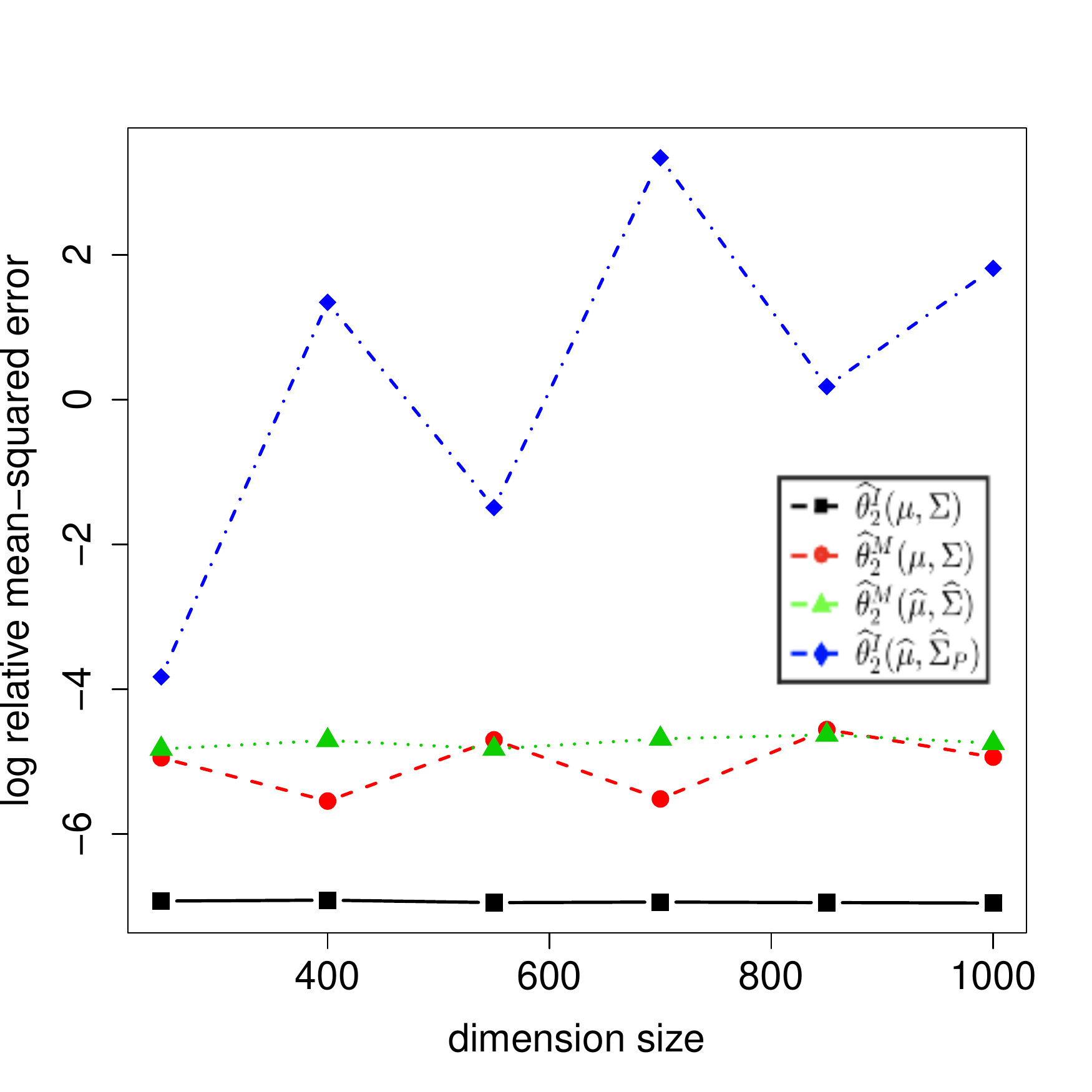}
        \caption{Experiment 1 (MAE). Top two panels: Experiment 1.1\&1.2 (multivariate Gaussian data). Bottom panels: Experiment 1.3\&1.4 (multivariate $t$ data).  Errors are the average of $200$ repetitions. Black-squared for the ideal-estimator $\hat \theta_2^I(\bmu, \bSigma)$; blue-diamond for the plug-in estimator $ \hat \theta_2^I(\hat \bmu, \hat \bSigma)$; red-dot for  the MAE  $\hat \theta_2^M(\bmu, \bSigma)$; green-triangle for the plug-in MAE $\hat \theta_2^M(\hat \bmu, \hat \bSigma)$}       \label{fig:Experiment1}
\end{figure}

\paragraph{Experiment 1: Performance of MAE.} Fix $m=2$. We consider four sub-experiments:
\begin{itemize} \itemsep -2pt
\item {\it Experiments 1.1 and 1.3}: We fix $p=500$ and let $n$ vary in $\{50, 100, 150, 200, 250, 300\}$. The data follow multivariate Gaussian distributions (Experiment 1.1) or  multivariate $t$-distributions with degrees of freedom equal to $4.5$ (Experiment 1.3).
\item {\it Experiments 1.2 and 1.4}: We fix $n=100$ and let $p$ vary in
$\{250, 400, 550, 700, 850, 1000\}$. The data follow multivariate Gaussian distributions (Experiment 1.2) or  multivariate $t$-distributions with degrees of freedom equal to $4.5$ (Experiment 1.4).
\end{itemize}
In all settings, $p>n$, so we focus on the challenging case of high-dimensionality. For each setting, we compare four estimators:
\begin{itemize} \itemsep -2pt
\item $\widehat{\theta}^{\, \rI}(\bmu,\bSigma)$: Ideal Estimator, which knowns $(\bmu,\bSigma)$.
\item $\widehat{\theta}^{\, \rM}(\bmu,\bSigma)$: MAE with given $(\bmu,\bSigma)$.
\item $\widehat{\theta}^{\, \rM}(\hbmu,\hbSigma)$: MAE, where $(\bmu,\bSigma)$ are estimated using the sample mean/covariance matrix in Experiment 1.1\&1.2 and using the aforementioned robust-M estimators for Experiment 1.3\&1.4.
\item $\widehat{\theta}^{\, \rI}(\hbmu,\hbSigma_P)$: Plug-in Ideal Estimator, with plugged-in estimators of $(\bmu,\bSigma)$. We use the sample mean to estimate $\hbmu$ and use POET \citep{fanmincheva} (with a default threshold) to estimate $\bSigma$.
\end{itemize}
The results are presented in Figure~\ref{fig:Experiment1}, where the $y$-axis is $\log\{(\widehat{\theta}_2 / \theta_2 - 1)^2\}$, based on the average over $200$ repetitions. As we have expected, the Ideal Estimator always gives the lowest error, however, such an estimator is not practically feasible. Instead, we plug estimates of $(\bmu,\bSigma)$ into the Ideal Estimator to make it practically feasible, then it has an unsatisfactory performance; this confirms our previous insight about the drawback of the plug-in estimator. Our proposed MAE works well, always significantly better than the plug-in estimator. The performance of MAE becomes better as the sample size $n$ grows, and its performance stays relatively stable as the dimension $p$ grows. This is desirable: our proposed estimator doesn't face any curse of dimensionality. The results are similar for the multivariate Gaussian data and the multivariate $t$-data, except that for Gaussian data, MAE with $(\hbmu,\hbSigma)$ even outperforms MAE with true $(\bmu,\bSigma)$. One possible reason is the self-normalization phenomenon: An estimator, when divided by its sample variance, gives better performance than that divided by the true variance.

\paragraph{Experiment 2: Confidence Interval.}
For each of the experiments above: {\it Experiments 1.1, 1.2, 1.3 and 1.4}, we calculate the probability that the true value of $\theta_2$ lies in the confidence interval derived in Theorem \ref{thm:normality} and presented in Equation \eqref{conf.interval}. In Table \ref{tab:CI}, we see that for a $95\%$ confidence interval, the empirical coverage probabilities are close to the confidence level.

\begin{table}[tbh]\centering
\caption{Empirical coverage probability that $\theta_2$ lies in the $95\%$ confidence interval by Equation \eqref{conf.interval} for data following multivariate Gaussian or multivariate $t$-distributions, across a variety of settings.}
\label{tab:CI}
\begin{tabular}{c|rrrrrrrrr}
\hline
 $n = 100$   & $ p=$ 250& 400 & 550  & 700  &  850   & 1000	\vspace{0.1cm} \\
 \hline 
Gaussian& 92.0\% & 95.0\% & 93.5\%&95.5\%& 95.5\% &96.5\% \\
Student's $t$&96.5\%  &  98.0\%  &  94.5\% & 97.0\%  & 96.0\% & 96.5\%    \\
\hline 
   $p = 500$&$n=$ 50 & 100 & 150  & 200  &  250   & 300   \vspace{0.1cm} \\
 \hline  
 Gaussian& 95.5\%  &  94.2\%  &  93.5\% & 93.0\% & 95.5\% & 94.0\%    \\
Student's $t$&  98.0\%  &  96.0\%  &  95.5\% & 93.5\%  & 94.5\% & 97.0\%    \\
\hline
\end{tabular}
\end{table}


\paragraph{Experiment 3: Performance of BAE.}
We study whether BAE, which uses a block of coordinates at a time and takes advantage of the correlation structure, can further improve the performance of MAE. The four sub-experiments, Experiments 3.1-3.4, have the same settings as those of Experiments 1.1-1.4. When implementing BAE, we use the second approach in Section~\ref{subsec:implement} to choose the blocks; note that the blocks all have a size $2$ and may overlap. We use the sample mean/covariance to estimate $(\bmu,\bSigma)$ for multivariate Gaussian data and the robust M-estimators for multivariate $t$ data. Since we focus on the comparison between MAE and BAE, we do not report the errors of the Ideal Estimator and plug-in estimator in this experiment.

\begin{figure}[!t]
        \centering
                \includegraphics[width=.48\textwidth,height=.42\textwidth]{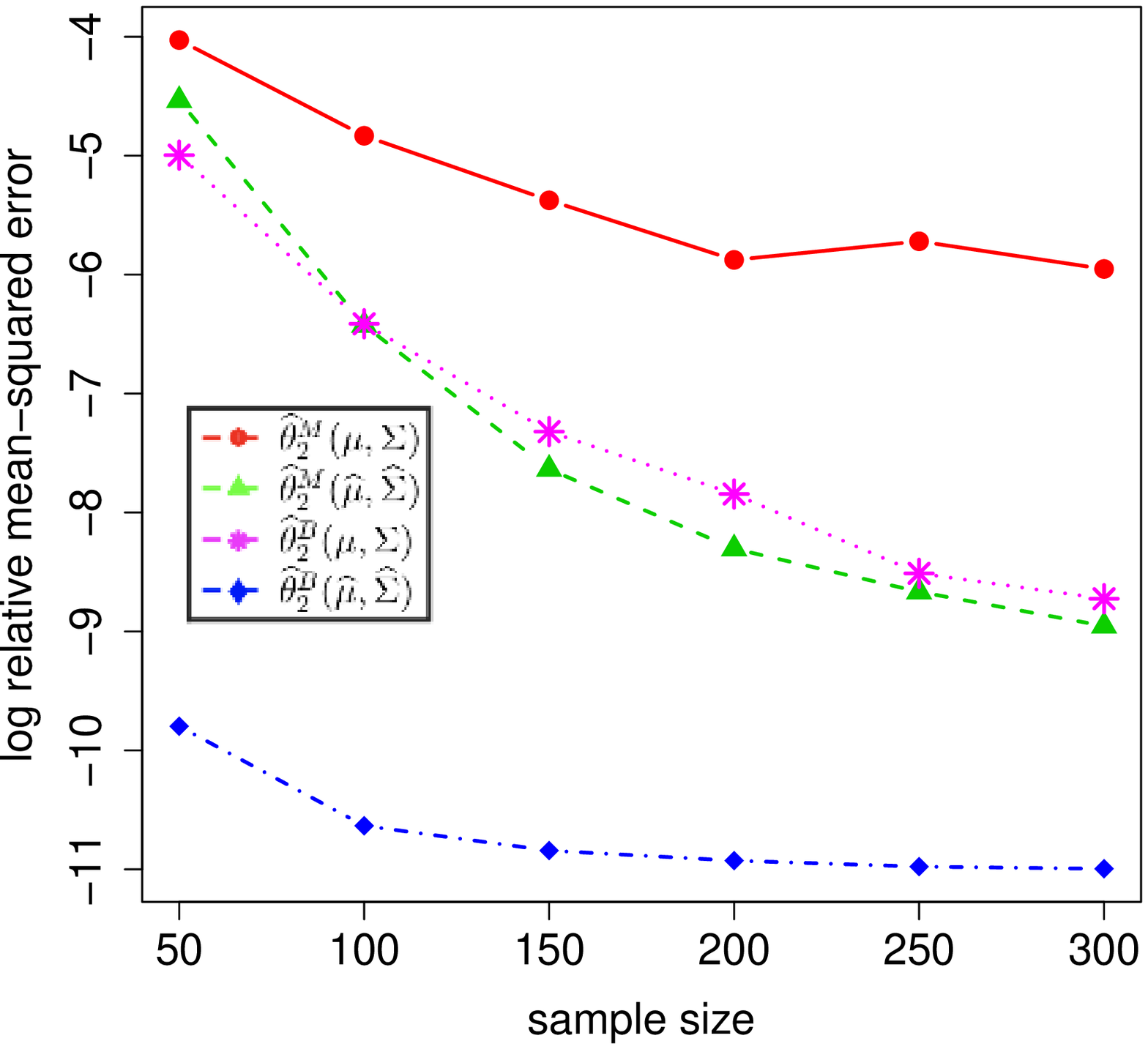}
                \includegraphics[width=.48\textwidth,height=.42\textwidth]{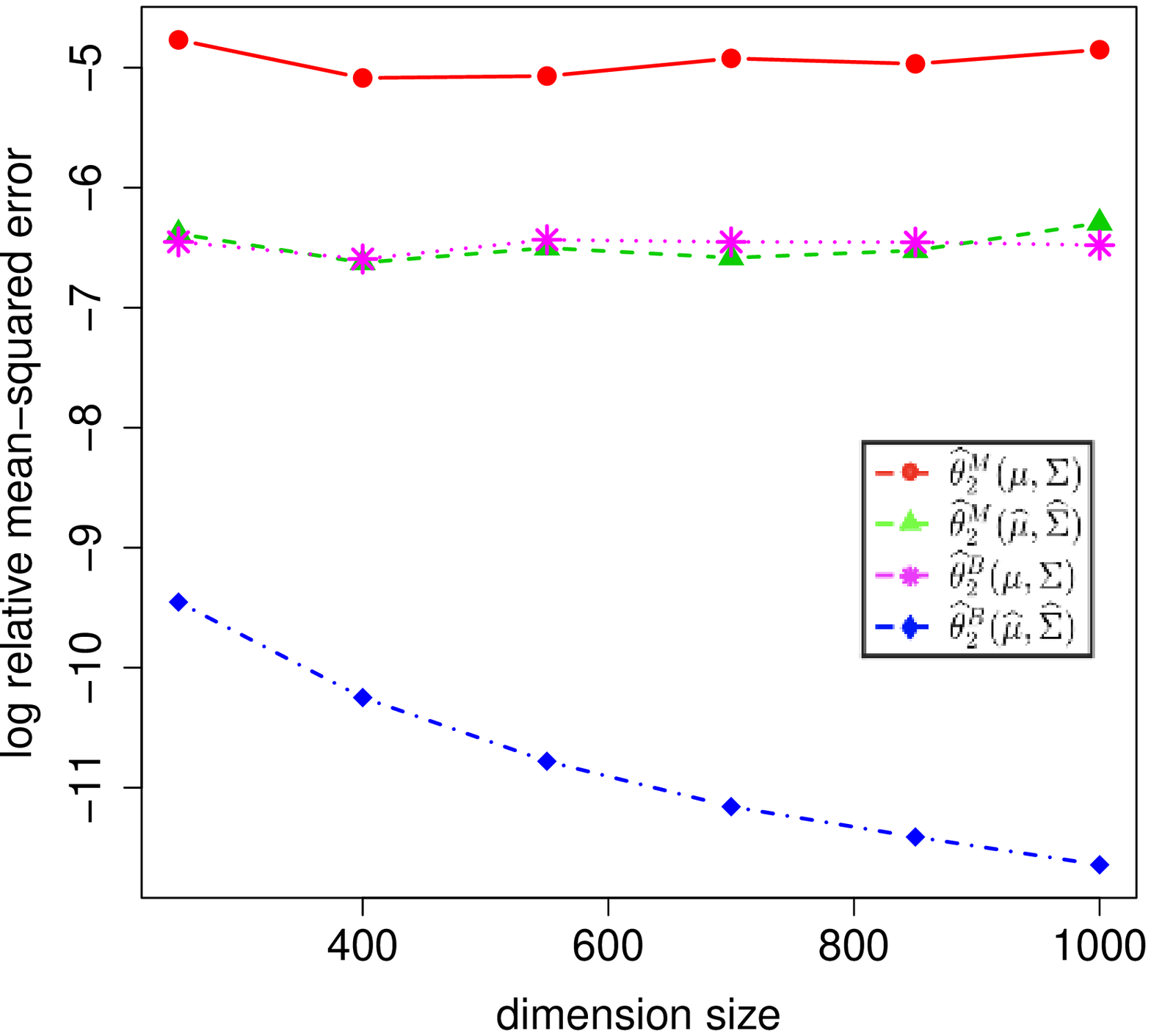}
                \includegraphics[width=.48\textwidth,height=.42\textwidth]{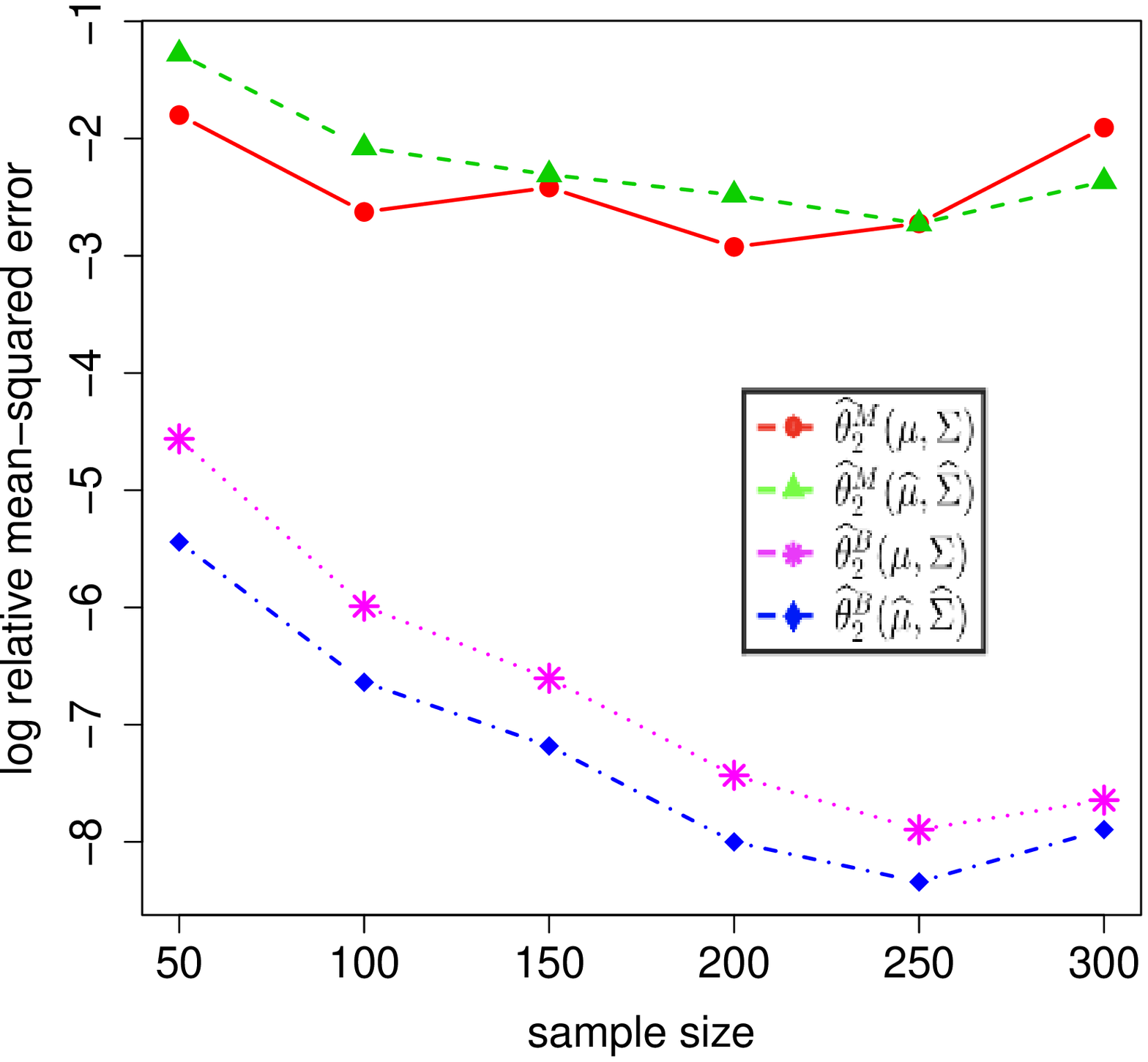}
                \includegraphics[width=.48\textwidth,height=.42\textwidth]{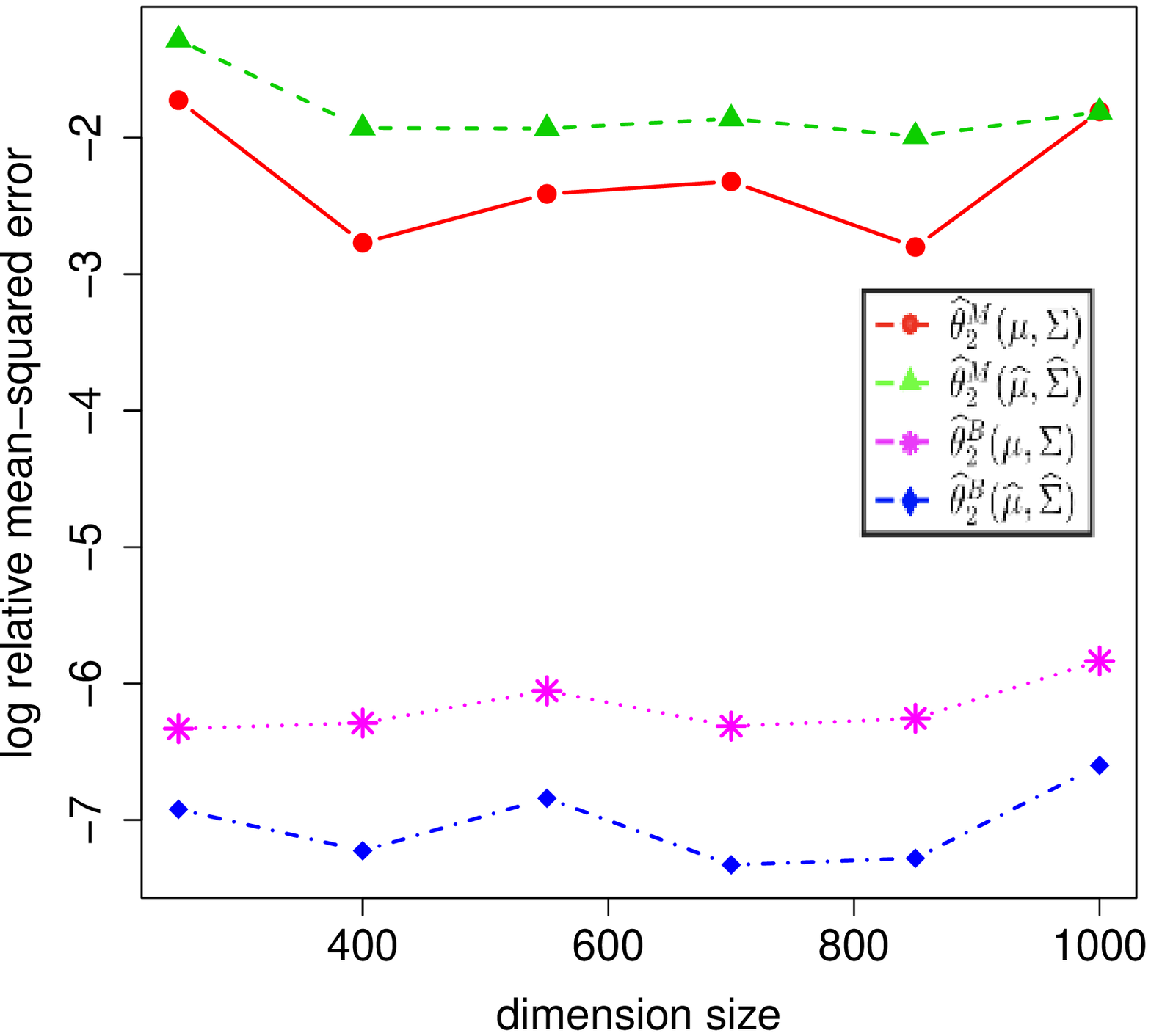}
  \caption{Experiment 3 (BAE). Top two panels: Experiment 3.1\&3.2 (multivariate Gaussian data). Bottom panels: Experiment 3.3\&3.4 (multivariate $t$ data).  Errors are the average of $200$ repetitions.
  Magenta-star for the BAE $ \hat \theta_2^{\rB}(\bmu,  \bSigma)$;
  blue-diamond for the plug-in BAE $\hat \theta_2^{\rB}(\hat \bmu, \hat \bSigma)$; red-dot for  the MAE  $\hat \theta_2^M(\bmu, \bSigma)$; green-triangle for the plug-in MAE $\hat \theta_2^M(\hat \bmu, \hat \bSigma)$}          \label{fig:Experiment3}
\end{figure}

The results are presented in Figure~\ref{fig:Experiment3}. First, we can see that BAE improves the performance of MAE, especially when $p$ is large. Second, the self-normalization phenomenon is also observed: BAE with $(\hbmu,\hbSigma)$ even outperforms BAE with true $(\bmu,\bSigma)$, especially for Gaussian data.

\section{Application: Estimating realized $\xi_t$ in a time series} \label{sec:realized_xi}
Given the returns of a panel of stocks, we are interested in extending the idea of MAE to provide a daily {\it risk index} for the whole panel of stocks. We cast it as the problem of estimating the realized $\xi_t$ in a multivariate time series with elliptically-distributed noise. Let $\Yb_1,\ldots,\Yb_T\in\mathbb{R}^p$ be the returns of $p$ stocks during a time period of $T$ days. We extend model \eqref{Ydecomposition} to an elliptical model for multivariate time series
\beq \label{stock-factor}
\Yb_t = \bmu_t + \bB  \fb_t + \xi_t \bSigma_t^{1/2}\Ub_t, \qquad t = 1,\cdots, T,
\eeq
where $\bmu_t$ is the time-varying mean, $\fb_t \in \mathbb{R}^K$ is a vector of $K$ factors, and $\bB$ is a $p \times K$ matrix of factor loadings. We are interested in estimating the daily realized $\xi_t$.

Our method has four steps:
\begin{description}
\item 1. Estimate $\bmu_t$. For daily or higher frequency data, we set  $\hbmu_t\equiv\bzero$, since it is commonly believed that the short-time returns are not predictable. For weekly or monthly data, we estimate $\bmu_t$ by the weekly or monthly average.
\item 2. Obtain the factor-adjusted returns $\widehat{\Zb}_t$. Let $\widehat{\fb}_t\in\mathbb{R}^K$ contain either observed factors or data-drive factors from PCA \citep{fanmincheva}. We then follow the approach in \cite{fanmincheva} to get $\widehat{\bB}$, the estimated factor loading matrix. Let 
\[
\widehat{\Zb}_t = \bY_t -\widehat{\bmu}_t- \widehat{\bB}  \widehat{\fb}_t, \qquad t = 1,\cdots, T.
\] 
\item 3. Estimate $\bSigma_t$.  We assume $\bSigma_t$ is a diagonal matrix and estimate its diagonal elements by fitting an ARCH model on each coordinate of $\Zb_t$. In detail, for each $1\leq j\leq p$, let $Z_t(j)$ be the $j$-th coordinate of $\Zb_t$. We assume there is idiosyncratic noise $\{\epsilon_t(j)\}_{t=1}^T$ such that
\[
Z_t(j)= \lambda_t(j) \epsilon_t(j), \qquad \mbox{where}\quad  \lambda^2_t(j) =a_0+ a_1Z^2_{t-1}(j)+ \ldots + a_kZ^2_{t-k}(j),
\]
where $k$ is the order of ARCH model and $(a_0,\ldots,a_k)$ are parameters. 
We estimate $(a_0,\ldots,a_k)$ using the conditional maximum likelihood estimator and then construct $\{\widehat{\lambda}_t(j)\}_{t=1}^T$. Let 
\[
\hbSigma_t=\mathrm{diag}\bigl(\widehat{\lambda}_t(1),\ldots,\widehat{\lambda}_t(p)\bigr).
\]
\item 4. Estimate $\xi_t$. We adapt the idea of MAE to the current setting. Let $\Zb_t=\Yb_t- \bB\fb_t$. Our model becomes $\Zb_t=\xi_t\bSigma_t^{1/2}\Ub_t$, i.e., the $j$-th component of $\Zb_t$ is $Z_t(j)=\xi_t (\bSigma^{1/2}\Ub_t)_j$. It follows that  
\[
\xi_t^{2}=\frac{Z^2_t(j)}{(\bSigma^{1/2}\Ub_t)_j^2}\approx \frac{Z^2_t(j)}{\E[(\bSigma^{1/2}\Ub_t)_j^2]}= \frac{pZ^2_t(j)}{\Sigma_t(j,j)},
\]
where $\Sigma_t(j,j)$ is the $j$-th diagonal of $\bSigma_t$. Here,  
the last equality is due to $c_1=1$ in Equation \eqref{cmp}. We approximate $(\Zb_t,\bSigma_t)$ by $(\widehat{\Zb}_t, \hbSigma_t)$ and get a marginal estimator of $\xi_t^2$: $\widehat{\xi}^2_{t,j}=p\widehat{Z}_t(j)/\widehat{\Sigma}_t(j,j)$. We then aggregate them: 
\beq \label{marginal_xi}
\widehat{\xi}_t^2= \sum_{j=1}^p \frac{\widehat{Z}^2_t(j)}{\widehat{\Sigma}_t(j,j)}, \qquad t=1,2,\ldots,T.
\eeq
\end{description}
In Section~\ref{sec:Sec5Simu} of the appendix, we investigate the performance of our estimator in simulations. Under a variety of settings, our estimated curve of $\widehat{\xi}_t$ fits  the true curve of $\xi_t$ very well. See details therein.  

\begin{figure}[!t]
        \centering
                \includegraphics[width=.6\textwidth]{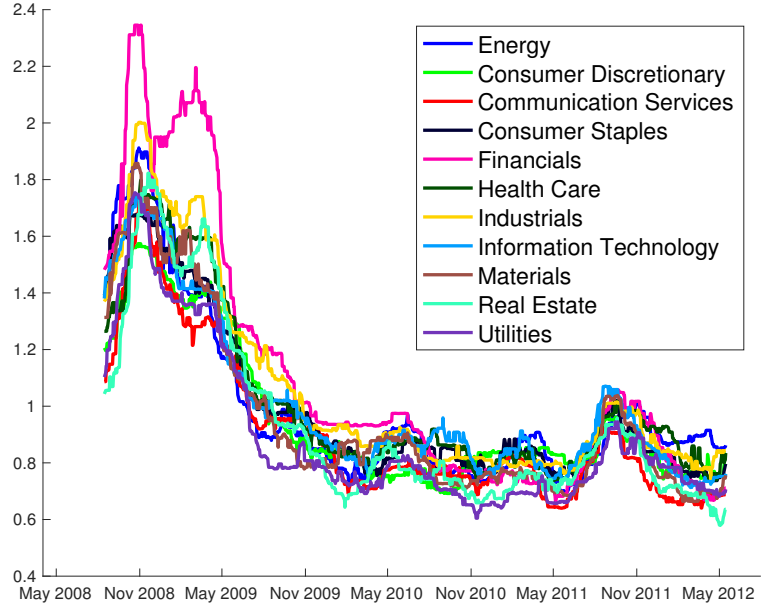}
               \caption{The estimated $\widehat{\xi}_t$ for 11 GCIS sectors. For a better representation, we have smoothed the curves by taking a moving average on a $65$-day window.}
                \label{fig:xi_GCIS}
\end{figure}

We applied our estimator to the S\&P 500 stock returns. We took the daily returns of $300$ stocks from the S\&P 500 index with the largest market capitalization, from July 1, 2008 to June 29, 2012. Each stock is assigned a Global Industry Classification Standard (GCIS) code. The GCIS code divides $300$ stocks into eleven sectors: Energy, Consumer Discretionary, Communication Services, Consumer Staples, Financials, Health Care, Industrials, Information Technology, Materials, Real Estate, and Utilities. We applied our estimator to stocks in each sector. When implementing our method, we set $\hbmu_t\equiv{\bf 0}$ in Step 1, used three observed Fama-French factors as $\widehat{\fb}_t$ in Step 2, and set the order of ARCH model to $k=2$ in Step 3. 

The curves of estimated $\widehat{\xi}_t$ for $11$ sectors are displayed in Figure~\ref{fig:xi_GCIS} (the curves are smoothed by taking a moving average on a $65$-day window). The estimated $\widehat{\xi}_t$ for all sectors largely synchronize, reaching their peaks during the 2008 financial crisis. In the crisis, the estimated $\widehat{\xi}_t$ for the Financials sector is significantly larger than that of other sectors. The large value of $\widehat{\xi}_t$ for the Financials sector remains in the post-crisis period until May, 2009. We also computed the pairwise correlations among $\widehat{\xi}_t$ of $11$ sectors, as shown in Table~\ref{tb:GCIS_correlations}. It suggests that the $\widehat{\xi}_t$ for the Energy sector and the Financials sector are highly correlated with each other. These two sectors are also highly correlated with sectors of Materials, Real Estate, and Utilities. In comparison, for the Consumer Discretionary sector and Information Technology sector, their $\widehat{\xi}_t$ are less correlated with those of other sectors.


\begin{table}[!bt]
\centering
\caption{Pairwise correlations of $\widehat{\xi}_t$ across GCIS sectors. Numbers $\geq .45$ are marked in circles.} \label{tb:GCIS_correlations}
\vspace{-5pt}
\scalebox{.8}{
\begin{tabular}{l | ccccccccccccccc}
& E & CD & CO & CS & F & HC & IN & IT & M & R & U \\
\hline
Energy (E) & --  &  .36  &  .43  &  .42  & \circled{.53}  &  .37  &  .44  &  .35  &  \circled{.48} & \circled{.47} &  \circled{.46}\\
Consumer Discretionary (CD) &  .36  & --  &  .30 &   .35  &  .43  &  .34  &  .37  &  .33  &  .33 & .36    & .35\\
Communication Services (CO) &   .43  &  .30  &  --   & .33  &  .43  &  .33  & .40  & .32  & .39&  .44  &  .41\\
Consumer Staples (CS) &  .42 &  .35  & .33 &  --  & .42  & .36  &  .37 &  .35 &  .38 &  .38 &     .43\\
Financials (F) &   \circled{.53}  &  .43  &  .43 &   .42  &  --  &  .42  &  \circled{.50} &    .40  &  \circled{.49} & \circled{.52}  &  \circled{.52} \\
Health Care (HC) &   .37  &  .34   & .33 &  .36  &  .42 &  --   & .40  &  .39  &  .37 & .36  &  .33\\
Industrials (IN) &  .44  &  .37  &  .40  &  .37  &  \circled{.50}  &  .40  &  --  &  .39  &  \circled{.49} & .44  &  \circled{.45}\\
Information Technology (IT) &  .35 &   .33 &  .32  &  .35 &  .40 &  .39   &  .39 &  -- &  .36 & .33  &  .34\\
Materials (M) &  \circled{.48}  &  .33  &  .39  &  .38  &  \circled{.49}  &  .37 &  .49 &  .36   & -- & .42  &  .43 \\
Real Estate (R) &   \circled{.47}  &  .36  &  .44  &  .38  &  \circled{.52}  &  .36  &  .44 &   .33  &  .42 & --  &  \circled{.46}\\
Utilities (U) & \circled{.46} &   .34 &  .41  &  .43  &  \circled{.52}  &  .33  & .45 &  .34  & .43 & .46 & --
\end{tabular}}
\end{table}

\section{Discussion} \label{sec:Discu}
In this paper, we consider the problem of estimating the even moments of $\xi$ in an elliptical distribution  $\Yb =  \bmu+\xi\bSigma^{1/2} \Ub$. A natural idea is the plug-in estimator \citep{maruyama,fan2013quadro}, which requires an estimator $\hbOmega$ of the precision matrix and whose performance crucially relies on structural assumptions on $\bOmega$ or $\bSigma$. Instead, we propose a marginal aggregation estimator (MAE) that only needs to estimate the diagonal of $\bSigma$. Our approach validates the insight that estimating a large precision matrix is statistically more challenging than estimating a moment parameter---it is unnecessary to use the sledge hammer to crack an egg. We prove that MAE is root-$n$ consistent, under {\it no} conditions on $\bSigma$ or $\bOmega$. We also show that MAE achieves the first-order efficiency, with an asymptotic variance matching with the variance of an ideal estimator when $(\bmu,\bSigma)$ are given. We further generalize MAE to a block-wise aggregation estimator (BAE) that needs to estimate small-size diagonal blocks of $\bSigma$. BAE takes advantage of correlations among coordinates and improves MAE on the second-order efficiency. Our proposed estimators are conceptually simple and easy to implement.

Estimating the moment parameters of an elliptical distribution is useful in quadratic discriminant analysis \citep{fan2013quadro} and estimating tail behavior of financial returns \citep{fama, Bollerslev, keller, frahm03, cizekbook}. In an application on the stock returns, we propose a multivariate time series model with factor structures and elliptically distributed idiosyncratic noise. We extend MAE to an estimator for estimating the day-to-day value of $\xi_t$. We apply the method to stocks of each industry sector. It produces an ``tail index" for each industry sector. These tail indices reveal interesting difference among industry sectors, especially during the financial crisis.

The study leaves a few open questions for future work. The first is how to improve the estimators for heavy tailed data. Our current approach plugs into MAE the robust estimators of mean and covariance matrix. Instead, we may construct a robust M-estimator for simultaneously estimating $(\theta_m, \mu_j, \sigma_{jj})$ with marginal data and then aggregate these marginal estimators of $\theta_m$ in a similar way. We hope such an approach helps remove the $\sqrt{\log(p)}$-factor in the error rate of Theorem~\ref{thm:consistency}. The second is the optimal strategy of constructing blocks in BAE. There is a trade-off in choosing the blocks: With larger blocks, it reduces the variance of the estimator when true $(\bmu,\bSigma)$ are plugged in, but at the same time, the errors of estimating diagonal blocks of $\bSigma$ increase. How to construct the blocks in a data-driven way is an interesting question. Third, the current theory for BAE assumes non-overlapping blocks. The results can be extended to overlapping blocks, with nontrivial efforts. We leave it for future work. The last problem is to extend our estimators to time dependent data, where the distribution of $\xi$ have change-points. For financial data, such change-points may relate to financial boom or crisis. We propose a kernel-smoothed version of MAE: Given data $\{\Yb_t\}_{t=1}^n$, for a kernel function $K_h(\cdot)$ with bandwidth $h$, let
\[
\widehat{\theta}_{m,t} =\frac{1}{\sum_{s=1}^n K_h(s-t)}\sum_{s=1}^n K_h(s-t)  \biggl[ \sum_{j=1}^p \frac{p^{-1} (Y_{s,j}-\widehat{\mu}_j)^{2m}}{c_m\widehat{\sigma}^{2m}_{jj}}\biggr].
\]
We can similarly define the one-sided versions of the kernel estimator.
We can combine these estimators with change-point detection methods, which we leave for future work.

\bibliographystyle{ims}
\small{\bibliography{kurtosis}}

\newpage

\appendix
\section{Proof of main results}

\subsection{Proof of Theorem~\ref{thm:root-n}}
Write for short $\widehat{\theta}^{\, \rM}_m=\widehat{\theta}^{\, \rM}_m(\hbmu,\mathrm{diag}(\hbSigma))$ and $\widetilde{\theta}^{\, \rM}_m=\widehat{\theta}^{\, \rM}_m(\bmu,\mathrm{diag}(\bSigma))$. By Theorem~\ref{thm:var}, $\widetilde{\theta}_m^{\,\rM}$ is unbiased and satisfies
\[
\var\bigl( \widetilde{\theta}^{\, \rM}_m\bigr)\leq \theta_m^2\cdot \biggl( \frac{1}{n}\frac{r_{2m}-r_m^2}{r_m^2} +\frac{1}{np}\frac{r_{2m}}{r^2_m} \frac{\eta_{2m}-\eta_m^2}{\eta_m^2}+ \frac{C_m}{n}\frac{r_{2m}}{r^2_m\eta_m^2}\frac{\|\bLambda-\bI\|_F^2}{p^2}\biggr).
\]
We note that $(\eta_m, C_m)$ are constants, $(\theta_m,r_m,r_{2m})$ are bounded above/below by constants,  and all entries of the correlation matrix $\bLambda$ are bounded by $1$. Hence, the right hand side is $O(n^{-1})$, and it implies
\[
|\widetilde{\theta}^{\, \rM}_m-\theta_m|=O_{\mathbb{P}}(n^{-1/2}).
\]
To show the claim, it suffices to show that
\beq \label{thm-rootn-goal}
|\widehat{\theta}^{\, \rM}_m - \widetilde{\theta}^{\, \rM}_m|= O_{\mathbb{P}}(n^{-1/2}).
\eeq

Below, we show \eqref{thm-rootn-goal}. Write for short $X_{ij} = (Y_{ij}-\mu_j)/\sqrt{\sigma_{jj}}$,
for $1\leq i\leq n,1\leq j\leq p$. For any $k\geq 0$, let $S_{kj} = \frac{1}{n}\sum_{i=1}^n X^k_{ij}$. Using these notations,
\[
\widetilde{\theta}^{\, \rM}_m = \frac{1}{npc_m}\sum_{j=1}^p \sum_{i=1}^n\biggl(  \frac{Y_{ij}-\mu_j}{\sqrt{\sigma_{jj}}}  \biggr)^{2m} = \frac{1}{c_m}\cdot \frac{1}{p}\sum_{j=1}^p S_{(2m)j}.
\]
At the same time, noticing that $(\widehat{\mu}_j-\mu_j)/\sqrt{\sigma_{jj}}=S_{1j}$ and $(Y_{ij}-\widehat{\mu}_j)/\sqrt{\sigma_{jj}}=X_{ij}-S_{1j}$, we have
\begin{align*}
 \widehat{\theta}^{\, \rM}_m &= \frac{1}{npc_m}\sum_{j=1}^p \sum_{i=1}^n\biggl( \frac{Y_{ij}-\widehat{\mu}_j}{\sqrt{\widehat{\sigma}_{jj}}}  \biggr)^{2m}\cr
 & =  \frac{1}{npc_m}\sum_{j=1}^p \biggl[ \frac{\sigma^m_{jj}}{\widehat{\sigma}^m_{jj}}\sum_{i=1}^n\biggl( \frac{Y_{ij}-\widehat{\mu}_j}{\sqrt{\sigma_{jj}}}  \biggr)^{2m} \biggr] \cr
 &=  \frac{1}{npc_m}\sum_{j=1}^p\biggl[ \frac{\sigma^m_{jj}}{\widehat{\sigma}^m_{jj}} \sum_{i=1}^n(X_{ij}-S_{1j})^{2m}\biggr]\cr
 &=  \frac{1}{npc_m}\sum_{j=1}^p \biggl[  \frac{\sigma^m_{jj}}{\widehat{\sigma}^m_{jj}} \sum_{i=1}^n \sum_{k=0}^{2m}\gamma_k S_{1j}^{k}X_{ij}^{2m-k}\biggr], \qquad \mbox{where}\;\; \gamma_k \equiv (-1)^{k}{2m\choose k}\cr
 &= \frac{1}{c_m} \sum_{k=0}^{2m}\gamma_{k} \biggl[\frac{1}{p}\sum_{j=1}^p   \frac{\sigma^m_{jj}}{\widehat{\sigma}^m_{jj}}S_{1j}^{k}\Bigl(\frac{1}{n}\sum_{i=1}^nX_{ij}^{2m-k}\Bigr)\biggr]\cr
 &= \frac{1}{c_m} \sum_{k=0}^{2m}\gamma_{k} \biggl[\frac{1}{p}\sum_{j=1}^p \frac{\sigma^m_{jj}}{\widehat{\sigma}^m_{jj}} S_{1j}^{k}S_{(2m-k)j}\biggr].
\end{align*}
Combining the above gives
\begin{align} \label{thm-rootn-decomposition}
\widehat{\theta}^{\, \rM}_m- \widetilde{\theta}^{\, \rM}_m &= \frac{2m}{c_m}\frac{1}{p}\sum_{j=1}^p\Bigl( \frac{\sigma^m_{jj}}{\widehat{\sigma}^m_{jj}}-1\Bigr)S_{(2m)j}  +\frac{2m}{c_m} \frac{1}{p}\sum_{j=1}^p \frac{\sigma^m_{jj}}{\widehat{\sigma}^m_{jj}} S_{1j}S_{(2m-1)j}\cr
& \qquad +\frac{1}{c_m} \sum_{k=2}^{2m}\gamma_{k} \biggl[\frac{1}{p}\sum_{j=1}^p \frac{\sigma^m_{jj}}{\widehat{\sigma}^m_{jj}} S_{1j}^{k}S_{(2m-k)j}\biggr]\cr
&=(I_1)+(I_2)+(I_3).
\end{align}

To bound the right hand side of \eqref{thm-rootn-decomposition}, we define an event. By \eqref{Ycoordinate}, $Y_{ij}=\mu_j +\xi_i(\bSigma^{1/2}\Ub_i)_j$. Let $\bLambda=[\mathrm{diag}(\bSigma)]^{-1/2}\bSigma[\mathrm{diag}(\bSigma)]^{-1/2}$. Then, $X_{ij}=\frac{Y_{ij}-\mu_j}{\sqrt{\sigma_{jj}}}=\xi_i (\bLambda^{1/2}\Ub_i)_j$. It follows that
\beq \label{thm-rootn-S}
S_{kj} = \frac{1}{n}\sum_{i=1}^n \xi^k_i (\bLambda^{1/2}\Ub_i)^k_j.
\eeq
Note that $\mathbb{E}X_{ij}=(\mathbb{E}\xi_i^k)\mathbb{E}[ (\bLambda^{1/2}\Ub_i)^k_j]$. At the same time, since $X_{ij}\sim N(0,1)$ when $\xi_i^2\sim \chi_p^2$, it holds that $\mathbb{E}[N^k(0,1)]=(\mathbb{E}\chi_p^k)\mathbb{E}[ (\bLambda^{1/2}\Ub_i)^k_j]$. Together, we have $\mathbb{E}(X^k_{ij})= \mathbb{E}[N^k(0,1)]\cdot [(\mathbb{E}\xi_i^k)/(\mathbb{E}\chi_p^k)]$. Our assumption of $\theta_{2m}\leq C$ guarantees $(\mathbb{E}\xi_i^k)/(\mathbb{E}\chi_p^k)\leq C$ for $1\leq k\leq 4m$. It follows that $\mathbb{E}(X_{ij}^k)\leq C$ and $\mathrm{var}(X_{ij}^k)\leq C$ for $1\leq k\leq 2m$. As a result,
\beq \label{thm-rootn-Sbound}
\mathbb{E}(|S_{kj}|^2)\leq C, \qquad \mathbb{E}\bigl( |S_{kj}-\mathbb{E}S_{kj}|^2 \bigr)=O(n^{-1}), \qquad 1\leq k\leq 2m.
\eeq
Using the marginal sub-Gaussianity,  for any $\epsilon>0$, there exists a constant $C>0$ such that, with probability $\geq 1-\epsilon$,
\beq  \label{thm-rootn-event}
\max_{\substack{1\leq k\leq 2m\\1\leq j\leq p}} |S_{kj}-\mathbb{E}S_{kj}|\leq C \sqrt{(\log p) / n}.
\eeq
Let $B$ be the event that \eqref{thm-rootn-event} holds.
To show \eqref{thm-rootn-goal}, it suffices to show that
\beq \label{thm-rootn-goal2}
|\widehat{\theta}^{\, \rM}_m- \widetilde{\theta}^{\, \rM}_m|\cdot I_B=O_{\mathbb{P}}(n^{-1/2}).
\eeq

We now show \eqref{thm-rootn-goal2}. Consider $(I_2)$ and $(I_3)$. By \eqref{thm-rootn-S} and using that $(\bLambda^{1/2}\Ub_i)_j$ has a symmetric distribution, we have $\mathbb{E}S_{kj}=0$ for any odd $k$. As a result, over the event $B$, $|S_{1j}|\leq C\sqrt{(\log p) / n}$, $|S_{(2m-1)j}|\leq C\sqrt{(\log p) / n}$ and $|S_{(2m-k)j}|\leq C$, for all $1\leq j\leq p$ and $1\leq k\leq 2m$. Additionally, since $\widehat{\sigma}_{jj}/\sigma_{jj}=\frac{1}{n}\sum_{i=1}^n \bigl(\frac{Y_{ij}-\widehat{\mu}_j}{\sqrt{\sigma_{jj}}}\bigr)^2=\frac{1}{n} \sum_{i=1}^n(X_{ij}-S_{1j})^2 = S_{2j}-S_{1j}^2$, where $\mathbb{E}S_{2j}=1$, it holds that $\sigma_{jj}/\widehat{\sigma}_{jj}\leq C$ over the event $B$.
It follows that
\begin{align} \label{thm-rootn-I2I3}
|(I_2)|& \leq  C\max_{1\leq j\leq p}\biggl\{ \frac{\sigma^m_{jj}}{\widehat{\sigma}^m_{jj}} |S_{1j}||S_{(2m-1)j}|\biggr\}=O(n^{-1}\log(p)).\cr
|(I_3)|& \leq  C\max_{1\leq j\leq p}\biggl\{\sum_{k=2}^{2m} \frac{\sigma^m_{jj}}{\widehat{\sigma}^m_{jj}} |S_{1j}|^k|S_{(2m-k)j}|\biggr\}=O(n^{-1}\log(p)).
\end{align}
Consider $(I_1)$. Since $\widehat{\sigma}_{jj}/\sigma_{jj}=S_{2j}-S_{1j}^2$, we write
\[
\widehat{\sigma}_{jj}/\sigma_{jj} - 1 = (S_{2j}-\mathbb{E}S_{2j}) - S^2_{1j}.
\]
Over the event $B$, $\max_j |S_{1j}|\leq C\sqrt{(\log p) / n}$ and $\max_{1\leq j\leq p}|\widehat{\sigma}_{jj}/\sigma_{jj}-1|\leq C\sqrt{(\log p) / n}$.
This means, for all $1\leq j\leq p$,  $\widehat{\sigma}_{jj}/\sigma_{jj}$ is contained in a diminishing neighborhood of $1$.
We use Taylor expansion of the function $(1+x)^{-m}-1$. It gives
\begin{align}  \label{thm-rootn-Taylor}
\frac{\sigma^m_{jj}}{\widehat{\sigma}^m_{jj}} -1 & = - m \frac{\widehat{\sigma}_{jj}-\sigma_{jj}}{\sigma_{jj}^m} + O(n^{-1}\log(p))\cr
& = -m[ (S_{2j}-\mathbb{E}S_{2j}) - S^2_{1j}]+ O(n^{-1}\log(p))\cr
&= -m(S_{2j}-\mathbb{E}S_{2j}) + O(n^{-1}\log(p))\cr
&= - \frac{m}{n}\sum_{i=1}^n \Bigl\{ \xi^2_i (\bLambda^{1/2}\Ub_i)^2_j - (\mathbb{E}\xi^2_i)\mathbb{E}[(\bLambda^{1/2}\Ub_i)^2_j] \Bigr\} + O(n^{-1}\log(p)).
\end{align}
where the third line is due to $\max_{1\leq j\leq p}|S_{1j}|\leq O(\sqrt{(\log p) / n})$ over the event $B$ and the fourth line is due to \eqref{thm-rootn-S}. By \eqref{thm-rootn-S} and \eqref{thm-rootn-Taylor},
\begin{align} \label{thm-rootn-I1simple}
(I_1) 
&= - \frac{2m^2}{c_mp}\sum_{j=1}^p\biggl[ \frac{1}{n}\sum_{i=1}^n
\bigl( \xi^2_i (\bLambda^{1/2}\Ub_i)^2_j - (\mathbb{E}\xi^2_i) \mathbb{E}[(\bLambda^{1/2}\Ub_i)^2_j] \bigr)\biggr] \biggl[\frac{1}{n}\sum_{k=1}^n \xi_k^{2m}(\bLambda^{1/2}\Ub_k)^{2m}_j \Biggr]+ o(n^{-1/2})\cr
& = - \frac{2m^2}{c_mpn^2} \sum_{i,k=1}^n \biggl\{\sum_{j=1}^p
\Bigl[ \xi^2_i (\bLambda^{1/2}\Ub_i)^2_j - (\mathbb{E}\xi^2_i) \mathbb{E}[(\bLambda^{1/2}\Ub_i)^2_j] \Bigr] \Bigl[\xi_k^{2m}(\bLambda^{1/2}\Ub_k)^{2m}_j \Bigr]\biggr\}+ o(n^{-1/2})\cr
& \equiv - \frac{2m^2}{c_mpn^2} \sum_{i,k=1}^nQ_{ik} + o(n^{-1/2}).
\end{align}
Write $R_{ij} = (\bLambda^{1/2}\Ub_i)_j$ for short. Then,
\beq \label{thm-rootn-Q}
Q_{ik} = \sum_{j=1}^p \bigl[\xi_i^2R_{ij}^2 - (\mathbb{E}\xi_i^2)(\mathbb{E}R_{ij}^2)\bigr]\xi_k^{2m}R_{kj}^{2m}.
\eeq
We introduce positive random variables $\{\omega_i\}_{i=1}^n$ such that $\omega_i^2\overset{iid}{\sim}\chi_p^2$ and that $\{\omega_i\}_{i=1}^n$ are independent of $\{(\xi_i,\Ub_i): 1\leq i\leq n\}$. Then, $Z_i\equiv \omega_i(\bLambda^{1/2}\Ub_i)\sim N(0, I_p)$. For even integers $s,t$ and $1\leq j,j'\leq p$,
\[
\mathbb{E}[Z^s_i(j)Z^t_i(j')] =\mathbb{E}(\omega_i^{s+t})\mathbb{E}(R_{ij}^sR_{ij'}^t).
\]
For all $s,t$ such that $s+t\leq 4m$, the left hand side is uniformly bounded by a constant. Additionally, by elementary probability, $\mathbb{E}(\omega_i^{s+t})\asymp p^{(s+t)/2}$. It follows that
\beq \label{thm-rootn-R}
\max_{1\leq j,j'\leq p}\mathbb{E}(R_{ij}^sR_{ij'}^t) \leq Cp^{-(s+t)/2}, \qquad \mbox{for even $s,t$ such that $s+t\leq 4m$}.
\eeq
In particular, by taking $s=2\ell $ and $t=0$ in the above, we have $\mathbb{E}R_{ij}^{2\ell}\leq Cp^{-\ell}$ for all $0\leq \ell\leq 2m$.
Additionally, $\theta_s=p^{-s}\mathbb{E}\xi^{2s}$ by definition, so the assumption $\theta_{2m}\leq C$ guarantees
\beq \label{thm-rootn-xi}
\mathbb{E}(\xi_i^{2s})\leq Cp^s, \qquad 0\leq s\leq 2m.
\eeq
Using \eqref{thm-rootn-R}-\eqref{thm-rootn-xi}, we first bound $|\sum_{i=1}^nQ_{ii}|$. It is seen that
\[
\mathbb{E}|Q_{ii}| \leq \sum_{j=1}^p \mathbb{E}(\xi_i^{2m+2})\mathbb{E}(R_{ij}^{2m+2}) + (\mathbb{E}\xi_i^2)(\mathbb{E}R_{ij}^2)\mathbb{E}(\xi_i^{2m})\mathbb{E}(R_{ij}^{2m})\leq Cp.
\]
As a result,
\beq \label{thm-rootn-QiiSum}
\mathbb{E}\Bigl( \frac{1}{pn^2}\Bigl| \sum_{i=1}^nQ_{ii}\Bigr|\Bigr) = O(n^{-1}) \qquad \Longrightarrow \qquad \frac{1}{pn^2} \Bigl|\sum_{i=1}^nQ_{ii}\Bigr| = o_{\mathbb{P}}(n^{-1/2}).
\eeq
We then bound $|\sum_{i\neq k}Q_{ik}|$. Consider $(i,k,i',k')$ such that $i\neq k$ and $i'\neq k'$. By \eqref{thm-rootn-Q}, $\mathbb{E}Q_{ik}=0$ for $i\neq k$. Therefore, if $(i,k,i',k')$ are mutually distinct, $\mathbb{E}(Q_{ik}Q_{i'k'})=0$. It follows that
\[
\mathbb{E}\Bigl[ \Bigl(\sum_{i\neq k}Q_{ik}\Bigr)^2\Bigr] = 6\sum_{\text{distinct } i, k,k'}\mathbb{E}(Q_{ik}Q_{ik'}) + 2\sum_{\text{distinct } i, k} \mathbb{E}(Q_{ik}^2).
\]
By \eqref{thm-rootn-Q} and \eqref{thm-rootn-R}-\eqref{thm-rootn-xi},
\begin{align*}
\mathbb{E}(Q_{ik}Q_{ik'}) &=\mathbb{E}\Bigl\{ \sum_{j,j'=1}^p \bigl[\xi_i^2R_{ij}^2 - (\mathbb{E}\xi_i^2)(\mathbb{E}R_{ij}^2)\bigr]\bigl[\xi_i^2R_{ij'}^2 - (\mathbb{E}\xi_i^2)(\mathbb{E}R_{ij'}^2)\bigr] \xi_k^{2m}\xi_{k'}^{2m}R_{kj}^{2m}R_{k'j'}^{2m}\Bigr\} \cr
&\leq \sum_{j,j'=1}^p \mathbb{E}(\xi_i^{4})\mathbb{E}(R_{ij}^2R_{ij'}^2)(\mathbb{E}\xi_k^{2m})(\mathbb{E}\xi_{k'}^{2m})\mathbb{E}(R_{kj}^{2m}R_{k'j'}^{2m})\quad \leq Cp^2,\cr
\mathbb{E}(Q_{ik}^2) &= \mathbb{E}\Bigl\{ \sum_{j,j'=1}^p \bigl[\xi_i^2R_{ij}^2 - (\mathbb{E}\xi_i^2)(\mathbb{E}R_{ij}^2)\bigr]\bigl[\xi_i^2R_{ij'}^2 - (\mathbb{E}\xi_i^2)(\mathbb{E}R_{ij'}^2)\bigr] \xi_k^{4m}R_{kj}^{2m}R_{kj'}^{2m}\Bigr\}\cr
&\leq \sum_{j,j'=1}^p \mathbb{E}(\xi_i^{4})\mathbb{E}(R_{ij}^2R_{ij'}^2)(\mathbb{E}\xi_k^{4m})\mathbb{E}(R_{kj}^{2m}R_{kj'}^{2m})\quad \leq Cp^2.
\end{align*}
Moreover, the total number of such distinct $(i,k,k')$ is $O(n^{3})$. It follows that
\beq \label{thm-rootn-QikSum}
\mathbb{E}\Bigl[ \Bigl(\frac{1}{pn^2}\sum_{i\neq k}Q_{ik}\Bigr)^2\Bigr]  = O(n^{-1}) \qquad \Longrightarrow \qquad \frac{1}{pn^2} \Bigl|\sum_{i\neq k}Q_{ik}\Bigr| = O_{\mathbb{P}}(n^{-1/2}).
\eeq
Pluging \eqref{thm-rootn-QiiSum} and \eqref{thm-rootn-QikSum} into \eqref{thm-rootn-I1simple} gives
\beq \label{thm-rootn-I1}
(I_1) = O_{\mathbb{P}}(n^{-1/2}).
\eeq
We further plug \eqref{thm-rootn-I2I3} and \eqref{thm-rootn-I1} into \eqref{thm-rootn-decomposition}. It gives \eqref{thm-rootn-goal2}. The proof is now complete.
\qed

\subsection{Proof of Theorem~\ref{thm:consistency}}
Similar to the proof of Theorem~\ref{thm:root-n}, let $\widetilde{\theta}^{\, \rM}_m$ and $\widehat{\theta}^{\, \rM}_m$ denote the MAE with true $(\bmu,\bSigma)$ and estimates $(\hbmu,\hbSigma)$; here, $(\hbmu,\hbSigma)$ are not necessarily the sample mean and sample covariance matrix. It follows from Theorem~\ref{thm:var} that $\E[(\widetilde{\theta}^{\, \rM}_m-\theta_m)^2]\leq Cn^{-1}$. By Markov's inequality, for any constant $C_1>0$,
\beq \label{thm-robust-1}
\mathbb{P}\Bigl(|\widetilde{\theta}^{\, \rM}_m-\theta_m|>C_1n^{-1/2}\Bigr)\leq \frac{\E[(\widehat{\theta}^{\, \rM}_m-\theta_m)^2]}{C_1^2 n^{-1}}\leq \frac{C}{C_1^2}.
\eeq
Hence, given $\epsilon>0$, we can choose an appropriate $C_1>0$ such that the above probability is bounded by $\epsilon/3$.

Below, we bound $|\widehat{\theta}^{\, \rM}_m-\widetilde{\theta}^{\, \rM}_m|$.
Letting $X_{ij}=(Y_{ij}-\mu_j)/\sqrt{\sigma_{jj}}$ and $\widehat{X}_{ij}=(Y_{ij}-\widehat{\mu}_j)/\sqrt{\widehat{\sigma}_{jj}}$, we have
\[
\widehat{\theta}^{\, \rM}_m-\widetilde{\theta}^{\, \rM}_m = \frac{1}{npc_m}\sum_{j=1}^p \sum_{i=1}^n\bigl(\widehat{X}_{ij}^{2m}-X_{ij}^{2m}),
\]
where
\[
\widehat{X}_{ij} = X_{ij} + X_{ij}\biggl(\frac{\sqrt{\sigma_{jj}}}{\sqrt{\widehat{\sigma}_{jj}}}-1  \biggr) +  \frac{\mu_j-\widehat{\mu}_j}{\sqrt{\widehat{\sigma}_{jj}}}\equiv X_{ij}+\Delta_{ij}.
\]
It follows that
\begin{align*}
\widehat{\theta}^{\, \rM}_m-\widetilde{\theta}^{\, \rM}_m &= \frac{1}{npc_m}\sum_{j=1}^p \sum_{i=1}^n\sum_{k=1}^{2m}{2m\choose k} X_{ij}^{2m-k}\Delta_{ij}^k\cr
& = \frac{1}{npc_m}\sum_{j=1}^p \sum_{i=1}^n2m X_{ij}^{2m-1}\Delta_{ij} + \sum_{k=2}^{2m} {2m\choose k}\cdot \frac{1}{npc_m}\sum_{j=1}^p \sum_{i=1}^nX_{ij}^{2m-k}\Delta^k_{ij} \cr
&\equiv (J_1)+(J_2).
\end{align*}
First, we consider $(J_1)$. By direct calculations,
\begin{align*}
(J_1) & = \frac{2m}{npc_m}\sum_{j=1}^p \sum_{i=1}^n X_{ij}^{2m}\Bigl(\frac{\sqrt{\sigma_{jj}}}{\sqrt{\widehat{\sigma}_{jj}}}-1\Bigr)+ \frac{2m}{npc_m}\sum_{j=1}^p \sum_{i=1}^n X_{ij}^{2m-1}\frac{\mu_j-\widehat{\mu}_j}{\sqrt{\widehat{\sigma}_{jj}}}\cr
&= \frac{2m}{pc_m}\sum_{j=1}^p S_{(2m)j} \Bigl(\frac{\sqrt{\sigma_{jj}}}{\sqrt{\widehat{\sigma}_{jj}}}-1  \Bigr)+ \frac{2m}{pc_m}\sum_{j=1}^p S_{(2m-1)j} \frac{\sqrt{\sigma_{jj}}}{\sqrt{\widehat{\sigma}_{jj}}}\frac{\mu_j-\widehat{\mu}_j}{\sqrt{\sigma_{jj}}},
\end{align*}
where $S_{kj}=\frac{1}{n}\sum_{i=1}^n X^k_{ij}$ for $k\geq 0$.
Under our assumption, $\max_j|\sigma_{jj}/\widehat{\sigma}_{jj}|\lesssim 1$, and $|\sqrt{\widehat{\sigma}_{jj}/\sigma_{jj}}-1|\leq C|\widehat{\sigma}_{jj}-\sigma_{jj}|$.  Moreover, by similar technique in the proof of Theorem~\ref{thm:var}, we can prove that, $\frac{1}{p}\sum_{j=1}^p \mathbb{E} |S_{kj}|\leq C$, for $1\leq k\leq 2m$. As a result, for any $\epsilon>0$, there exists $C_2>0$ such that, $\frac{1}{p}\sum_{j=1}^p |S_{(2m)j}|\leq C_2$ simultaneously for $1\leq k\leq 2m$, with probability $1-\epsilon/3$. On this event,
\beq \label{thm-robust-J1}
|(J_1)| \leq C\Bigl(\frac{1}{p}\sum_{j=1}^p |S_{(2m)j}|\Bigr) |\widehat{\sigma}_{jj}-\sigma_{jj}| + C\Bigl(\frac{1}{p}\sum_{j=1}^p |S_{(2m-1)j}|\Bigr)|\widehat{\mu}_j-\mu_j| \leq C\max\{\alpha_n, \beta_n\}.
\eeq
Next, we consider $(J_2)$. By our assumption, $|\Delta_{ij}|\leq \alpha_n +\beta_n |X_{ij}|$. It follows that
\[
|\Delta_{ij}|^k \leq C\alpha_n^k + C\beta_n^k|X_{ij}|^k.
\]
Plugging it into the definition of $(J_2)$, we have
\begin{align*}
|(J_2)|& \leq C\sum_{k=2}^{2m}\frac{1}{np}\sum_{j=1}^p\sum_{i=1}^n|X_{ij}|^{2m-k}(\alpha_n^k + \beta_n^k|X_{ij}|^k)\cr
&\leq C\sum_{k=2}^{2m} \alpha_n^k \Bigl(\frac{1}{np}\sum_{j=1}^p\sum_{i=1}^n|X_{ij}|^{2m-k}\Bigr) + C\sum_{k=2}^{2m} \beta_n^k \Bigl(\frac{1}{np}\sum_{j=1}^p\sum_{i=1}^n|X_{ij}|^{2m}\Bigr).
\end{align*}
Again, we can easily prove that $\frac{1}{np}\sum_{j=1}^p\sum_{i=1}^n\mathbb{E}|X_{ij}|^{k}\leq C$ for all $1\leq k\leq 2m$. It follows that, for any $\epsilon>0$, there exists $C_3>0$, such that
$\frac{1}{np}\sum_{j=1}^p\sum_{i=1}^n|X_{ij}|^{k}\leq C_3$ simultaneously for all $1\leq k\leq 2m$. On this event,
\beq   \label{thm-robust-J2}
|(J_2)|\leq C\sum_{k=2}^{2m}(\alpha_n^k + \beta_n^k)\leq C\max\{\alpha_n^2,\beta_n^2\}.
\eeq
Combining \eqref{thm-robust-J1}-\eqref{thm-robust-J2} gives $|\widehat{\theta}^{\, \rM}_m-\widetilde{\theta}^{\, \rM}_m|\leq C\max\{\alpha_n,\beta_n\}$. We further combine it with \eqref{thm-robust-1}. It gives the claim.

\subsection{Proof of Theorem~\ref{thm:var}}
Write for short $\widehat{\theta}^{\, \rM}_m=\widehat{\theta}^{\, \rM}_m(\bmu, \bOmega)$ and $\widehat{\theta}^{\, \rM}_{ m, j}=\widehat{\theta}^{\, \rM}_{ m, j}(\mu_j,\sigma_{jj})$.
First, we show that $\widehat{\theta}^{\, \rM}_m$ is unbiased. Recall that $\widehat{\theta}^{\, \rM}_m=p^{-1}\sum_{j=1}^p\widehat{\theta}^{\, \rM}_{ m, j}$. It suffices to show $\widehat{\theta}^{\, \rM}_{ m, j}$ is unbiased for each $1\leq j\leq p$. Recall that
\beq \label{thm-MAErisk-1}
 \widehat{\theta}^{\, \rM}_{ m, j} = \frac{1}{n c_{m}} \sum_{i=1}^n \  \frac{( Y_{ij}- \mu_j )^{2m}}{\sigma_{jj}^m}, \qquad \mbox{where}\quad c_{m} = (2m-1)!! \, (p/2)^m \frac{ \Gamma(p/2)}{  \Gamma(p/2+m)}.
\eeq
By the form of elliptical distribution, $\Yb_i-\bmu = \xi_i\widetilde{\Ub}_i$, where $\xi_i$ and $\widetilde{\Ub}_i$ are independent of each other. We have seen in Section~\ref{subsec:MAE} that $\mathbb{E}\widetilde{U}_{ij}^{2m}=p^{-m}c_m\sigma_{jj}^m$. It follows that
\[
\mathbb{E}[( Y_{ij}- \mu_j )^{2m}]=(\mathbb{E}\xi_i^{2m})(\mathbb{E}\widetilde{U}_i^{2m})=(p^m\theta_m)(p^{-m}c_m\sigma_{jj}^m)=c_m\theta_m\sigma_{jj}^m.
\]
Plugging it into \eqref{thm-MAErisk-1} gives
\beq \label{thm-MAErisk-2}
\mathbb{E}\widehat{\theta}^{\, \rM}_{ m, j} = \frac{1}{n c_{m}} \sum_{i=1}^n \  \frac{\mathbb{E}[( Y_{ij}- \mu_j )^{2m}]}{\sigma_{jj}^m} =\frac{1}{nc_m}\sum_{i=1}^n\frac{c_m\theta_m\sigma_{jj}^m}{\sigma_{jj}^m}=\theta_m.
\eeq
This proves that each $\widehat{\theta}^{\, \rM}_{ m, j}$ is unbiased. It follows that $\widehat{\theta}^{\, \rM}_m$ is also unbiased.

Next, we calculate the variance of $\widehat{\theta}^{\, \rM}_m$. For each $1\leq i\leq n$, let $\Wb^{(i)}=(W_1^{(i)},\ldots,W_p^{(i)})^T$, where $W_j^{(i)}=(Y_{ij}-\mu_j)^{2m}/\sigma_{jj}^m$, $1\leq j\leq p$. Noting that $\{\bW^{(i)}\}_{i=1}^n$ are $iid$ random vectors, we have
\beq \label{thm-MAErisk-3}
 \mbox{var}\bigl(\widehat{\theta}^{\, \rM}_m \bigr) = \mbox{var}\biggl(\frac{1}{npc_m}\sum_{j=1}^p\sum_{i=1}^n W_j^{(i)} \biggr) =\frac{1}{n} \mbox{var}\biggl( \frac{1}{pc_m}\sum_{j=1}^p W_j^{(1)} \biggr).
\eeq
It suffices to calculate the variance in the case of $n=1$. From now on, we fix $n=1$. Let $\Yb=\bmu + \xi\Ub$ be the observed realization of the elliptical distribution. Write
\[
\widehat{\theta}^{\, \rM}_m=\frac{1}{c_mp}\sum_{j=1}^p W_j, \qquad\mbox{where}\quad W_j \equiv \frac{(Y_j-\mu_j)^{2m}}{\sigma_{jj}^m}.
\]
We now calculate $\mathrm{var}(W_j)$ and $\mathrm{cov}(W_j,W_k)$. Recalling that $\widetilde{\Ub}=\bSigma^{1/2}\Ub$, we define random vectors
\beq \label{thm-MAErisk-defineZ}
\Zb\equiv\chi_p^2\cdot \widetilde{\Ub} \qquad\mbox{and}\qquad \widetilde{\Zb}\equiv \mathrm{diag}(\bSigma)^{-1/2}\Zb,
\eeq
where $\chi_p^2$ is a chi-square random variable independent of $\widetilde{\Ub}$. Since the multivariate normal distribution is a special elliptical distribution with $\xi\sim \chi_p^2$, we immediately have $\Zb\sim N(\bzero, \bSigma)$. It follows that $\mathbb{E}\widetilde{Z}_j^{2m}=\sigma_{jj}^{-m}(\mathbb{E}Z_j^{2m})=\sigma_{jj}^{-m}(\mathbb{E}\chi_p^{2m})(\mathbb{E}\widetilde{U}_j^{2m})$. At a result, for all $m\geq 1$,
\[
\mathbb{E}[(Y_j-\mu_j)^{2m}] = (\mathbb{E}\xi^{2m})(\mathbb{E}\widetilde{U}^{2m}_j) = \mathbb{E}\xi^{2m}\cdot \frac{\sigma_{jj}^m (\mathbb{E}\widetilde{Z}_j^{2m})}{\mathbb{E}\chi_p^{2m}} =\sigma_{jj}^m \cdot r_m \mathbb{E}\widetilde{Z}_j^{2m}.
\]
It follows that
\begin{align} \label{thm-MAErisk-varW}
\var(W_j)& = \frac{\mathbb{E}[(Y_j-\mu_j)^{4m}]}{\sigma_{jj}^{2m}} - \biggl(\frac{\mathbb{E}[(Y_j-\mu_j)^{2m}]}{\sigma_{jj}^{m}}\biggr)^2\cr
& = r_{2m} (\mathbb{E}\widetilde{Z}_j^{4m}) - r_m^2 (\mathbb{E}\widetilde{Z}_j^{2m})^2\cr
& = r_{2m}\cdot \var(\widetilde{Z}_j^{2m}) + (r_{2m}-r_m^2)\cdot (\mathbb{E}\widetilde{Z}_j^{2m})^2.
\end{align}
Similarly, since $Y_j-\mu_j = \xi\widetilde{U}_j$ and $Z_j = \chi_p^2\widetilde{U}_j$, we have
\begin{align*}
\mathbb{E}\bigl[(Y_j-\mu_j)^{2m}(Y_k-\mu_k)^{2m}\bigr] & = (\mathbb{E}\xi^{4m})\bigl(\mathbb{E}[\widetilde{U}^{2m}_j\widetilde{U}^{2m}_k]\bigr)\cr
& = (r_{2m}\mathbb{E}\chi_p^{4m})\bigl(\mathbb{E}[\widetilde{U}^{2m}_j\widetilde{U}^{2m}_k]\bigr)\cr
& = r_{2m}\mathbb{E}[Z_j^{2m}Z^{2m}_k]\cr
& =  \sigma_{jj}^m\sigma_{kk}^m\cdot r_{2m} \mathbb{E}[\widetilde{Z}_j^{2m}\widetilde{Z}^{2m}_k].
\end{align*}
Therefore,
\begin{align} \label{thm-MAErisk-covW}
\cov(W_j, W_k)& = \frac{\mathbb{E}\bigl[(Y_j-\mu_j)^{2m}(Y_k-\mu_k)^{2m}\bigr] }{\sigma_{jj}^{2m}\sigma_{kk}^{2m}} - \frac{\mathbb{E}[(Y_j-\mu_j)^{2m}]}{\sigma_{jj}^{m}}\frac{\mathbb{E}[(Y_k-\mu_k)^{2m}]}{\sigma_{kk}^{m}}\cr
& = r_{2m}  \mathbb{E}[\widetilde{Z}_j^{2m}\widetilde{Z}^{2m}_k] - r_m^2 (\mathbb{E}\widetilde{Z}_j^{2m})(\mathbb{E}\widetilde{Z}_k^{2m})\cr
& = r_{2m}\cdot \cov(\widetilde{Z}_j^{2m}, \widetilde{Z}_k^{2m}) + (r_{2m}-r_m^2)\cdot (\mathbb{E}\widetilde{Z}_j^{2m})(\mathbb{E}\widetilde{Z}_k^{2m}).
\end{align}
Combining \eqref{thm-MAErisk-varW} and \eqref{thm-MAErisk-covW} and noting that $\tilde{Z}_j\sim N(0,1)$ for all $1\leq j\leq p$, we rewrite
\[
\cov(W_j, W_k) = r_{2m}\cov(\widetilde{Z}_j^{2m}, \widetilde{Z}_k^{2m}) + (r_{2m}-r_m^2)\eta_m^2, \qquad\mbox{where}\quad \eta_m = \mathbb{E}[N(0,1)^{2m}].
\]
As a result,
\begin{align}  \label{thm-MAErisk-4}
\var(\widehat{\theta}^{\, \rM}_m) &= \frac{1}{c_m^2p^2}\sum_{1\leq j,k\leq p}\cov(W_j, W_k)\cr
& = \frac{1}{c_m^2p^2}\Bigl[ r_{2m}\sum_{1\leq j,k\leq p}\cov(\widetilde{Z}_j^{2m}, \widetilde{Z}_k^{2m}) + (r_{2m}-r_m^2) p^2\eta_m^2\Bigr]\cr
& = \frac{1}{c_m^2}\biggl[ r_{2m}\cdot \frac{1}{p^2}\var\Bigl(\sum_{j=1}^p\widetilde{Z}_j^{2m} \Bigr) + (r_{2m}-r_m^2)\cdot \eta_m^2  \biggr].
\end{align}
Moreover, since $\mathbb{E}\widetilde{U}_{ij}^{2m}=p^{-m}c_m\sigma_{jj}^m$ and $\mathbb{E}\widetilde{Z}_j^{2m}=\sigma_{jj}^{-m}(\mathbb{E}\chi_p^{2m})(\mathbb{E}\widetilde{U}_j^{2m})$, we have
\[
c_m = \frac{p^m \mathbb{E}\widetilde{Z}_j^{2m}}{\mathbb{E}\chi_p^{2m}} = \frac{p^m \mathbb{E}[N(0,1)^{2m}]}{r_m^{-1}\mathbb{E}\xi^{2m}} = \frac{p^m\eta_m}{r_m^{-1}(p^m\theta_m)}= \frac{\eta_mr_m}{\theta_m}.
\]
Plugging it into \eqref{thm-MAErisk-4} gives
\[
\var(\widehat{\theta}^{\, \rM}_m) =\theta_m^2\biggl[ \frac{r_{2m}}{r^2_m}\frac{\var\bigl(\sum_{j=1}^p\widetilde{Z}_j^{2m}\bigr)}{p^2\eta_m^2}+ \frac{r_{2m}-r_m^2}{r_m^2}\biggr].
\]
This is for the case of $n=1$. For a general $n$, we combine it with \eqref{thm-MAErisk-3} to get
\beq \label{thm-MAErisk-5}
\frac{\var(\widehat{\theta}^{\, \rM}_m)}{\theta_m^2} = \frac{1}{n}\biggl[ \frac{r_{2m}}{r^2_m}\frac{\var\bigl(\sum_{j=1}^p\widetilde{Z}_j^{2m}\bigr)}{p^2\eta_m^2}+ \frac{r_{2m}-r_m^2}{r_m^2}\biggr].
\eeq

What remains is to calculate the variance of $\sum_{j=1}^p\widetilde{Z}_j^{2m}$. By definition,
\[
\widetilde{\Zb}\sim N(\bzero, \bLambda), \qquad \mbox{where}\quad \bLambda=[\mathrm{diag}(\bSigma)]^{-1/2}\bSigma[\mathrm{diag}(\bSigma)]^{-1/2}.
\]
Here $\bLambda$ coincides with the correlation matrix of the elliptical distribution. It is seen that
\begin{align*}
\var\Bigl(\sum_{j=1}^p\widetilde{Z}_j^{2m}\Bigr)& =\sum_{j=1}^p \var(\widetilde{Z}_j^{2m}) + 2\sum_{1\leq j<k\leq p}\cov(\widetilde{Z}_j^{2m}, \widetilde{Z}_k^{2m}) \cr
&= p(\eta_{2m}-\eta^2_m) + 2\sum_{1\leq j<k\leq p}\beta_m(\Lambda_{jk}),
\end{align*}
where $\beta_m(\Lambda_{jk})$ denotes the covariance between $X_1^{2m}$ and $X_2^{2m}$ when $(X_1,X_2)^{\T}$ follows a bivariate normal distribution with covariances $\var(X_1)=\var(X_2)=1$ and $\cov(X_1, X_2)=\Lambda_{jk}$. The following lemma is proved in Section~\ref{subsec:bivar-proof}:
\begin{lemma} \label{lem:bivar-normal}
Let $\Xb = (X_1, X_2)^T$ be a bivariate normal random vector satisfying $\E(X_1^2)=\E(X_2^2) =1$ and $\cov(X_1, X_2) = \rho$. Let $\eta_m=\mathbb{E}[N(0,1)^{2m}]$ and $\beta_m(\rho) = \cov(X_1^{2m}, X_2^{2m})$ for $m\geq 2$. Define
\[
B_m(s) = \sum_{\substack{1\leq k_1,k_2\leq m\\ k_1+k_2=s}} {2m \choose 2k_1}{2m \choose 2k_2}\cdot \eta_{m-k_1}\eta_{m-k_2}(\eta_s -\eta_{k_1}\eta_{k_2}), \quad s=2,3,\ldots,m
\]
Then, for all $m\geq 2$,
\[
\beta_m(\rho) =  \sum_{s=2}^{m} B_m(s) (1-|\rho|)^{m-s}|\rho|^{s}.
\]
As a result, $\beta_m(\rho)=72\rho^2$ for $m=2$, and $\beta_m(\rho)\leq C_m \rho^2$ for $m\geq 3$, where $C_m>0$ is a constant that only depends on $m$.
\end{lemma}

By Lemma~\ref{lem:bivar-normal},
\beq \label{thm-MAErisk-6}
\var\Bigl(\sum_{j=1}^p\widetilde{Z}_j^{2m}\Bigr) \leq
p(\eta_{2m}-\eta^2_m) +2C_m \sum_{1\leq j<k\leq p}\Lambda_{jk}^2 \leq p(\eta_{2m}-\eta^2_m)+C_m\|\bLambda-\bI \|_F^2.
\eeq
Plugging it into \eqref{thm-MAErisk-5} gives
\[
\frac{\var(\widehat{\theta}^{\, \rM}_m)}{\theta_m^2} \leq \frac{1}{n}\frac{r_{2m}-r_m^2}{r_m^2} +\frac{1}{np}\frac{r_{2m}}{r^2_m} \biggl(\frac{\eta_{2m}-\eta_m^2}{\eta_m^2}+ \frac{C_m}{\eta_m^2}\frac{\|\bLambda-\bI\|_F^2}{p}\biggr).
\]
Moreover, for $m=2$, the equality holds for $C_m=72$. Since $\eta_m=3$ and $\eta_{2m}=105$, we have
\[
\frac{\var(\widehat{\theta}^{\, \rM}_m)}{\theta_m^2} =\frac{1}{n}\frac{r_{2m}-r_m^2}{r_m^2} +\frac{1}{np}\frac{r_{2m}}{r^2_m} \Bigl(\frac{32}{3}+ \frac{8\|\bLambda-\bI\|_F^2}{p}\Bigr), \qquad \mbox{for }m=2.
\]

\subsection{Proof of Proposition~\ref{prop:ideal}}
Write $\widehat{\theta}^{\,{\rm I}}_m=\widehat{\theta}^{\,{\rm I}}_m(\bmu, \bOmega)$ for short. By definition, $\widehat{\theta}^{\,{\rm I}}_m=\frac{1}{np^m}\sum_{i=1}^n\xi^{2m}_i$, and $\theta_m=p^{-m}\E(\xi^{2m})$. Therefore
\[
\var( \widehat{\theta}^{\,{\rm I}}_m) = \frac{1}{np^{2m}}\var(\xi^{2m}) = \frac{1}{n}(\theta_{2m}-\theta_m^2).
\]
We divide both sides by $\theta^2_m$ and note that $\theta_m=p^{-m}(\e\xi^{2m})=p^{-m}r_m(\e\chi_p^{2m})$. It follows that
\begin{align}  \label{prop-compare-1}
\frac{\var( \widehat{\theta}^{\,{\rm I}}_m)}{\theta_m^2}
&= \frac{1}{n} \cdot \frac{r_{2m}(\mathbb{E}\chi_p^{4m})-r_m^2 (\mathbb{E}\chi_p^{2m})^2}{r_m^2 (\mathbb{E}\chi_p^{2m})^2}\cr
& =  \frac{1}{n}\cdot \frac{r_{2m}\mathrm{var}(\chi_p^{2m})+(r_{2m}-r_m^2) (\mathbb{E}\chi_p^{2m})^2}{r_m^2 (\mathbb{E}\chi_p^{2m})^2}\cr
& = \frac{1}{n}\left[\frac{r_{2m}}{r_m^2}\frac{\mathrm{var}(\chi_p^{2m})}{(\mathbb{E}\chi_p^{2m})^2} + \frac{r_{2m}-r_m^2}{r_m^2}\right],
\end{align}
By elementary statistics, $\mathbb{E}\chi_p^{2m}=\prod_{j=0}^{m-1}(p+2j)$.
As a result,
\begin{align}  \label{prop-compare-2}
\frac{\mathrm{var}(\chi_p^{2m})}{(\mathbb{E}\chi_p^{2m})^2}
&=\frac{\prod_{j=0}^{2m-1}(p+2j) -\prod_{j=0}^{m-1}(p+2j)^2}{(\mathbb{E}\chi_p^{2m})^2}  \cr
&= \frac{\prod_{j=0}^{m-1}(p+2j)}{(\mathbb{E}\chi_p^{2m})^2}\Bigl\{ \prod_{j=m}^{2m-1}(p+2j) -\prod_{j=0}^{m-1}(p+2j)  \Bigr\}\cr
& = \frac{1}{ \mathbb{E}\chi_p^{2m}}\biggl\{ \Bigl[p^m + \bigl(p^{m-1}\sum_{j=m}^{2m-1}2j\bigr)\Bigr] - \Bigr[p^m + \bigl(p^{m-1}\sum_{j=0}^{m-1}2j\bigr)\Bigr] +O(p^{m-2}) \biggr\}\cr
& =  \frac{1}{ \mathbb{E}\chi_p^{2m}}\cdot  \bigl[ 2m^2 p^{m-1} + O(p^{m-2})\bigr] \cr
&=\frac{2m^2}{p}[1+o(1)].
\end{align}
Plugging \eqref{prop-compare-2} into \eqref{prop-compare-1} gives the claim.

%

\subsection{Proof of Theorem~\ref{thm:normality}}
Fix $1\leq j\leq p$. Using the Slutsky's lemma, we only need to prove
\beq \label{thm-normality-0}
\frac{\widehat{\theta}^{\, \rM}_{ m, j}(\widehat{\mu}_j,\widehat{\sigma}_{jj})-\theta_m}{\sqrt{ \frac{c_{2m}}{c_m^2}\theta_{2m} -\theta_m^2 }} \to_d N(0,1).
\eeq
Write for short $\widehat{\theta}^{\, \rM}_{ m, j}=\widehat{\theta}^{\, \rM}_{ m, j}(\widehat{\mu}_j,\widehat{\sigma}_{jj})$.
Let $X_{ij} = (Y_{ij}-\mu_j)/\sqrt{\sigma_{jj}}$ and $S_{kj} = \frac{1}{n}\sum_{i=1}^n X^k_{ij}$, for $1\leq i\leq n$ and $k\geq 0$. Then, $\widehat{\mu}_j = S_{1j}$, $\widehat{\sigma}_{jj}=S_{2j}-S_{1j}^2$, and
\[
\frac{Y_{ij}-\widehat{\mu}_j}{\sqrt{\widehat{\sigma}_{jj}}} = \frac{\sqrt{\sigma_{jj}}}{\sqrt{\widehat{\sigma}_{jj}}}\frac{Y_{ij}-\widehat{\mu}_j}{\sqrt{\sigma_{jj}}}
= \frac{\sqrt{\sigma_{jj}}}{\sqrt{\widehat{\sigma}_{jj}}}(X_{ij}-S_{1j}).
\]
It follows that
\begin{align}  \label{thm-normality-1}
 \widehat{\theta}^{\, \rM}_{m,j} &= \frac{1}{nc_m} \sum_{i=1}^n\biggl( \frac{Y_{ij}-\widehat{\mu}_j}{\sqrt{\widehat{\sigma}_{jj}}}  \biggr)^{2m}
 =  \frac{1}{nc_m}\frac{\sigma^m_{jj}}{\widehat{\sigma}^m_{jj}} \sum_{i=1}^n(X_{ij}-S_{1j})^{2m}\cr
 &=  \frac{1}{nc_m}\frac{\sigma^m_{jj}}{\widehat{\sigma}^m_{jj}} \sum_{i=1}^n \sum_{k=0}^{2m}\gamma_k S_{1j}^{k}X_{ij}^{2m-k}, \qquad \mbox{where}\;\; \gamma_k \equiv (-1)^{k}{2m\choose k}\cr
 &= \frac{1}{c_m}\frac{\sigma^m_{jj}}{\widehat{\sigma}^m_{jj}}  \sum_{k=0}^{2m}\gamma_{k} S_{1j}^{k}S_{(2m-k)j}.
\end{align}
Let $\Sb=(S_{1j}, S_{2j},\ldots,S_{(2m)j})^{\T}$. Below, we first derive the asymptotic normality of $\Sb$, then we use the delta method to prove \eqref{thm-normality-0}.

First, we study the random vector $\Sb$. It is not hard to see that $\E S_{kj}=\E X_{ij}^k$. By \eqref{Ycoordinate}, $X_{ij}=\xi_i (\bLambda^{1/2}\Ub_i)_j$, where $\{(\xi_i, \Ub_i)\}_{i=1}^n$ are mutually indepependent and $\bLambda=[\mathrm{diag}(\bSigma)]^{-1/2}\bSigma[\mathrm{diag}(\bSigma)]^{-1/2}$ is the correlation matrix. Since $X_{ij}\sim N(0,1)$ when $\xi_i\sim \chi_p^2$, the symmetry of $N(0,1)$ implies that $(\bLambda^{1/2}\Ub_i)_j$ has a symmetric distribution. Hence, $\E X_{ij}^k=0$ for an odd $k$. For an even $k=2s$, by definition of $c_m$ in \eqref{cmp}, $\E[(\bLambda^{1/2}\Ub)^{2s}_j]=p^{-s}c_s$; also, $\E(\xi_i^{2s})= p^{s}\theta_s$; combining them gives $\E X_{ij}^{2s}=\E(\xi_i^{2s})\E[(\bLambda^{1/2}\Ub)^{2s}_j]=c_s\theta_s$. It follows that
\beq \label{thm-normality-meanS}
\E(S_k)  = \begin{cases}
0, & \mbox{$k$ is odd}, \cr
c_{k/2}\theta_{k/2}, & \mbox{$k$ is even}.
\end{cases}
\eeq
Moreover, $\cov(S_{kj}, S_{\ell j})= \frac{1}{n}\cov(X_{ij}^k, X_{ij}^\ell)=\frac{1}{n}[\E X_{ij}^{k+\ell} - (\E X_{ij}^k)(\E X_{ij}^\ell)]$. It follows that
\beq \label{thm-normality-covS}
\mathrm{Cov}(S_k, S_\ell) = \frac{1}{n}\begin{cases}
0, & \mbox{$k$ is odd, $\ell$ is even},\cr
c_{(k+\ell)/2}\theta_{(k+\ell)/2}, & \mbox{$k$ and $\ell$ are odd},\cr
c_{(k+\ell)/2}\theta_{(k+\ell)/2} - c_{k/2}\theta_{k/2}c_{\ell/2}\theta_{\ell/2}, & \mbox{$k$ and $\ell$ are even}.
\end{cases}
\eeq
By classical central limit theorem,
\beq \label{thm-normality-CLT}
\sqrt{n}[\cov(\Sb)]^{-1/2}(\Sb - \E\Sb) \to_d N\bigl(\bzero, \bI_{2m}\bigr).
\eeq

Next, we prove \eqref{thm-normality-0}. Define a function $h: \mathbb{R}^{2m}\to\mathbb{R}$ by $h(\bx) = \sum_{k=0}^{2m}\gamma_k x_1^kx_{2m-k}$. By \eqref{thm-normality-meanS},
\[
h(\E\Sb) =  \sum_{k=0}^{2m}\gamma_k (\E S_{1j})^k \E S_{(2m-k)j} = \E[S_{(2m)j}] = c_m\theta_m.
\]
Note that $\frac{\partial}{\partial x_1}h(\bx) = \sum_{k=1}^{2m}k \gamma_k x_1^{k-1} x_{2m-k}$, and  $\frac{\partial}{\partial x_k}h(\bx) = \gamma_{2m-k} x_1^{2m-k}$ for $k\neq 1$. Combining them with \eqref{thm-normality-meanS} and \eqref{thm-normality-covS} gives
\[
\bigtriangledown h(\E\Sb) = (0,0,\ldots,0, 1)^{\T}, \qquad  [\bigtriangledown h(\E\Sb)]^{\T}\cov(\Sb)\; [\bigtriangledown h(\E\Sb)] = c_{2m}\theta_{2m} - c_m^2\theta_m^2.
\]
We then apply the delta method and obtain
\beq \label{thm-normality-2}
\frac{\sqrt{n}[h(\Sb) - c_m\theta_m]}{\sqrt{c_{2m}\theta_{2m}-c_m^2\theta_m^2}}\to_d N(0, 1).
\eeq
By \eqref{thm-normality-1}, $\widehat{\theta}^{\, \rM}_{m,j}= \frac{\sigma_{jj}^m}{\widehat{\sigma}_{jj}^m}\cdot c_m^{-1}h(\Sb)$. Since $\frac{\sigma_{jj}^m}{\widehat{\sigma}_{jj}^m}\to 1$ in probability, using the Slutsky's lemma, we have
\[
\frac{\sqrt{n}(\widehat{\theta}^{\, \rM}_{m,j} - \theta_m)}{\frac{1}{c_m}\sqrt{c_{2m}\theta_{2m}-c_m^2\theta_m^2}} \to_d N(0, 1).
\]
This proves \eqref{thm-normality-0}.

\subsection{Proof of Theorem~\ref{thm:root-n(BAE)}}
Write for short $\widehat{\theta}^{\, \rB}_m=\widehat{\theta}^{\, \rB}_m\bigl(\hbmu,\;\mathrm{diag}_{\cal A}(\hbSigma)\bigr)$ and $\widetilde{\theta}^{\, \rB}_m=\widehat{\theta}^{\, \rB}_m\bigl(\bmu,\;\mathrm{diag}_{\cal A}(\bSigma)\bigr)$. It follows from Theorem~\ref{thm:BAEideal} that $\E[(\widetilde{\theta}^{\, \rB}_m-\theta_m)^2]=O(n^{-1/2})$. This implies $|\widetilde{\theta}^{\, \rB}_m-\theta_m|=O_{\mathbb{P}}(n^{-1/2})$. Hence, it suffices to show
\beq \label{BAErootn-key}
|\widehat{\theta}^{\, \rB}_m-\widetilde{\theta}^{\, \rB}_m| = O_{\mathbb{P}}(n^{-1/2}).
\eeq

First, we derive an expression of $\widehat{\theta}^{\, \rB}_m-\widetilde{\theta}^{\, \rB}_m$. Let $\Xb_{i,J}=\bSigma_{J,J}^{-1/2}(\Yb_{i,J}-\bmu_J)$ and $\widehat{\Xb}_{i,J}=\hbSigma_{J,J}^{-1/2}(\Yb_{i,J}-\hbmu_J)$ for all $1\leq i\leq n$ and $J\in{\cal A}$. Then,
\beq \label{BAErootn-1}
\widehat{\theta}^{\, \rB}_m = \frac{1}{n|{\cal A}|}\sum_{J\in {\cal A}}\sum_{i=1}^n\frac{\|\Xb_{i,J}\|^2}{c^*_{m,|J|}}, \qquad \widetilde{\theta}^{\, \rB}_m = \frac{1}{n|{\cal A}|}\sum_{J\in {\cal A}}\sum_{i=1}^n\frac{\|\widehat{\Xb}_{i,J}\|^2}{c^*_{m,|J|}}.
\eeq
Let $\Sb_{1,J}=\frac{1}{n}\sum_{i=1}^n \Xb_{i,J}$ and $\Sb_{2,J}=\frac{1}{n}\sum_{i=1}^n\Xb_{i,J}\Xb_{i,J}^{\T}$. By direct calculations,
\[
\bSigma^{-1/2}(\hbmu_{J}-\bmu_J)= \Sb_{1,J}, \qquad \bSigma^{-1/2}_{J,J}\hbSigma_{J,J}\bSigma^{-1/2} = \Sb_{2,J} -\Sb_{1,J}\Sb_{1,J}^{\T}.
\]
Define an event $B$ such that
\beq \label{BAErootn-event}
\left\{ \begin{array}{l}
\max_{1\leq i\leq n, J\in{\cal A}}\|\Xb_{i,J}\|\leq C\sqrt{\log(n\vee p)},\\
\max_{1\leq i\leq n, J\in{\cal A}, 1\leq k\leq 4m}\bigl| \|\Xb_{i,J}\|^k-\E\|\Xb_{i,J}\|^k\bigr| \leq C\sqrt{\log(n\vee p)},\\
\max_{J\in{\cal A}}\|\Sb_{1,J}\|\leq C\sqrt{(\log p) / n},\\
\max_{J\in{\cal A}}\|\Sb_{2,J}-\E\Sb_{2,J}\|\leq C\sqrt{(\log p) / n}.
\end{array}
\right.
\eeq
It is not hard to see that the event $B$ holds with probability $1-o(1)$ (see the proof of Theorem~\ref{thm:root-n} for similar arguments). On the event $B$, noting that $\E\Sb_{2,J}=\bI_{|J|}$, we have
\begin{align*}
 (\bSigma^{-1/2}_{J,J}\hbSigma_{J,J}\bSigma^{-1/2})^{-1} &= \bigl[ \bI_{|J|} + (\Sb_{2,J}-\E\Sb_{2,J}) -\Sb_{1,J}\Sb_{1,J}^{\T}\bigr]^{-1}\cr
 &=  \bI_{|J|} -  (\Sb_{2,J}-\E\Sb_{2,J}) + O(n^{-1}\log(p)).
\end{align*}
It follows that
\begin{align}  \label{BAErootn-2}
\|\widehat{\Xb}_{i,J}\|^2 &= (\Yb_{i,J}-\hbmu_J)^{\T}\hbSigma_{J,J}^{-1}(\Yb_{i,J}-\hbmu_J)\cr
&= \bigl[ \bSigma_{J,J}^{-1/2}(\Yb_{i,J}-\hbmu_J)\bigr]^{\T} \bigl[ \bSigma^{-1/2}_{J,J}\hbSigma_{J,J}\bSigma^{-1/2} \bigr]^{-1}
\bigl[ \bSigma_{J,J}^{-1/2}(\Yb_{i,J}-\hbmu_J)^{\T} \bigr]\cr
&= (\Xb_{i,J}- \Sb_{1,J})^{\T}\bigl\{\bI_{|J|} + (\Sb_{2,J}-\E\Sb_{2,J})\bigr\} (\Xb_{i,J}- \Sb_{1,J})+O(n^{-1}\log^2(n\vee p))\cr
&= \|\Xb_{i,J}\|^2\underbrace{-2\Sb_{1,J}^{\T}\Xb_{i,J} + \Xb_{i,J}^{\T}(\Sb_{2,J}-\E\Sb_{2,J})\Xb_{i,J}^{\T} + O(n^{-1}\log^2(n\vee p))}_{\equiv \Delta_{i,J}}.
\end{align}
Over the event $B$, $|\Delta_{i,J}|\leq Cn^{-1/2}\log(n\vee p)$.
As a result,
\begin{align*}
\|\widehat{\Xb}_{i,J}\|^{2m} & = (\|\Xb_{i,J}\|^2 +\Delta_{i,J})^{2m}\cr
&= \sum_{k=0}^m {m\choose k} \|\Xb_{i,J}\|^{2(m-k)}\Delta_{i,J}^k\cr
&= \|\Xb_{i,J}\|^{2m} +m \|\Xb_{i,J}\|^{2m-2}\Delta_{i,k} + O(n^{-1}\log^{m}(n\vee p)).
\end{align*}
Plugging it into \eqref{BAErootn-1}, we obtain
\begin{align} \label{BAErootn-3}
\widehat{\theta}^{\, \rB}_m - \widetilde{\theta}^{\, \rB}_m &=
 \frac{m}{n|{\cal A}|}\sum_{J\in {\cal A}}\sum_{i=1}^n\frac{1}{c^*_{m,|J|}}\|\Xb_{i,J}\|^{2m-2}\Delta_{i,J} + O(n^{-1}\log^{m}(n\vee p))\cr
 &= \frac{m}{n|{\cal A}|}\sum_{i=1}^n\sum_{J\in {\cal A}}\frac{1}{c^*_{m,|J|}}\|\Xb_{i,J}\|^{2m-2}\Xb_{i,J}^{\T}(\Sb_{2,J}-\E\Sb_{2,J})\Xb_{i,J}^{\T}\cr
 &\qquad - \frac{2m}{n|{\cal A}|}\sum_{i=1}^n\sum_{J\in {\cal A}}\frac{1}{c^*_{m,|J|}}\|\Xb_{i,J}\|^{2m-2}\Sb_{1,J}^{\T}\Xb_{i,J} + O(n^{-1}\log^{m}(n\vee p))\cr
 &= (K_1) + (K_2) + o(n^{-1/2}).
\end{align}

Next, we bound $(K_1)$ and $(K_2)$. Note that $\Sb_{2,J}-\E\Sb_{2,J}=\frac{1}{n}\sum_{k=1}^n [\Xb_{k,J}\Xb_{k,J}-\E(\Xb_{k,J}\Xb_{k,J})]$. This allows us to re-write
\begin{align*}
(K_1) =& \frac{m}{n^2|{\cal A}|}\sum_{i,k=1}^n\underbrace{\sum_{J\in {\cal A}}\frac{1}{c^*_{m,|J|}}\|\Xb_{i,J}\|^{2m-2}\Xb_{i,J}^{\T}\bigl[\Xb_{k,J}\Xb_{k,J}-\E(\Xb_{k,J}\Xb_{k,J})\bigr]\Xb_{i,J}^{\T}}_{\equiv Q_{ik}}.
\end{align*}
It is not hard to see that $\E|Q_{ii}|\leq C|{\cal A}|$ and that $\E|Q_{ik}Q_{i'k'}|\leq C|{\cal A}|^2$ when $\{i,k,i',k'\}$ has at least two distinct values (see the proof of Theorem~\ref{thm:root-n} for similar arguments). As a result,
\[
\E\Bigl|\frac{m}{n^2|{\cal A}|}\sum_{i=1}^n Q_{ii}\Bigr|=O(n^{-1})\qquad \Longrightarrow\qquad  \Bigl|\frac{m}{n^2|{\cal A}|}\sum_{i=1}^n Q_{ii}\Bigr| = o_{\mathbb{P}}(n^{-1/2}).
\]
Moreover, noting that $\E Q_{ik}=0$ for $i\neq k$, we have $\E (Q_{ik}Q_{i'k'})=0$ for $\{i,k,i',k'\}$ that are mutually distinct. It follows that
\begin{align*}
\E\Bigl(\frac{m}{n^2|{\cal A}|}\sum_{1\leq i\neq k\leq n} Q_{ik}\Bigr)^2 &= \frac{m^2}{n^4|{\cal A}|^2}\sum_{\substack{(i,k,i',k'):\text{at least }\\\text{two are equal}}} \E(Q_{ik}Q_{i'k'})\leq \frac{m^2}{n^4|{\cal A}|^2}\cdot n^3\cdot C|{\cal A}|^2 = O(n^{-1})\cr
 \qquad \Longrightarrow \qquad &  \Bigl|\frac{m}{n^2|{\cal A}|}\sum_{1\leq i\neq k\leq n} Q_{ik}\Bigr|=O_{\mathbb{P}}(n^{-1/2}).
\end{align*}
Combining the above gives
\beq \label{BAErootn-4}
(K_1) = O_{\mathbb{P}}(n^{-1/2}).
\eeq
Similarly, since $\Sb_{1,J}=\frac{1}{n}\sum_{k=1}^n \Xb_{i,J}$, we re-write
\[
(K_2) = - \frac{2m}{n|{\cal A}|}\sum_{i,k=1}^n\underbrace{\sum_{J\in {\cal A}}\frac{1}{c^*_{m,|J|}}\|\Xb_{i,J}\|^{2m-2}\Xb_{k,J}^{\T}\Xb_{i,J}}_{R_{ik}}.
\]
Then, $\E R_{ik}=0$ for $i\neq k$, $\E |R_{ii}|\leq C|{\cal A}|$, and $\E(R_{ik}R_{i'k'})\leq C|{\cal A}|^2$ when $\{i,k,i',k'\}$ has at least two distinct values. As a result,
\begin{align*}
\E\Bigl|\frac{m}{n^2|{\cal A}|}\sum_{i=1}^n R_{ii}\Bigr|=O(n^{-1})\qquad &\Longrightarrow\qquad  \Bigl|\frac{m}{n^2|{\cal A}|}\sum_{i=1}^n R_{ii}\Bigr| = o_{\mathbb{P}}(n^{-1/2})\cr
 \E\Bigl(\frac{m}{n^2|{\cal A}|}\sum_{1\leq i\neq k\leq n} R_{ik}\Bigr)^2 =O\Bigl(\frac{n^3|{\cal A}|^2}{n^4|{\cal A}|^2}\Bigr)=O(n^{-1}) \quad &\Longrightarrow \quad   \Bigl|\frac{m}{n^2|{\cal A}|}\sum_{1\leq i\neq k\leq n} R_{ik}\Bigr|=O_{\mathbb{P}}(n^{-1/2}).
\end{align*}
We immediately have
\beq  \label{BAErootn-5}
(K_2) = O_{\mathbb{P}}(n^{-1/2}).
\eeq
Plugging \eqref{BAErootn-4}-\eqref{BAErootn-5} into \eqref{BAErootn-3} gives \eqref{BAErootn-key}. The claim then follows.

\subsection{Proof of Theorem~\ref{thm:consistency(BAE)}}
Similar to the proof of Theorem~\ref{thm:root-n}, let $\widetilde{\theta}^{\, \rB}_m$ and $\widehat{\theta}^{\, \rB}_m$ denote the BAE with true $(\bmu,\bSigma)$ and estimates $(\hbmu,\hbSigma)$; here, $(\hbmu,\hbSigma)$ may not be the sample mean and sample covariance matrix. By Theorem~\ref{thm:BAEideal}, $\E[(\widetilde{\theta}^{\, \rM}_m-\theta_m)^2]\leq Cn^{-1}$. It follows from the Markov's inequality that, for any $\epsilon>0$, there is a constant $C_\epsilon>0$ such that,  with probability $1-\epsilon/2$,
\[
|\widetilde{\theta}^{\, \rM}_m-\theta_m|\leq C_\epsilon n^{-1/2}.
\]
To show the claim, it suffices to show that, there is a constant $C'_\epsilon>0$ such that with probability $1-\epsilon/2$,
\beq \label{BAEconsistency-1}
|\widehat{\theta}^{\, \rB}_m-\widetilde{\theta}^{\, \rB}_m|\leq C_\epsilon \max\{\alpha_n,\beta_n\}.
\eeq

We now show \eqref{BAEconsistency-1}. Let $\Xb_{i,J}=\bSigma_{J,J}^{-1/2}(\Yb_{i,J}-\bmu_J)$ and $\widehat{\Xb}_{i,J}=\hbSigma_{J,J}^{-1/2}(\Yb_{i,J}-\hbmu_J)$. Then,
\[
\widehat{\theta}^{\, \rB}_m - \widetilde{\theta}^{\, \rB}_m = \frac{1}{n|{\cal A}|}\sum_{J\in {\cal A}}\sum_{i=1}^n\frac{\|\widehat{\Xb}_{i,J}\|^{2m}- \|\Xb_{i,J}\|^{2m}}{c^*_{m,|J|}}.
\]
By direct calculations,
\begin{align} \label{BAEconsistency-Delta}
\Delta_{i,J}& \equiv \|\widehat{\Xb}_{i,J}\|^2 - \|\Xb_{i,J}\|^2\cr
& = (\Yb_{i,J}-\bmu_J)^{\T}(\hbSigma_{J,J}^{-1}-\bSigma_{J,J}^{-1})(\Yb_{i,J}-\bmu_J)+ 2(\bmu_J-\hbmu_J)^{\T}\hbSigma_{J,J}^{-1}(\Yb_{i,J}-\bmu_J) \cr
&\qquad + (\bmu_J-\hbmu_J)^{\T}\hbSigma_{J,J}^{-1}(\bmu_J-\hbmu_J)\cr
&= \Xb_{i,J}^{\T} \bigl(\bSigma_{J,J}^{1/2}\hbSigma_{J,J}^{-1}\bSigma_{J,J}^{1/2}-\bI_{|J|}\bigr) \Xb_{i,J}-  2 \bigl[\bSigma^{-1/2}_{JJ}(\hbmu_J-\bmu_J)\bigr]^{\T}\bigl(\bSigma_{J,J}^{1/2}\hbSigma_{J,J}^{-1}\bSigma_{J,J}^{1/2}\bigr)\Xb_{i,J}\cr
&\qquad + \bigl[\bSigma^{-1/2}_{JJ}(\hbmu_J-\bmu_J)\bigr]^{\T}\bigl(\bSigma_{J,J}^{1/2}\hbSigma_{J,J}^{-1}\bSigma_{J,J}^{1/2}\bigr) \bigl[\bSigma^{-1/2}_{JJ}(\hbmu_J-\bmu_J)\bigr].
\end{align}
As a result,
\begin{align}  \label{BAEconsistency-2}
& \widehat{\theta}^{\, \rB}_m - \widetilde{\theta}^{\, \rB}_m \cr
=&\;\; \frac{1}{n|{\cal A}|}\sum_{J\in {\cal A}}\frac{1}{c^*_{m,|J|}}\sum_{i=1}^n\biggl[\sum_{k=1}^{m}{m\choose k}\|\Xb_{i,J}\|^{2(m-k)}\Delta_{i,J}^k\biggr]\cr
=&\;\; \frac{m}{n|{\cal A}|}\sum_{J\in {\cal A}}\frac{1}{c^*_{m,|J|}}\sum_{i=1}^n \|\Xb_{i,J}\|^{2(m-1)}\Delta_{i,J} + rem\cr
=&\;\;  \frac{m}{n|{\cal A}|}\sum_{J\in {\cal A}}\frac{1}{c^*_{m,|J|}}\sum_{i=1}^n \|\Xb_{i,J}\|^{2m-2}\Xb_{i,J}^{\T} \bigl(\bSigma_{J,J}^{1/2}\hbSigma_{J,J}^{-1}\bSigma_{J,J}^{1/2}-\bI_{|J|}\bigr) \Xb_{i,J} +  rem\cr
&\qquad  -
\frac{2m}{n|{\cal A}|}\sum_{J\in {\cal A}}\frac{1}{c^*_{m,|J|}}\sum_{i=1}^n \|\Xb_{i,J}\|^{2m-2}  [\bSigma^{-1/2}_{JJ}(\hbmu_J-\bmu_J)]^{\T}(\bSigma_{J,J}^{1/2}\hbSigma_{J,J}^{-1}\bSigma_{J,J}^{1/2})\Xb_{i,J}.
\end{align}
Introduce
\[
\widetilde{\Sb}^{(m)}_{2,J} = \frac{1}{n}\sum_{i=1}^n \|\Xb_{i,J}\|^{2m-2}\Xb_{i,J}\Xb_{i,J}^{\T}, \qquad \widetilde{\Sb}_{1,J}^{(m)} = \frac{1}{n}\sum_{i=1}^n \|\Xb_{i,J}\|^{2m-2}\Xb_{i,J}.
\]
Then, \eqref{BAEconsistency-2} can be rewritten as
\begin{align}  \label{BAEconsistency-3}
\widehat{\theta}^{\, \rB}_m - \widetilde{\theta}^{\, \rB}_m &
= \frac{m}{|{\cal A}|}\sum_{J\in {\cal A}}\frac{1}{c^*_{m,|J|}} \tr\Bigl[\bigl(\bSigma_{J,J}^{1/2}\hbSigma_{J,J}^{-1}\bSigma_{J,J}^{1/2}-\bI_{|J|}\bigr)\widetilde{\Sb}^{(m)}_{2,J}\Bigr] \cr
&\qquad - \frac{2m}{|{\cal A}|}\sum_{J\in {\cal A}}\frac{1}{c^*_{m,|J|}}[\bSigma^{-1/2}_{JJ}(\hbmu_J-\bmu_J)]^{\T}\widetilde{\Sb}_{1,J}^{(m)} + rem.
\end{align}

First, we study the main terms in \eqref{BAEconsistency-3}.
Note that $\widetilde{\Sb}^{(m)}_{2,J}$ is the sample covariance matrix of $\{\|\Xb_{i,J}\|^{m-1}\Xb_{i,J}:1\leq i\leq n\}$, and $\widetilde{\Sb}_{1,J}^{(m)}$ is the sample mean of $\{\|\Xb_{i,J}\|^{2m-2}\Xb_{i,J}:1\leq i\leq n\}$. Using similar calculations as in the proof of Theorem~\ref{thm:BAEideal}, we can prove that
\[
\Bigl\| \frac{1}{|{\cal A}|}\sum_{J\in {\cal A}}\mathbb{E}\widetilde{\Sb}^{(m)}_{2,J}\Bigr\|\leq C, \qquad \Bigl\| \frac{1}{|{\cal A}|}\sum_{J\in {\cal A}}\mathbb{E}\widetilde{\Sb}^{(m)}_{1,J}\Bigr\|\leq C.
\]
Combining it with the Markov inequality, for any $\epsilon>0$, there is $C>0$ such that, with probability $1-\epsilon/4$, $\bigl\| \frac{1}{|{\cal A}|}\sum_{J\in {\cal A}}\widetilde{\Sb}^{(m)}_{2,J}\bigr\|\leq C$ and $\bigl\| \frac{1}{|{\cal A}|}\sum_{J\in {\cal A}}\widetilde{\Sb}^{(m)}_{1,J}\bigr\|\leq C$. On this event, the sum of the first two terms in \eqref{BAEconsistency-3} is bounded in absolute value by
\begin{align} \label{BAEconsistency-main}
& C\Bigl\| \frac{1}{|{\cal A}|}\sum_{J\in {\cal A}}\widetilde{\Sb}^{(m)}_{2,J}\Bigr\|\cdot \max_{J\in{\cal A}}\bigl\| \bSigma_{J,J}^{1/2}\hbSigma_{J,J}^{-1}\bSigma_{J,J}^{1/2}-\bI_{|J|}\bigr\| + C\Bigl\| \frac{1}{|{\cal A}|}\sum_{J\in {\cal A}}\widetilde{\Sb}^{(m)}_{1,J}\Bigr\|\cdot \max_{J\in{\cal A}} \bigl\| \bSigma^{-1/2}_{JJ}(\hbmu_J-\bmu_J)  \bigr\|\cr
\leq\; & C\max_{J\in{\cal A}}\bigl\| \bSigma_{J,J}^{1/2}\hbSigma_{J,J}^{-1}\bSigma_{J,J}^{1/2}-\bI_{|J|}\bigr\| + C\max_{J\in{\cal A}} \bigl\| \bSigma^{-1/2}_{JJ}(\hbmu_J-\bmu_J)  \bigr\|
\leq  C\max\{\alpha_n,\beta_n\}.
\end{align}

Next, we study the remainder terms in \eqref{BAEconsistency-3}. By \eqref{BAEconsistency-Delta} and our assumption on $(\hbmu,\hbSigma)$, we have
\[
\|\Delta_{i,J}\|\leq C\beta_n\|\Xb_{i,J}\|^2 + C\alpha_n\|\Xb_{i,J}\|.
\]
It follows that $\|\Delta_{i,J}\|^k\leq C\beta^k_n\|\Xb_{i,J}\|^{2k} + C\alpha^k_n\|\Xb_{i,J}\|^k$. Then,
\begin{align*}
|rem| &\leq C\sum_{k=2}^{m} \frac{1}{n|{\cal A}|}\sum_{J\in{\cal A}}\sum_{i=1}^n \beta_n^k \|\Xb_{i,J}\|^{2n} + C\sum_{k=2}^{m} \frac{1}{n|{\cal A}|}\sum_{J\in{\cal A}}\sum_{i=1}^n \alpha_n^k \|\Xb_{i,J}\|^{2n-k}\cr
&\leq C\sum_{k=2}^m \beta_n^k\Bigl( \frac{1}{n|{\cal A}|}\sum_{J\in{\cal A}}\sum_{i=1}^n \|\Xb_{i,J}\|^{2n} \Bigr) +  C\sum_{k=2}^m \alpha_n^k\Bigl( \frac{1}{n|{\cal A}|}\sum_{J\in{\cal A}}\sum_{i=1}^n \|\Xb_{i,J}\|^{2n-k} \Bigr).
\end{align*}
Using similar calculations as in the proof of Theorem~\ref{thm:BAEideal}, we can prove that $\frac{1}{n|{\cal A}|}\sum_{J\in{\cal A}}\sum_{i=1}^n\mathbb{E}\|\Xb_{i,J}\|^k\leq C$, for all $1\leq k\leq 4m$. It follows from the Markov inequality that, for a constant $C>0$, with probability $1-\epsilon/4$, $\frac{1}{n|{\cal A}|}\sum_{J\in{\cal A}}\sum_{i=1}^n\|\Xb_{i,J}\|^k\leq C$, for all $1\leq k\leq 2m$. On this event,
\beq \label{BAEconsistency-rem}
|rem| \leq C(\alpha_n^2+\beta_n^2).
\eeq
Combining \eqref{BAEconsistency-main} and \eqref{BAEconsistency-rem} gives $|\widehat{\theta}^{\, \rB}_m - \widetilde{\theta}^{\, \rB}_m|\leq C\max\{\alpha_n,\beta_n\}$. This proves \eqref{BAEconsistency-1}, and the claim follows immediately.

\subsection{Proof of Theorem~\ref{thm:BAEideal}}
Fix a collection ${\cal A}$ of blocks. Write for short $\widehat{\theta}_m^B=\widehat{\theta}_m^B(\bmu, \mathrm{diag}_{\cal A}(\bSigma))$. For preparation, first, we verify that $\widehat{\theta}_m^B$ is an unbiased estimator. For any $J\in {\cal A}$, by \eqref{FangZhang1} and the fact that $\|\Ub_{|J|}\|=1$, we have
\[
\bigl[(\Yb_J - \bmu_J )^{\T} \bSigma_{J,J}^{-1} (\Yb_J - \bmu_J )\bigr]^m=\|\bSigma_{JJ}^{-1/2}(\Yb_J - \bmu_J)\|^{2m} = \|\xi B^{1/2} \Ub_{|J|}\|^{2m}= \xi^{2m}B^m.
\]
As a result,
\beq \label{thm-BAE-1}
\E\bigl[(\Yb_J - \bmu_J )^{\T} \bSigma_{J,J}^{-1} (\Yb_J - \bmu_J )\bigr]^m =  (\E\xi^{2m})(\E B^m)=\theta_m\cdot c^*_{m,|J|}.
\eeq
In particular, it implies that
\[
\theta_m = \frac{1}{|{\cal A}| n}\sum_{J\in{\cal A}} \bigg[ \frac{1}{ c^*_{m,|J|}} \sum_{i=1}^n \E\bigl\{  (\Yb_{i, J}-\bmu_J)^\T \bSigma_{J J}^{-1}(\Yb_{i, J}-\bmu_J) \bigl\}^{m} \bigg].
\]
Therefore, $\widehat{\theta}_m^B$ is unbiased. Additionally, we have
\beq  \label{thm-BAE-2}
\frac{\var(\widehat{\theta}_m^B)}{\theta_m^2} =\frac{1}{n|{\cal A}|^2} \var \left( \sum_{J \in {\cal A}}\frac{\bigl[(\Yb_J - \bmu_J )^{\T}  \bSigma_{J,J}^{-1}(\Yb_J - \bmu_J )\bigr]^m}{\theta_m\cdot c^*_{m,|J|}}\right).
\eeq
Second, we introduce an alternative expression of $\theta_m\cdot c^*_{m,|J|}$. Consider the special case $\xi\sim \chi_p^2$. Since $\Yb\sim N(\bmu,\bSigma)$ in this case, we then have $\Yb_J\sim N(\bmu_J, \bSigma_{J,J})$ and $(\Yb_J - \bmu_J )^{\T} \bSigma_{J,J}^{-1} (\Yb_J - \bmu_J )\sim \chi^2_{|J|}$. Hence, in \eqref{thm-BAE-1}, the left hand side equals to $\E\chi^{2m}_{|J|}$. At the same time, the right hand side is equal to $\theta_m\cdot c_{m,|J|}^*=p^{-m}\E\xi^{2m}\cdot c_{m,|J|}^*=p^{-m}\E\chi_p^{2m}\cdot c_{m,|J|}^*$. Equating the left/right hand sides gives
\[
c^*_{m,|J|} =\frac{p^m\E\chi^{2m}_{|J|}}{\E \chi_p^{2m}}.
\]
We combine it with the definition of $\theta_m=p^{-m}\E\xi^{2m}$ and $r_m=\E\xi^{2m}/\E\chi_p^{2m}$. It implies that
\beq \label{thm-BAE-3}
\theta_m\cdot c^*_{m,|J|} = \E\xi^{2m}\cdot \frac{\E\chi^{2m}_{|J|}}{\E \chi_p^{2m}} = r_m \cdot \E\chi^{2m}_{|J|}.
\eeq

We now show the claim. For $J\in {\cal A}$, let $W_J= [(\Yb_J - \bmu_J )^{\T} \bSigma_{J,J}^{-1}(\Yb_J - \bmu_J )]^m$. By \eqref{thm-BAE-2}-\eqref{thm-BAE-3},
\begin{align} \label{thm-BAE-4}
\frac{\var(\widehat{\theta}_m^B)}{\theta_m^2}& =\frac{1}{n|{\cal A}|^2}
 \var \left( \sum_{J \in {\cal A}}\frac{W_J}{r_m\cdot  \E\chi^{2m}_{|J|}}\right)\cr
 & = \frac{1}{n|{\cal A}|^2r_m^2} \sum_{J \in {\cal A} } \frac{\var (W_J )}{(\E\chi^{2m}_{|J|})^2}+ \frac{1}{n|{\cal A}|^2r_m^2} \sum_{\substack{I,J\in {\cal A}\\  I\neq J}} \frac{\cov (W_I, W_J)}{(\E\chi^{2m}_{|I|})(\E\chi^{2m}_{|J|})}\cr
 &\equiv (I) + (II).
\end{align}
Consider $(I)$. Combining \eqref{thm-BAE-1} and \eqref{thm-BAE-3}, we have
\beq \label{thm-BAE-5}
\E W_J = r_m \cdot \E\chi^{2m}_{|J|}, \qquad \E W_J^2 =  r_{2m} \cdot \E\chi^{4m}_{|J|}.
\eeq
Hence,
\begin{align} \label{thm-BAE-(I)}
(I) &= \frac{1}{n|{\cal A}|^2r_m^2} \sum_{J \in {\cal A} }\frac{r_{2m}\E\chi^{4m}_{|J|} - r_m^2 (\E\chi^{2m}_{|J|})^2}{(\E\chi^{2m}_{|J|})^2}\cr
& =  \frac{1}{n|{\cal A}|^2r_m^2} \sum_{J \in {\cal A} }\frac{r_{2m}\var(\chi^{2m}_{|J|}) + (r_{2m} - r_m^2) (\E\chi^{2m}_{|J|})^2}{(\E\chi^{2m}_{|J|})^2}\cr
& = \frac{1}{n|{\cal A}|^2} \sum_{J \in {\cal A} }\left[ \frac{r_{2m}}{r_m^2}\frac{\var(\chi^{2m}_{|J|})}{(\E\chi^{2m}_{|J|})^2} + \frac{(r_{2m} - r_m^2)}{r_m^2}\right]\cr
& = \frac{1}{np}\cdot  \frac{r_{2m}}{r_m^2}\cdot \underbrace{\frac{p}{|{\cal A}|^2}\sum_{J \in {\cal A} }\frac{h_m(|J|)}{|J|}}_{\bar{h}_m({\cal A})} +  \frac{1}{n|{\cal A}|}\cdot \frac{(r_{2m} - r_m^2)}{r_m^2},
\end{align}
where the last two lines are from Definition~\ref{def:h-factor}.

Consider $(II)$. Fix $I$ and $J$. Note that
\[
\cov(W_I, W_J) =  \E (W_IW_J) - (\E W_I)(\E W_J).
\]
We have had an expression of $\E W_I$ as in \eqref{thm-BAE-5}. We still need to get an expression of $\E (W_IW_J)$. For the set $I\cup J$, we apply \eqref{FangZhang1} and find that
\[
\begin{pmatrix}
\Yb_I\\\Yb_J
\end{pmatrix} = \begin{pmatrix}
\bmu_I\\\bmu_J
\end{pmatrix}  +\xi\cdot B^{1/2}\cdot \bSigma^{1/2}_{I\cup J,I\cup J}\Ub_{|I|+|J|},
\]
where $B$ is a Beta distribution with parameters $\frac{|I|+|J|}{2}$ and $\frac{p-(|I|+|J|)}{2}$.
Let $\widetilde{\Ub}_I$ and $\widetilde{\Ub}_J$ be the vectors formed by the first $|I|$ coordinates and the last $|J|$ coordinates of $\bSigma^{1/2}_{I\cup J,I\cup J}\Ub_{|I|+|J|}$, respectively. We then have $W_I=\xi^{2m} B^m \|\bSigma_{II}^{-1/2}\widetilde{\Ub}_I\|^{2m}$ and $W_J=\xi^{2m} B^m \|\bSigma_{JJ}^{-1/2}\widetilde{\Ub}_J\|^{2m}$. As a result,
\beq \label{thm-BAE-cross1}
\E(W_I W_J) = \E\xi^{4m}\cdot \E B^{2m} \cdot \E\bigl( \|\bSigma_{II}^{-1/2}\widetilde{\Ub}_I\|^{2m}\|\bSigma_{JJ}^{-1/2}\widetilde{\Ub}_J\|^{2m} \bigr).
\eeq
We then use the cross-moments of multivariate normal distributions to get the last term above.
Let $\xi_0^2\sim \chi_p^2$ be a random variable independent of $B$ and $\Ub_{|I|+|J|}$. The random vector
\[
\begin{pmatrix}
\Zb_I\\\Zb_J
\end{pmatrix} \equiv \xi_0\cdot B^{1/2}\cdot \begin{pmatrix}
\widetilde{\Ub}_I\\\widetilde{\Ub}_J
\end{pmatrix} \sim N\bigl({\bf 0},\; \bSigma_{I\cup J,I\cup J}\bigr).
\]
It follows that
\beq \label{thm-BAE-cross2}
\E\bigl(\|\bSigma_{II}^{-1/2}\Zb_I\|^{2m}\| \bSigma_{JJ}^{-1/2}\Zb_J\|^{2m}\bigr) = \E\chi_p^{4m}\cdot \E B^{2m} \cdot \E\bigl( \|\bSigma_{II}^{-1/2}\widetilde{\Ub}_I\|^{2m}\|\bSigma_{JJ}^{-1/2}\widetilde{\Ub}_J\|^{2m} \bigr).
\eeq
Write $\widetilde{\Zb}_1=\bSigma_{II}^{-1/2}\Zb_I$ and $\widetilde{\Zb}_2=\bSigma_{JJ}^{-1/2}\Zb_J$.
Note that
\beq  \label{thm-BAE-Zdist}
\begin{pmatrix}
\widetilde{\Zb}_1\\\widetilde{\Zb}_2
\end{pmatrix} \sim N\left( {\bf 0}, \;\;
 \begin{bmatrix}
 \bI_{|I|} & \bGamma\\
 \bGamma^{\T} & \bI_{|J|}
 \end{bmatrix}
\right), \qquad \mbox{where}\quad \bGamma = \bSigma_{II}^{-1/2}\bSigma_{IJ}\bSigma_{JJ}^{-1/2}.
\eeq
Combining \eqref{thm-BAE-cross1} and \eqref{thm-BAE-cross2} gives
\beq \label{thm-BAE-cross}
\E(W_I W_J) = \E(\|\widetilde{\Zb}_1\|^{2m}\|\widetilde{\Zb}_2\|^{2m} )\cdot \frac{\E\xi^{4m}}{\E\chi_p^{4m}}=  \E(\|\widetilde{\Zb}_1\|^{2m}\|\widetilde{\Zb}_2\|^{2m} )\cdot r_{2m}.
\eeq
We now combine \eqref{thm-BAE-5} and \eqref{thm-BAE-cross} and note that $\|\widetilde{\Zb}_1\|^2 \sim \chi_{|I|}^2$ and $\|\widetilde{\Zb}_2\|^2 \sim \chi_{|J|}^2$. It yields
\begin{align*}
\frac{\cov(W_I, W_J)}{(\E\chi_{|I|}^{2m})(\E\chi_{|J|}^{2m})} &= \frac{ r_{2m}\E(\|\widetilde{\Zb}_1\|^{2m}\|\widetilde{\Zb}_2\|^{2m} ) - r_m^2 (\E\chi_{|I|}^{2m})(\E\chi_{|J|}^{2m})}{(\E\chi_{|I|}^{2m})(\E\chi_{|J|}^{2m})}\cr
& = \frac{ r_{2m} \E(\|\widetilde{\Zb}_1\|^{2m}\|\widetilde{\Zb}_2\|^{2m} ) - r_m^2 (\E\|\widetilde{\Zb}_1\|^{2m})(\E\|\widetilde{\Zb}_1\|^{2m})}{(\E\chi_{|I|}^{2m})(\E\chi_{|J|}^{2m})}\cr
& = \frac{r_{2m} \cov( \|\widetilde{\Zb}_1\|^{2m}, \|\widetilde{\Zb}_2\|^{2m}  ) +(r_{2m}-r_m^2) (\E\|\widetilde{\Zb}_1\|^{2m})(\E\|\widetilde{\Zb}_1\|^{2m})}{(\E\chi_{|I|}^{2m})(\E\chi_{|J|}^{2m})} \cr
& = r_{2m}  \frac{\cov( \|\widetilde{\Zb}_1\|^{2m}, \|\widetilde{\Zb}_2\|^{2m} )}{ (\E\|\widetilde{\Zb}_1\|^{2m})(\E\|\widetilde{\Zb}_1\|^{2m})} + (r_{2m}-r_m^2).
\end{align*}
As a result,
\begin{align} \label{thm-BAE-(II)}
(II) &= \frac{1}{n|{\cal A}|^2r_m^2} \sum_{\substack{I,J\in {\cal A}\\  I\neq J}} \left[ r_{2m} \frac{\cov( \|\widetilde{\Zb}_1\|^{2m}, \|\widetilde{\Zb}_2\|^{2m} )}{ (\E\|\widetilde{\Zb}_1\|^{2m})(\E\|\widetilde{\Zb}_1\|^{2m})} + (r_{2m}-r_m^2)\right]\cr
& = \frac{1}{n}\cdot \frac{r_{2m}}{r_m^2}\cdot \frac{1}{|{\cal A}|^2} \sum_{\substack{I,J\in {\cal A}\\  I\neq J}} \frac{\cov( \|\widetilde{\Zb}_1\|^{2m}, \|\widetilde{\Zb}_2\|^{2m} )}{ (\E\|\widetilde{\Zb}_1\|^{2m})(\E\|\widetilde{\Zb}_1\|^{2m})} + \frac{1}{n}\cdot \frac{(r_{2m}-r_m^2)}{r_m^2} \Bigl(1 - \frac{1}{|{\cal A}|}\Bigr).
\end{align}

We now plug \eqref{thm-BAE-(I)} and \eqref{thm-BAE-(II)} into \eqref{thm-BAE-4}. It gives
\begin{align} \label{thm-BAE-final}
\frac{\var(\widehat{\theta}_m^B)}{\theta_m^2} & \leq  \frac{1}{n}\cdot \frac{(r_{2m} - r_m^2)}{r_m^2} + \frac{1}{np}\cdot  \frac{r_{2m}}{r_m^2} \bar{h}_m({\cal A}) \cr
&+ \frac{1}{n}\cdot \frac{r_{2m}}{r_m^2}\cdot \frac{1}{|{\cal A}|^2} \sum_{\substack{I,J\in {\cal A}\\  I\neq J}} \frac{\cov( \|\widetilde{\Zb}_1\|^{2m}, \|\widetilde{\Zb}_2\|^{2m} )}{ (\E\|\widetilde{\Zb}_1\|^{2m})(\E\|\widetilde{\Zb}_1\|^{2m})}.
\end{align}
What remains is to bound the last term. Since the random vectors $\widetilde{\Zb}_1$ and $\widetilde{\Zb}_2$ jointly follow a multivariate normal distribution as dictated in \eqref{thm-BAE-Zdist}, we can apply the following lemma:
\begin{lemma} \label{lem:multivar-normal}
Let $\Zb_1$ and $\Zb_2$ be two random vectors such that
\[
\begin{pmatrix}
\Zb_1\\ \Zb_2
\end{pmatrix} \sim N\left( {\bf 0}, \;\;
 \begin{bmatrix}
 \bI_{k_1} & \bGamma\\
 \bGamma' & \bI_{k_2}
 \end{bmatrix}
\right).
\]
Then, for a constant $\widetilde{C}_m>0$ that only depends on $m$ but is independent of $(k_1,k_2)$,
\[
0\leq \frac{\cov( \|\Zb_1\|^{2m}, \|\Zb_2\|^{2m} )}{ (\E\|\Zb_1\|^{2m})(\E\|\Zb_1\|^{2m})}\leq \widetilde{C}_m \|\bGamma\|^2.
\]
\end{lemma}
We combine Lemma~\ref{lem:multivar-normal} with \eqref{thm-BAE-Zdist} and them plug it into \eqref{thm-BAE-final}. It follows that
\begin{align*}
\frac{\var(\widehat{\theta}_m^B)}{\theta_m^2} \leq  \frac{1}{n}\;\frac{(r_{2m} - r_m^2)}{r_m^2} + \frac{1}{np}\; \frac{r_{2m}}{r_m^2} \bar{h}_m({\cal A}) + \frac{1}{n}\frac{r_{2m}}{r_m^2}\; \frac{\widetilde{C}_m}{|{\cal A}|^2} \sum_{\substack{I,J\in {\cal A}\\  I\neq J}}\|\bSigma_{II}^{-1/2}\bSigma_{IJ}\bSigma_{JJ}^{-1/2}\|^2.
\end{align*}
This proves the claim.

\section{Supplementary proofs}

\subsection{Proof of Lemma~\ref{lem:bivar-normal}} \label{subsec:bivar-proof}
Let $\theta=\arcsin\bigl(\sgn(\rho)\cdot \sqrt{|\rho|}\bigr)\in [-\frac{\pi}{2},\frac{\pi}{2}]$. We then have $\sin\theta=\sgn(\rho)\cdot\sqrt{|\rho|}$ and $\cos\theta=\sqrt{1-|\rho|}$.
Let $U_1, U_2, V$ be iid $N(0,1)$ random variables. It is easy to see that
\begin{align*}
(Z_1, Z_2) \overset{(d)}{=} \Bigl( (\cos\theta) U_1  + (\sin\theta)  V,  \ \  (\cos\theta) U_2 + (\sin\theta) V\Bigr).
\end{align*}
For notation simplicity, we omit the superscript $(d)$ in all equations. It follows that
\begin{align*}
Z_1^{2m} &= \sum_{k_1=0}^{2m}{2m \choose k_1}(\cos\theta)^{2m-k_1}(\sin\theta)^{k_1}U_1^{2m-k_1}V^{k_1}, \cr
Z_2^{2m} &= \sum_{k_2=0}^{2m}{2m \choose k_2}(\cos\theta)^{2m-k_1}(\sin\theta)^{k_2}U_2^{2m-k_1}V^{k_2}.
\end{align*}
Then,
\[
\cov(Z_1^{2m}, Z_2^{2m})= \sum_{k_1,k_2=0}^{2m} {2m \choose k_1}{2m \choose k_2}(\cos\theta)^{4m-k_1-k_2}(\sin\theta)^{k_1+k_2}\bigl[\cov(U_1^{2m-k_1}V^{k_1},\;U_2^{2m-k_2}V^{k_2})\bigr].
\]
Note that for random variables $(X,Y,W_1,W_2)$, when $X$, $Y$ and $(W_1,W_2)$ are mutually independent, $\cov(XW_1, YW_2)=\E X\cdot \E Y\cdot \cov(W_1, W_2)$. Plugging it into the above expression, we obtain
\begin{align*}
& \cov(Z_1^{2m}, Z_2^{2m})\cr
 =& \sum_{\substack{2\leq k_1,k_2\leq 2m\\k_1,k_2 \text{ even}}} {2m \choose k_1}{2m \choose k_2}(\cos\theta)^{4m-k_1-k_2}(\sin\theta)^{k_1+k_2}(\E U_1^{2m-k_1})(\E U_2^{2m-k_2})\cov(V^{k_1}, V^{k_2})\cr
 =&\sum_{s=2}^{m} (\cos\theta)^{2m-2s}(\sin\theta)^{2s} \sum_{\substack{1\leq k_1,k_2\leq m\\ k_1+k_2=s}}{2m \choose 2k_1}{2m \choose 2k_2}\bigl[\E U_1^{2(m-k_1)}\bigr]\bigl[\E U_2^{2(m-k_2)}\bigr]\bigl(\E V^{2s} - \E V^{2k_1}\E V^{2k_2}\bigr).
\end{align*}
Using our previous notations, $\eta_m$ is the $2m$-th moment of $N(0,1)$. By elementary statistics,
$\eta_m=(2m-1)!!=\prod_{j=0}^{m-1}(1+2j)$. Using this formula, we can prove $\E V^{2s} - \E V^{2k_1}\E V^{2k_2}\geq 0$. Hence,
\[
\cov(Z_1^{2m}, Z_2^{2m})\geq 0.
\]
At the same time, we note that $\cos^2\theta = 1-|\rho|$ and $\sin^2\theta=|\rho|$.
It follows that
\begin{align*}
\cov(Z_1^{2m}, Z_2^{2m})
& \leq \sum_{s=2}^{m} (1-|\rho|)^{m-s}|\rho|^{s}\cdot \underbrace{\sum_{\substack{1\leq k_1,k_2\leq m\\ k_1+k_2=s}} {2m \choose 2k_1}{2m \choose 2k_2}\cdot \eta_{m-k_1}\eta_{m-k_2}\eta_s}_{B_m(s)}\cr
 &\leq \bigl[\max_sB_m(s)\bigr]\cdot \sum_{s=2}^{m} (1-|\rho|)^{m-s}|\rho|^{s}\leq \bigl[\max_sB_m(s)\bigr]\cdot |\rho|^2.
\end{align*}
The claim then follows.

\subsection{Proof of Lemma~\ref{lem:multivar-normal}}
Suppose the rank of $\bGamma$ is $k\leq \min\{k_1,k_2\}$.
Let $\bGamma=\bH_1\bLambda\bH_2^{\T}$ be the singular value decomposition of $\bGamma$. We note that all singular values have an absolute value no larger than $1$. For $\ell=1,2$, let $\widetilde{\bH}_\ell\in\mathbb{R}^{k_\ell,k_\ell-k}$ be such that $[\bH_\ell,\widetilde{\bH}_\ell]$ form an orthogonal basis of $\mathbb{R}^{k_\ell}$. Define
\[
\bA_\ell = \bigl[\bH_\ell(\bI-\bLambda)^{1/2}, \;\; \widetilde{\bH}_\ell\bigr], \qquad \ell=1,2.
\]
It is easy to see that $\bA_\ell\bA'_\ell=\bI-\bH_\ell\bLambda\bH_\ell'$.
Let $\Xb_1\sim N({\bf 0}, \bI_{k_1})$, $\Xb_2\sim N({\bf 0}, \bI_{k_2})$, and $\Wb\sim N({\bf 0}, \bI_{k})$ be mutually independent random variables. We claim that
\[
\begin{pmatrix}
\Zb_1\\
\Zb_2
\end{pmatrix} \overset{(d)}{=}
\begin{pmatrix}
\bA_1\Xb_1 + \bH_1\bLambda^{1/2}\Wb\\
\bA_2\Xb_2 + \bH_2\bLambda^{1/2}\Wb
\end{pmatrix}.
\]
This can be verified by computing the covariance matrix of the right hand side. We shall omit the superscript $(d)$ in all equations for notation simplicity. Write $\Xb_\ell=(\Xb_{\ell1}^{\T},\Xb_{\ell 2}^{\T})^{\T}$, corresponding to the first $k_\ell$ and the last $(k_\ell-k)$ coordinates, respectively, $\ell=1,2$. It follows that
\begin{align} \label{lem-MN-key}
\|\Zb_\ell\|^2 &= \| \bA_\ell\Xb_\ell + \bH_\ell\bLambda^{1/2}\Wb \|^2 \cr
& = \| \bH_\ell(\bI-\bLambda)^{1/2}\Xb_{\ell 1} + \widetilde{\bH}_\ell\Xb_{\ell 2} + \bH_\ell\bLambda^{1/2}\Wb \|^2 \cr
&= \| \bH_\ell(\bI-\bLambda)^{1/2}\Xb_{\ell 1}\|^2 + \|\widetilde{\bH}_\ell\Xb_{\ell 2}\|^2 + \|\bH_\ell\bLambda^{1/2}\Wb \|^2, \cr
& = \underbrace{\|(\bI-\bLambda)^{1/2}\Xb_{\ell 1}\|^2 + \|\Xb_{\ell 2}\|^2}_{\equiv U_\ell} + \underbrace{\|\bLambda^{1/2}\Wb \|^2}_{\equiv V},
\end{align}
where the third line is from the zero mean and mutual independence of $(\Xb_{\ell 1},\Xb_{\ell 2}, \Wb)$ and the last line is due to that $\bH_\ell'\bH_\ell=\bI_{k}$ and $\widetilde{\bH}_\ell\widetilde{\bH}_\ell'=\bI_{k_\ell-k}$. Since $(U_1,U_2,V)$ are mutually independent, it follows that
\begin{align} \label{lem-MN-1}
\cov\bigl( \|\Zb_1\|^{2m}, \|\Zb_2\|^{2m}  \bigr) & = \cov\left(
\sum_{j_1=1}^{m}{m\choose j_1}U_1^{m-j_1}V^{j_1},\;\;\; \sum_{j_2=1}^{m}{m\choose j_2}U_2^{m-j_2}V^{j_1}\right)\cr
& =  \sum_{j_1,j_2=1}^{m}{m\choose j_1}{m\choose j_2} \cov(U_1^{m-j_1}V^{j_1}, U_2^{m-j_2}V^{j_2})\cr
&= \sum_{j_1,j_2=1}^{m}{m\choose j_1}{m\choose j_2} (\E U_1^{m-j_1})(\E U_2^{m-j_2})\cov(V^{j_1}, V^{j_2}).
\end{align}
It is not hard to see that $\cov(V^{j_1}, V^{j_2})\geq 0$. Hence, $\cov\bigl( \|\Zb_1\|^{2m}, \|\Zb_2\|^{2m}  \bigr)\geq 0$. Furthermore,
since all entries of the diagonal matrix $\bLambda$ are between $0$ and $1$, we have
\[
U_\ell \leq \sum_{j=1}^{k_\ell}X^2_\ell(j), \qquad V\leq \|\bLambda\|\sum_{j=1}^kW^2(j),
\]
where $X_\ell(j)$'s and $W(j)$'s are all $iid$ standard normal variables. In particular,
\[
0\leq \E U_\ell^{m-j_\ell} \leq \E\chi^{2(m-j_\ell)}_{k_\ell}, \qquad \cov(V^{j_1}, V^{j_2})\leq \E V^{j_1+j_2}\leq \|\bLambda\|^{j_1+j_2} \E\chi_k^{2(j_1+j_2)}.
\]
Plugging these results into \eqref{lem-MN-1} gives
\begin{align*} \label{lem-MN-2}
\frac{\cov( \|\Zb_1\|^{2m}, \|\Zb_2\|^{2m} )}{ (\E\|\Zb_1\|^{2m})(\E\|\Zb_1\|^{2m})}&=\frac{\cov( \|\Zb_1\|^{2m}, \|\Zb_2\|^{2m} )}{ (\E\chi_{k_1}^{2m})(\E\chi_{k_2}^{2m})} \cr
&\leq  \sum_{j_1,j_2=1}^{m}\|\bLambda\|^{j_1+j_2} {m\choose j_1}{m\choose j_2} \frac{(\E \chi_{k_1}^{2m-2j_1})(\E \chi_{k_2}^{2m-2j_2})(\E\chi_k^{2(j_1+j_2)})}{ (\E\chi_{k_1}^{2m})(\E\chi_{k_2}^{2m})}
\end{align*}
We note that $m$ is bounded, but $(k_1,k_2,k)$ can grow with $(n,p)$.
Note that $\E\chi_k^{2m}=\prod_{j=0}^{m-1}(k+2j)$ for all $k,m\geq 1$. As a result,
\begin{align*}
 \frac{(\E \chi_{k_1}^{2m-2j_1})(\E \chi_{k_2}^{2m-2j_2})(\E\chi_k^{2(j_1+j_2)})}{ (\E\chi_{k_1}^{2m})(\E\chi_{k_2}^{2m})} &=\frac{\prod_{j=0}^{m-j_1-1}(k_1+2j)\prod_{j=0}^{m-j_2-1}(k_2+2j)\prod_{j=0}^{j_1+j_2-1}(k+2j)}{\prod_{j=0}^{m-1}(k_1+2j)\prod_{j=0}^{m-1}(k_2+2j)}\cr
 &= \frac{\prod_{j=0}^{j_1+j_2-1}(k+2j)}{\prod_{j=m-j_1}^{m-1}(k_1+2j)\prod_{j=m-j_2}^{m-1}(k_2+2j)}\leq 1.
\end{align*}
Therefore,
\beq
\frac{\cov( \|\Zb_1\|^{2m}, \|\Zb_2\|^{2m} )}{ (\E\|\Zb_1\|^{2m})(\E\|\Zb_1\|^{2m})}\leq  \sum_{j_1,j_2=1}^{m}\|\bLambda\|^{j_1+j_2} {m\choose j_1}{m\choose j_2}=O(\|\bLambda\|^2).
\eeq
Noticing that $\bLambda$ is a diagonal matrix containing the singular values of $\bGamma$, we have proved the claim.

\section{The case of multivariate Gaussian distributions}
We present a corollary about the errors of MAE and BAE for the special case of multivariate Gaussian distributions. Here $R_n(\widehat{\theta}_2)=\E[(\widehat{\theta}_2-\theta_2)^2/\theta_2^2]$. The proof is elementary and omitted.
\begin{coro} \label{coro:Gaussian}
Let $\Yb_1,\cdots,\Yb_n$ be i.i.d. samples of $N(\bmu,\bSigma)$. For a constant integer $k\geq 2$, we assume the blocks in BAE are $J_i=\{(i-1)k+1, (i-1)k+2,\cdots, \min\{ik, p\}\}$, $1\leq i\leq \lceil p/k\rceil$.
\begin{itemize}
\item Suppose $\bSigma=\bI_p$. Then, $R_n(\widehat{\theta}_2^{\,\rI})\sim \frac{8}{np}$, $R_n(\widehat{\theta}_2^{\,\rM})\sim \frac{32}{3np}$, and $R_n(\widehat{\theta}^{\,\rB}_2)\sim \frac{8(k+3)}{(k+2)np}$.
\item Suppose $\bSigma$ is a block-wise diagonal matrix with $2\times 2$ blocks, where each block has diagonals $1$ and off-diagonals $\rho\in (-1,1)$. Let $k=2$ in BAE. Then, $R_n(\widehat{\theta}_2^{\,\rI})\sim \frac{8}{np}$, $R_n(\widehat{\theta}_2^{\,\rM})\sim\frac{ 8(4+3\rho^2+\rho^4)}{3np}$,  and $R_n(\widehat{\theta}^{\,\rB}_2)\sim \frac{10}{np}$.
\end{itemize}
\end{coro}

\section{Simulations for the estimator in Section~\ref{sec:realized_xi}} \label{sec:Sec5Simu}

We conducted simulations to investigate the performance of the estimator of realized $\xi_t$ in Section~\ref{sec:realized_xi}. 

\begin{figure}[!t]
        \centering
                \includegraphics[width=.6\textwidth]{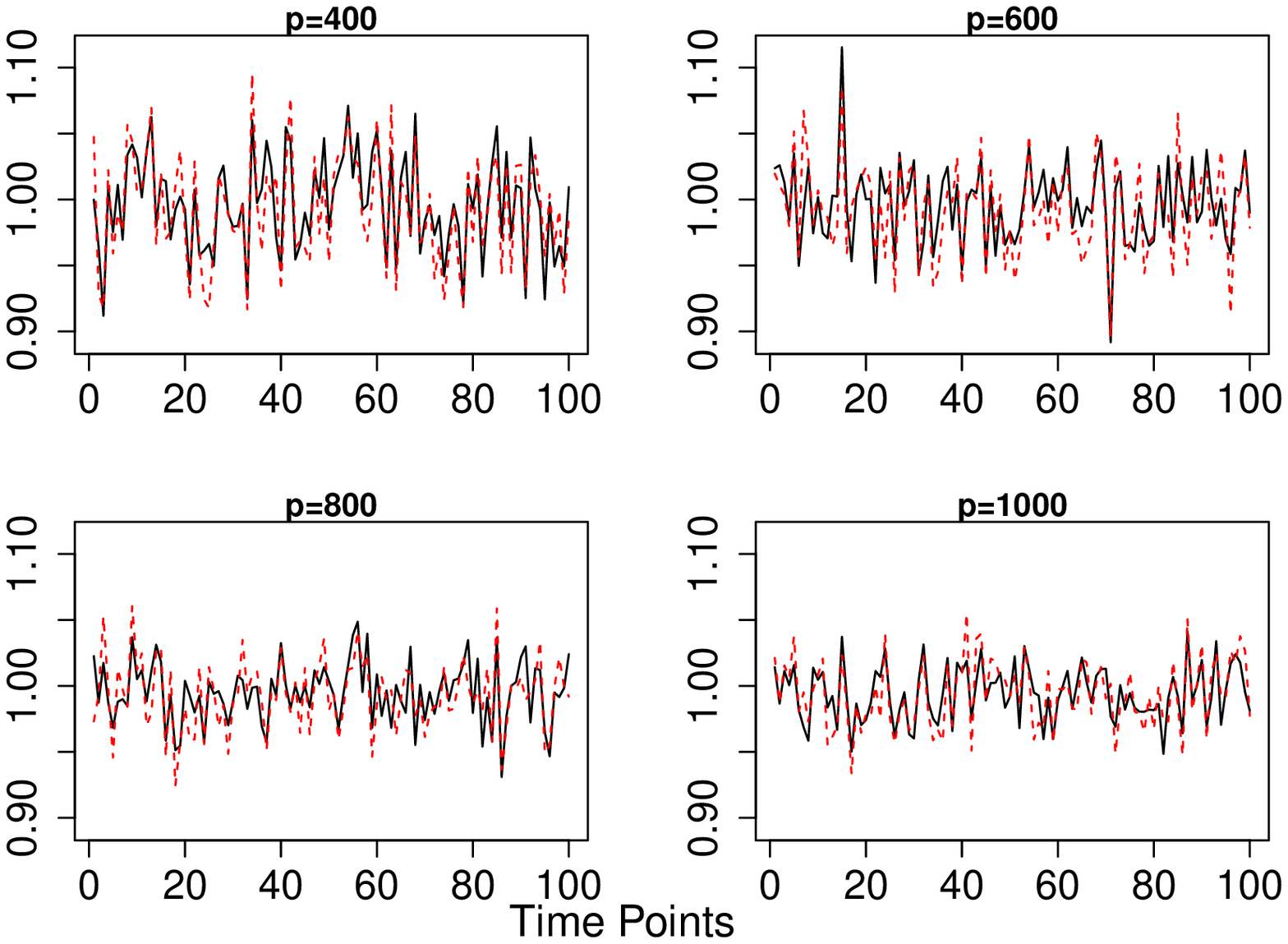}
           \includegraphics[width=.6\textwidth]{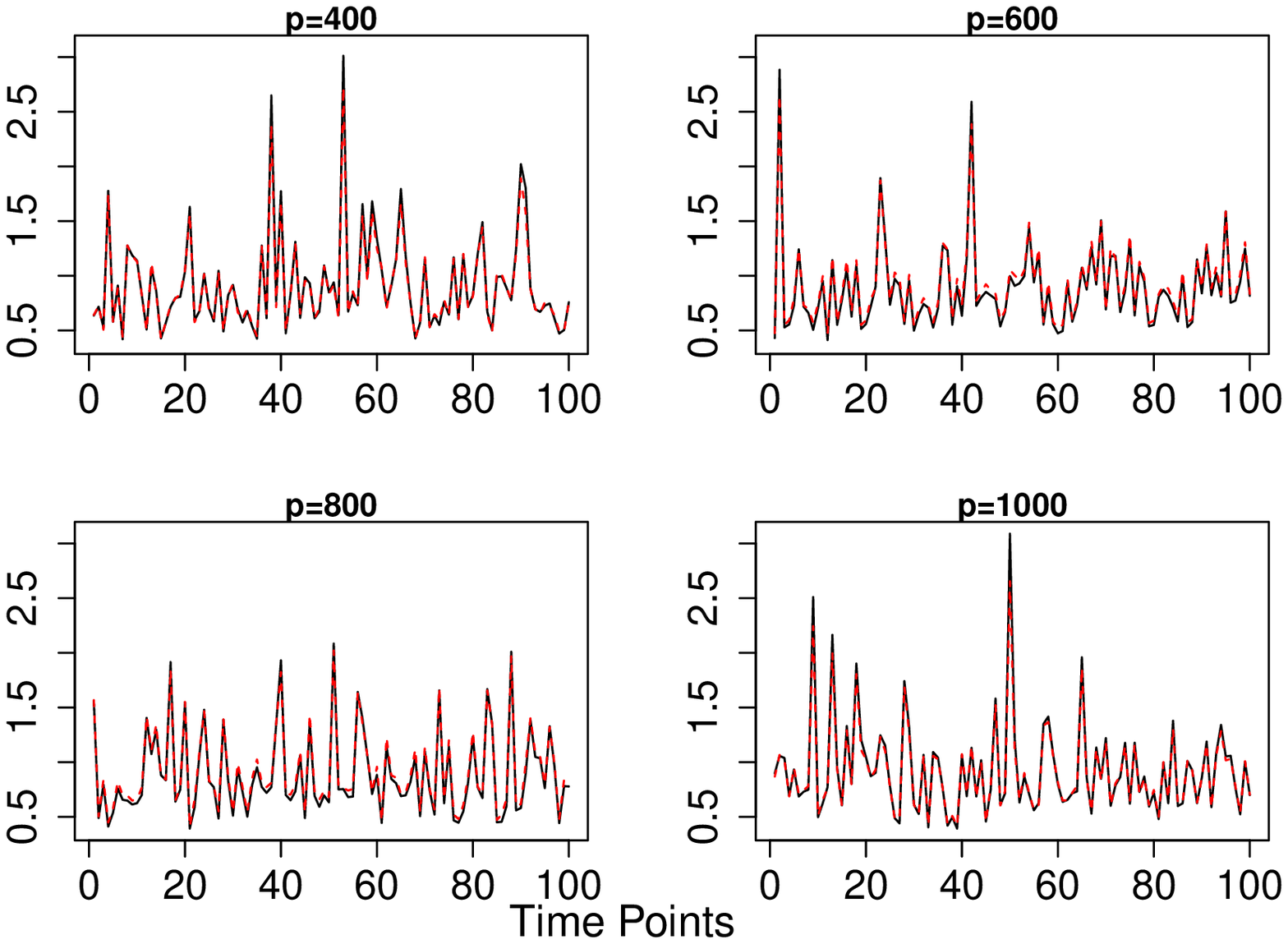}
               \caption{Estimated $\xi_t$ (red broken line) versus true $\xi_t$ (solid black line). The covariance matrix $\bSigma$ is sparse. Top four panels: multivariate Gaussian data. Bottom four panels: multivariate $t$ data.}
                \label{fig:xi_t_Experiment1}
\end{figure}

In the first experiment, we generate $\{\Yb_t\}_{t=1}^T$ iid from model \eqref{Ydecomposition} with a constant covariance matrix $\bSigma$. The covariance is set to be $\Sigma_{ij}=0.3^{|i-j|}$, which is approximately banded. We fix $T=100$ and let $p$ varies. The results are displayed in Figure \ref{fig:xi_t_Experiment1}, where we study both cases of multivariate Gaussian data and multivariate $t_{4.5}$ data. We see that the estimated values are very close to the true values in all the cases.

In the second experiment, we generate data using the calibrated covariance matrix from S\&P500 stock returns as in Section~\ref{sec:simulations}. In this case, the covariance matrix is heavily non-sparse, however, our estimator still works very well, no matter for Gaussian data or heavy-tailed data with multivariate $t$-distributions.

\begin{figure}[!t]
        \centering
\includegraphics[width=.65\textwidth]{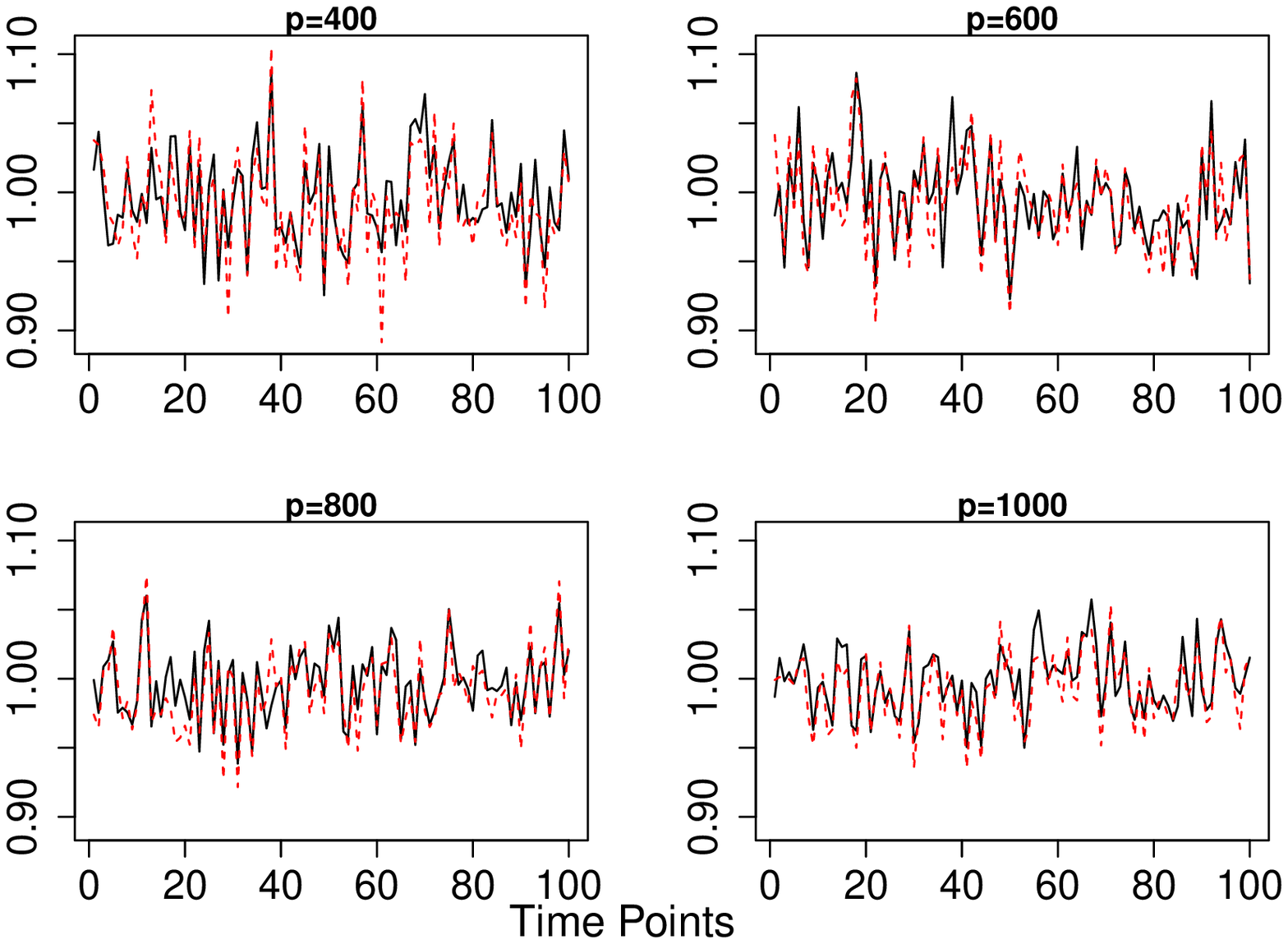}
\includegraphics[width=.65\textwidth]{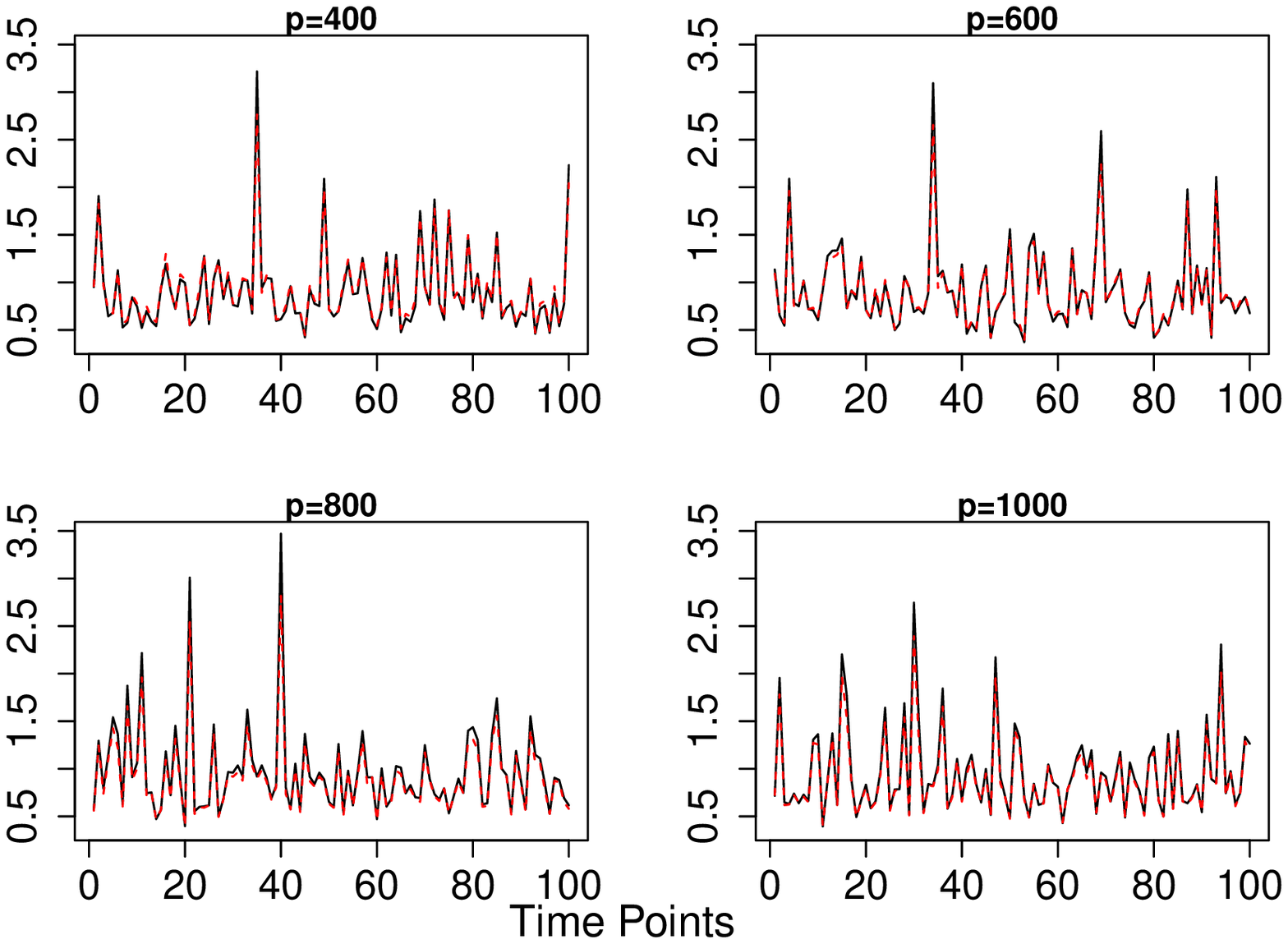}
\caption{Estimated $\xi_t$ (red broken line) versus true $\xi_t$ (solid black line). The covariance matrix $\bSigma$ is calibrated from S\&P stock returns and is dense. Top four panels: multivariate Gaussian data. Bottom four panels: multivariate $t$ data.}
                \label{fig:xi_t_Experiment2}
\end{figure}

\end{document}